\documentclass[10pt,twocolumn, journal]{IEEEtran}
\usepackage{amssymb,amsmath}
\usepackage{subfigure}
\usepackage{stfloats}
\usepackage{graphicx}
\usepackage{multicol}
\usepackage{times}
\usepackage{algorithm}
\usepackage{algorithmic}
\usepackage{array}
\graphicspath{{figures/}}

\usepackage{color}
\def\DLaddition#1{\noindent\ {\color{black} #1}}

\begin{document}
\title{Buffer Sizing for 802.11 Based Networks}
\author{\IEEEauthorblockN{Tianji Li, Douglas Leith, David Malone} \\
\IEEEauthorblockA{ Hamilton Institute, National University of Ireland Maynooth, Ireland} \\
Email: {\{tianji.li, doug.leith, david.malone\}}@nuim.ie
\thanks{This work is supported by Irish Research Council for Science, Engineering and Technology and Science Foundation Ireland Grant 07/IN.1/I901.}}
\maketitle

\begin{abstract}
We consider the sizing of network buffers in 802.11 based networks.   Wireless networks
face a number of fundamental issues that  do not arise in wired networks.  We demonstrate
that the use of fixed size buffers in 802.11 networks inevitably leads to either
undesirable channel under-utilization or unnecessary high delays. We present two novel
dynamic buffer sizing algorithms that achieve high throughput while maintaining low delay
across a wide range of network conditions. Experimental measurements demonstrate the
utility of the proposed algorithms in a production WLAN and a lab testbed.
\end{abstract}


\section{Introduction}\label{sec_intr}

In communication networks, buffers are used to accommodate short-term packet bursts so as
to mitigate packet drops and to maintain high link efficiency. Packets are queued if too
many packets arrive in a sufficiently short interval of time during which a network
device lacks the capacity to process all of them immediately.

For wired routers, the sizing of buffers is an active research topic
(\cite{Vallamizar_CCR_1994} \cite{Appenzeller_SIGCOMM_2004} \cite{Rade_Letters_2006}
\cite{Vu-Brugier_CCR_2007} \cite{Dhamdhere_ccr_2006}). The classical rule of thumb for
sizing wired buffers is to set buffer sizes to be the product of the \emph{bandwidth} and the average \emph{delay} of the flows utilizing this link, namely
the {\em Bandwidth-Delay Product} (BDP) rule \cite{Vallamizar_CCR_1994}. See Section
\ref{sec_related_work} for discussion of other related work.

Surprisingly, however the sizing of buffers in wireless networks (especially those based
on 802.11/802.11e) appears to have received very little attention within the networking
community. Exceptions include the recent work in \cite{Malone_BufferSizing_voip}
relating to buffer sizing for voice traffic in 802.11e \cite{802.11e} WLANs, work in
\cite{Pilosof_INFOCOM_2003} which considers the impact of buffer sizing on TCP
upload/download fairness, and work in \cite{Thottan_wicon_2006} which is related to
802.11e parameter settings.

Buffers play a key role in 802.11/802.11e wireless networks. To illustrate this, we
present measurements from the production WLAN of the Hamilton Institute, which
show that the current state of the art which makes use of fixed size buffers, can easily
lead to poor performance. The topology of this WLAN is shown in Fig. \ref{fig_hi_topo}.
See the Appendix for further details of the configuration used. We recorded RTTs before and after one
wireless station started to download a 37MByte file from a web-site. Before starting the
download, we pinged the access point (AP) from a laptop 5 times, each time sending 100 ping packets. The RTTs reported by the ping program was between 2.6-3.2 ms.  However, after starting the
download and allowing it to continue for a while (to let the congestion control algorithm
of TCP probe for the available bandwidth), the RTTs to the AP hugely increased to
\textbf{2900-3400 ms}. During the test, normal services such as web browsing experienced
obvious pauses/lags on wireless stations using the network.  Closer inspection revealed
that the buffer occupancy at the AP exceeded 200 packets most of the time and reached 250
packets from time to time during the test.  Note that the increase in measured RTT could be almost
entirely attributed to the resulting queuing delay at the AP, and indicates that a more sophisticated approach to buffer sizing is required.  Indeed, using the A* algorithm proposed in this
paper, the RTTs observed when repeating the same experiment fall to only \textbf{90-130 ms}. This
reduction in delay does not come at the cost of reduced throughput, i.e., the measured
throughput with the A* algorithm and the default buffers is similar.

In this paper, we consider the sizing of buffers in 802.11/802.11e (\cite{802.11}
\cite{802.11e}) based WLANs. We focus on single-hop WLANs since these are rapidly becoming
ubiquitous as the last hop on home and office networks as well as in so-called ``hot
spots'' in airports and hotels, but note that the proposed schemes can be easily applied
in multi-hop wireless networks. Our main focus in this paper is on TCP traffic since this
continues to constitute the bulk of traffic in modern networks (80--90\%
\cite{Zhao_Ton_2004} of current Internet traffic and also of WLAN traffic
\cite{Tang_MobiCom_2000}), although we extend consideration to UDP traffic at various
points during the discussion and also during our experimental tests.

Compared to sizing buffers in wired routers, a number of fundamental new issues arise
when considering 802.11-based networks.   Firstly, unlike wired networks, wireless transmissions are inherently broadcast in nature which leads to the packet service times at different stations in a WLAN being strongly coupled.   For example, the basic 802.11 DCF ensures that the wireless stations in a WLAN win a roughly equal number of transmission opportunities \cite{David_TON_2006}, hence, the mean packet service time at a station is an order of magnitude longer when 10 other stations are active than when only a single station is active.  Consequently, the buffering requirements at each station would also differ, depending on the number of other active stations in the WLAN.   In addition to variations in the mean service time, the distribution of packet service times is also strongly dependent on the WLAN offered load.  This directly affects the burstiness of transmissions and so buffering requirements (see Section \ref{sec_fixed} for details).   Secondly, wireless stations dynamically adjust the physical transmission rate/modulation used in order to regulate non-congestive channel losses.   This rate adaptation, whereby the transmit rate may change by a factor of 50 or more (e.g. from 1Mbps to 54Mbps in 802.11a/g), may induce large and rapid variations in required buffer sizes.   Thirdly, the ongoing 802.11n standards process proposes to improve throughput efficiency by the use of large frames formed by aggregation of multiple packets (\cite{802.11n} \cite{Li_ton_2009}).  This acts to couple throughput efficiency and buffer sizing in a new way since the latter directly affects the availability of sufficient packets for aggregation into large frames.

It follows from these observations that, amongst other things, there does not exist a fixed buffer size which can be used for sizing buffers in WLANs.   This leads naturally to consideration of dynamic buffer sizing strategies that adapt to changing conditions.   In this paper we demonstrate the major performance costs associated with the use of fixed buffer sizes in 802.11 WLANs (Section \ref{sec_fixed}) and present two novel dynamic buffer sizing algorithms (Sections \ref{sec_eBDP} and \ref{sec_astar}) that achieve significant performance gains.  The stability of the feedback loop induced by the adaptation is analyzed, including when cascaded with the feedback loop created by TCP congestion control action. The proposed dynamic buffer sizing algorithms are computationally cheap and suited to implementation on standard hardware. Indeed, we have implemented the algorithms in both the NS-2 simulator and the Linux MadWifi driver \cite{madwifi}. In this paper, in addition to extensive simulation results we also present experimental measurements demonstrating the utility of the proposed algorithms in a testbed located in office environment and with realistic traffic. This latter includes a mix of TCP and UDP traffic, a mix of uploads and downloads, and a mix of connection sizes.

The remainder of the paper is organized as follows. Section \ref{sec_background}
introduces the background of this work. In Section \ref{sec_fixed} simulation results
with fixed size buffers are reported to further motivate this work. The proposed algorithms are then detailed in Sections \ref{sec_eBDP} and \ref{sec_astar}. Experiment details are presented in Section \ref{sec_expt}. After introducing related work in Section \ref{sec_related_work}, we summarize our conclusions
in Section \ref{sec_concl}.

\section{Preliminaries}\label{sec_background}

\subsection{IEEE 802.11 DCF}
IEEE 802.11a/b/g WLANs all share a common MAC algorithm called the Distributed Coordinated Function (DCF) which is a CSMA/CA based algorithm.   On detecting the wireless medium to be idle
for a period $DIFS$, each wireless station initializes a backoff counter to a random
number selected uniformly from the interval [0, CW-1] where CW is the contention window.
Time is slotted and the backoff counter is decremented each slot that the medium is idle.
An important feature is that the countdown halts when the medium is detected busy and
only resumes after the medium is idle again for a period $DIFS$.  On the counter reaching
zero, a station transmits a packet.  If a collision occurs (two or more stations transmit
simultaneously), CW is doubled and the process repeated. On a successful transmission, CW
is reset to the value $CW_{min}$ and a new countdown starts.

\subsection{IEEE 802.11e EDCA}
The 802.11e standard extends the DCF algorithm (yielding the EDCA) by allowing the
adjustment of MAC parameters that were previously fixed.    In particular, the values of
$DIFS$ (called $AIFS$ in 802.11e) and $CW_{min}$ may be set on a per class basis for each
station.    While the full 802.11e standard is not implemented in current commodity
hardware, the EDCA extensions have been widely implemented for some years.

\subsection{Unfairness among TCP Flow}\label{subsec_tcp_unfairness}
Consider a WLAN consisting of $n$ client stations each carrying one TCP upload flow.  The TCP ACKs are transmitted by the wireless AP.  In this case TCP ACK packets can be easily queued/dropped due to the fact that the basic 802.11 DCF ensures that stations win a roughly equal number of transmission opportunities.   Namely, while the data packets for the $n$ flows have an aggregate $n/(n+1)$ share of the transmission opportunities the TCP ACKs for the $n$ flows have only a $1/(n+1)$ share.   Issues of this sort are known to lead to significant unfairness amongst TCP flows but can be readily resolved using 802.11e functionality by treating TCP ACKs as a separate traffic class which is assigned higher priority \cite{Leith_CommLetter_2005}.   With regard to throughput efficiency, the algorithms in this paper perform similarly when the DCF is used and when TCP ACKs are prioritized using the EDCA as in \cite{Leith_CommLetter_2005}.   Per flow behavior does, of course, differ due to the inherent unfairness in the DCF and we therefore mainly present results using the EDCA to avoid flow-level unfairness.

\begin{table}[tb]
\centering
                    \begin{tabular}[b]{|l|l|}%
                        \hline
                        $T_{SIFS}$ ($\mu s) $  & 10  \\
                        \hline
                        Idle slot duration ($\sigma$) ($\mu s$) & 9  \\
                        \hline
                        Retry limit  & 11 \\
                        \hline
                        Packet size (bytes)    & 1000 \\
                        \hline
                        PHY data rate (Mbps)   & 54 \\
                        \hline
                        PHY basic rate (Mbps)  & 6 \\
                        \hline
                        PLCP rate (Mbps)  & 6 \\
                        \hline
                    \end{tabular}
                \caption{MAC/PHY parameters used in simulations, corresponding to 802.11g.}
                \label{parameters_buffersizing}
\end{table}

\subsection{Simulation Topology}
\begin{figure}[tb]
    \centering
    \includegraphics[width=0.7\columnwidth]{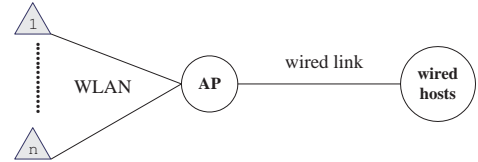}
    \caption{WLAN topology used in simulations. Wired backhaul link bandwidth 100Mbps.
    MAC parameters of the WLAN are listed in Table \ref{parameters_buffersizing}.}
    \label{fig_topo_simu}
\end{figure}
In Sections \ref{sec_fixed}, \ref{sec_eBDP} and \ref{subsec_performance}, we use the simulation
topology shown in Fig. \ref{fig_topo_simu} where the AP acts as a wireless
router between the WLAN and the Internet. Upload flows originate from stations in the
WLAN on the left and are destined to wired host(s) in the wired network on the right.
Download flows are from the wired host(s) to stations in the WLAN. We ignore differences
in wired bandwidth and delay from the AP to the wired hosts which can cause TCP
unfairness issues on the wired side (an orthogonal issue) by using the same wired-part
RTT for all flows.  Unless otherwise stated, we use the IEEE 802.11g PHY parameters shown in Table
\ref{parameters_buffersizing} and the wired backhaul link bandwidth is 100Mbps with RTT 200ms.  For TCP traffic, the widely deployed TCP Reno with SACK extension is used.  The advertised window size is set to be 4096 packets (each has a payload of 1000 bytes) which is the default size of current Linux kernels. The maximum value of the TCP smoothed RTT measurements (sRTT) is used as the measure of the delay experienced by a flow.

\section{Motivation and Objectives}\label{sec_fixed}

\begin{figure}[tb]
	\centering
	\subfigure[2 stations]{\includegraphics[width=0.47\columnwidth]{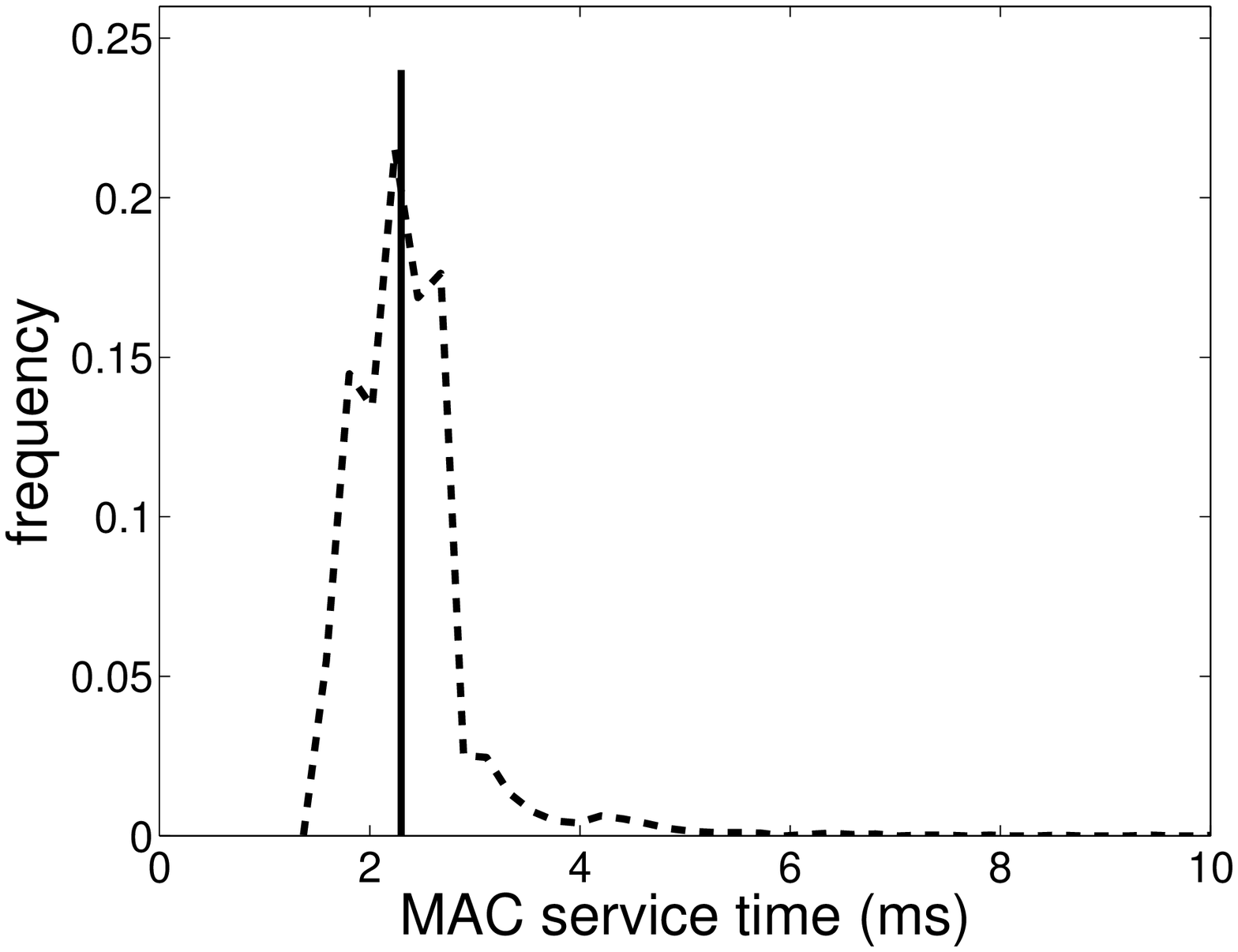}}
	\subfigure[12 stations]{\includegraphics[width=0.47\columnwidth]{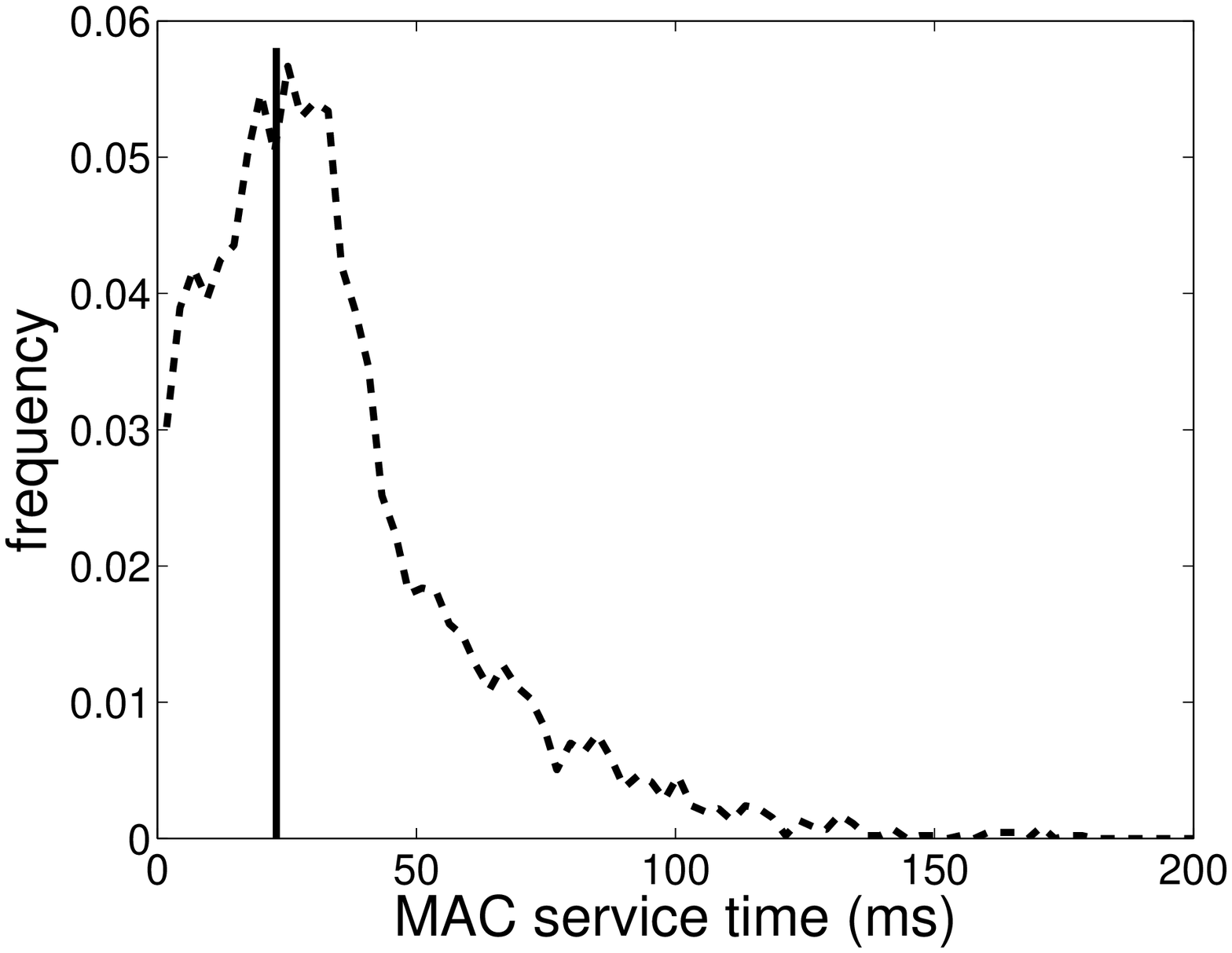}}
	\caption{Measured distribution of per packet MAC service time. Solid vertical lines mark the mean values of distributions. Physical layer data/basic rates are 11/1 Mbps. }\label{fig:802.11_0}
\end{figure}

\begin{figure*}[tb]
   \centering
   \subfigure[1/1Mbps, throughput]{\includegraphics[width=.45\columnwidth]{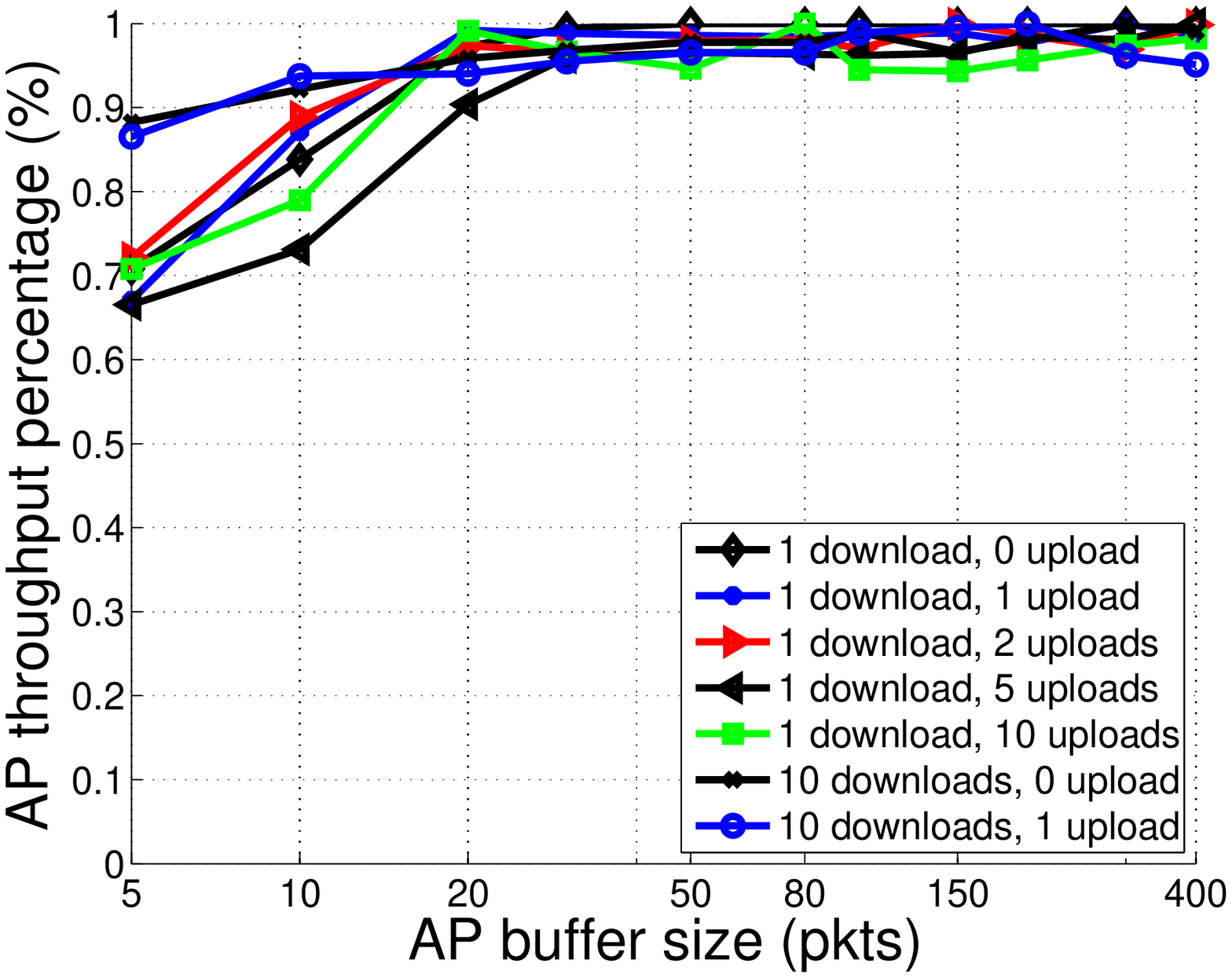}
              \label{fig_motivation_thru_1}}
   \subfigure[1/1Mbps, delay]{\includegraphics[width=.47\columnwidth]{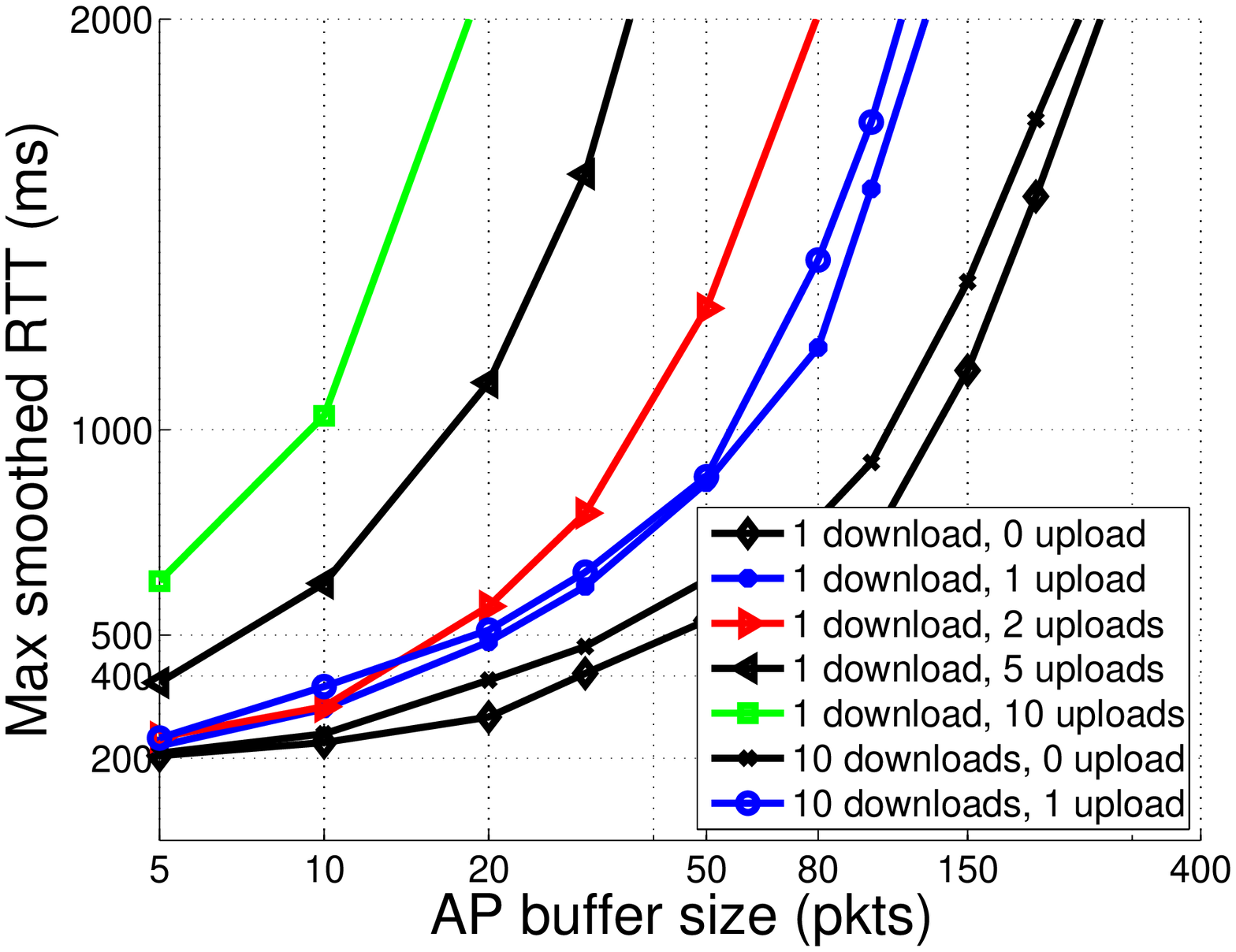}
              \label{fig_motivation_srtt_1}}
   \subfigure[11/1Mbps, throughput]{\includegraphics[width=.45\columnwidth]{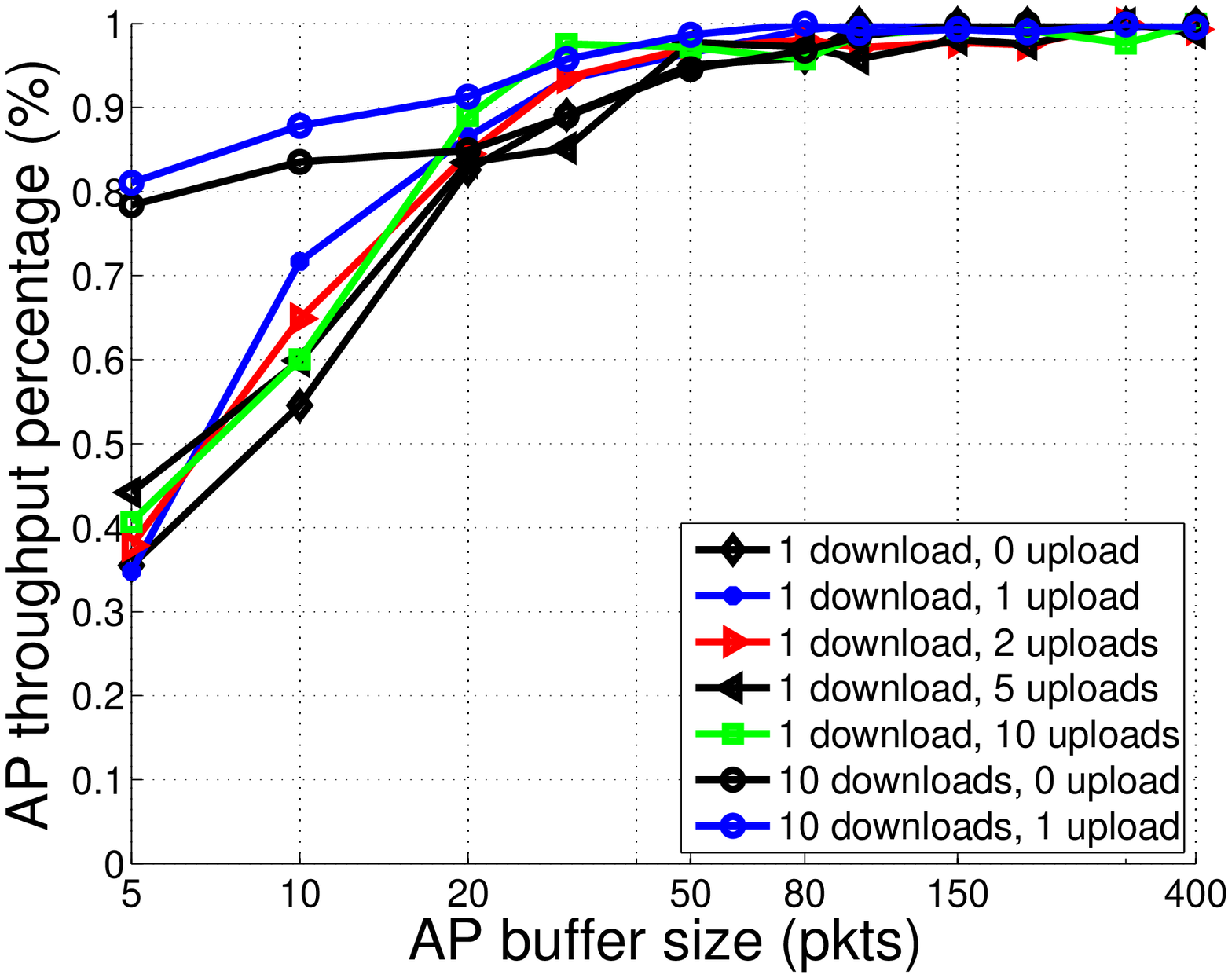}
              \label{fig_motivation_thru_11}}
   \subfigure[11/1Mbps, delay]{\includegraphics[width=.47\columnwidth]{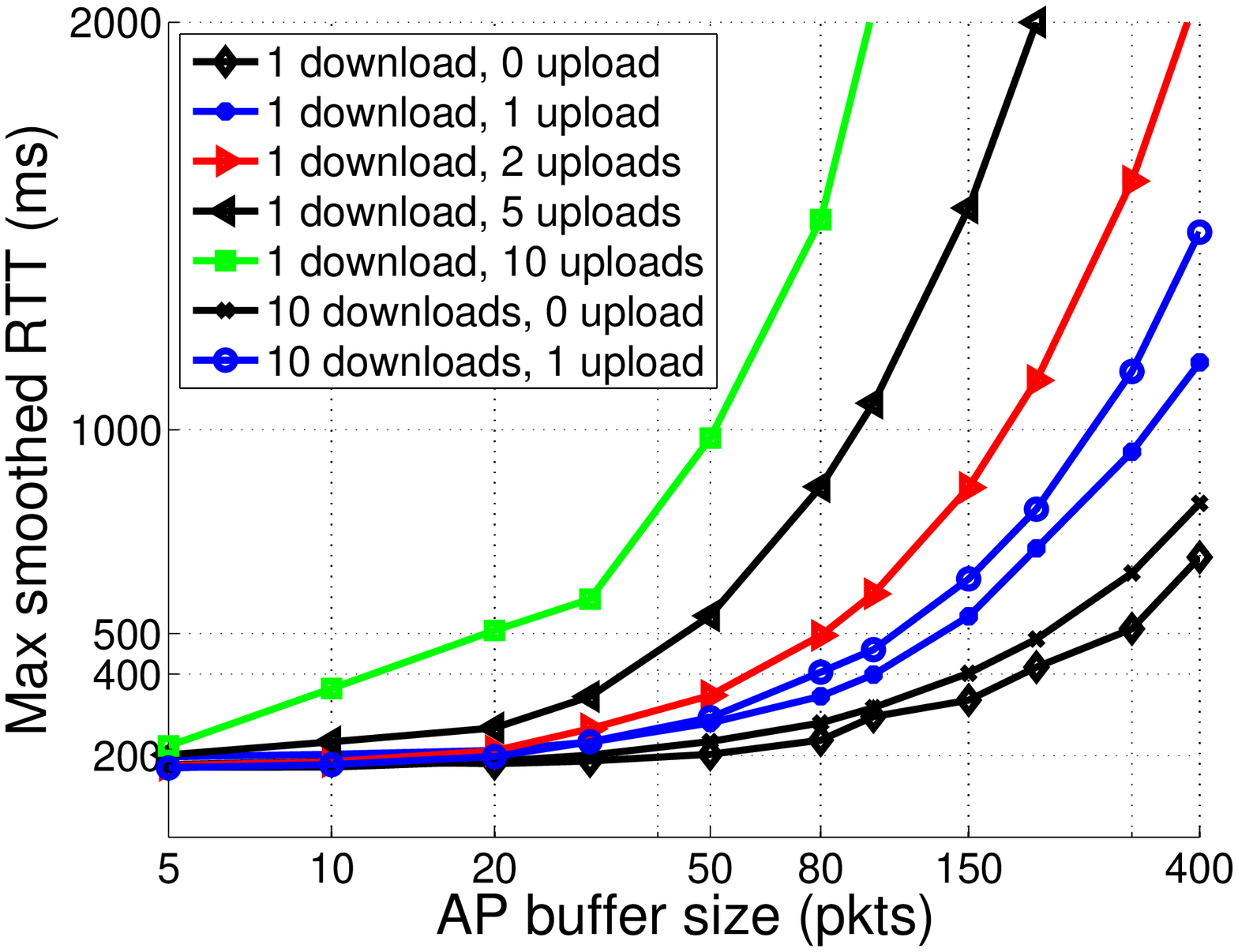}
              \label{fig_motivation_srtt_11}}
   \subfigure[54/6Mbps, throughput]{\includegraphics[width=.45\columnwidth]{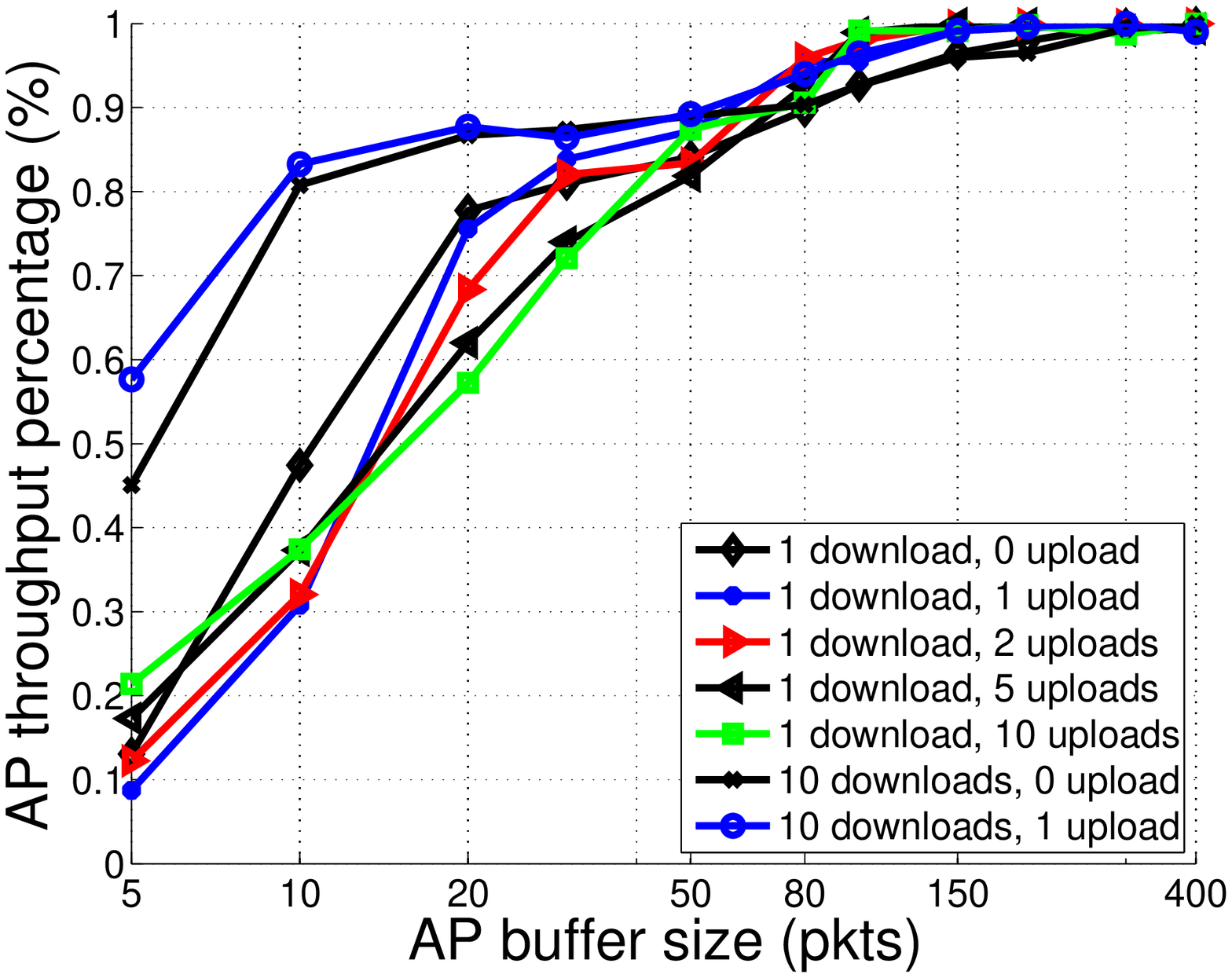}
              \label{fig_motivation_thru_54}}
   \subfigure[54/6Mbps, delay]{\includegraphics[width=.47\columnwidth]{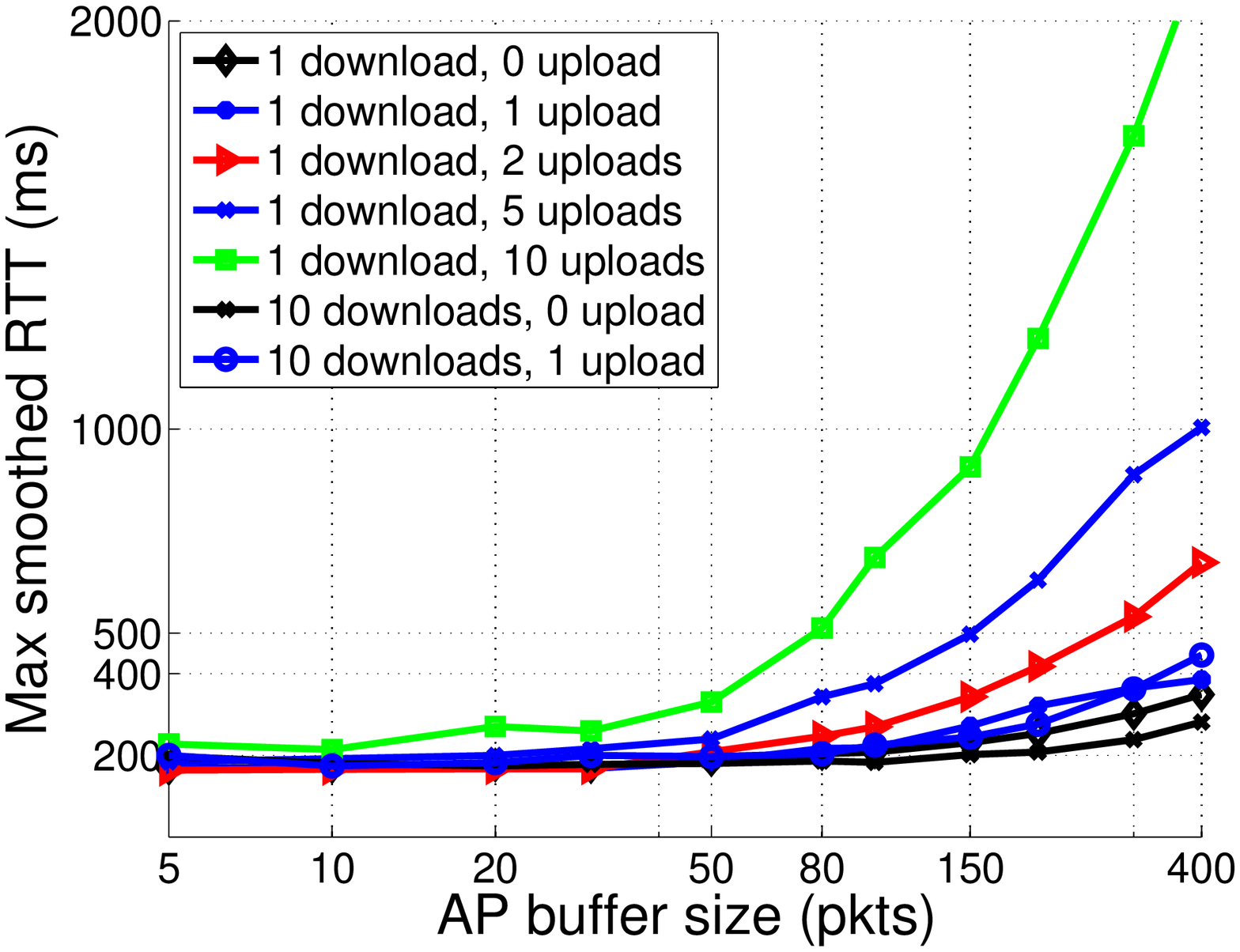}
              \label{fig_motivation_srtt_54}}
   \subfigure[216/54Mbps, throughput]{\includegraphics[width=.45\columnwidth]{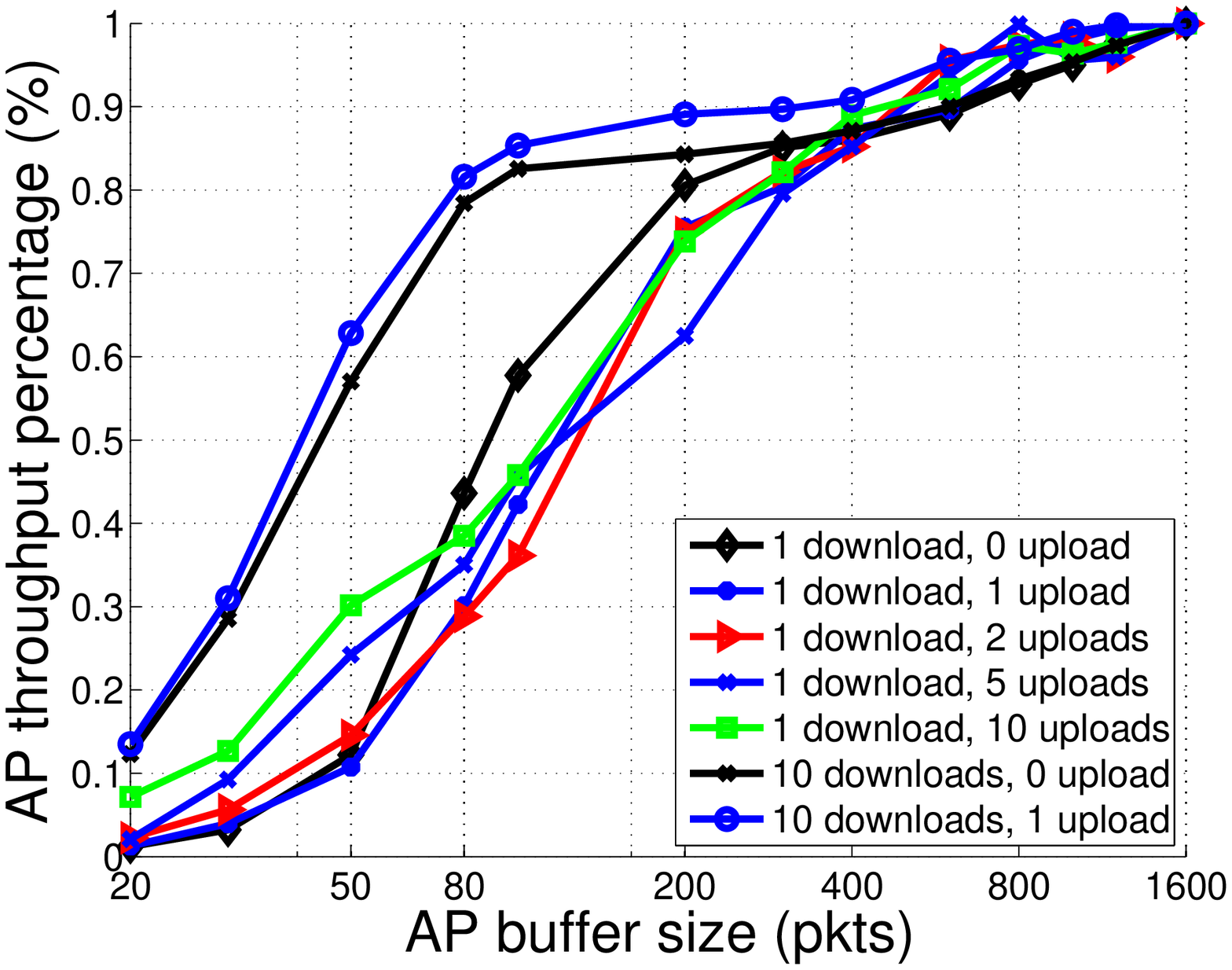}
              \label{fig_motivation_thru_216}}
   \subfigure[216/54Mbps, delay]{\includegraphics[width=.47\columnwidth]{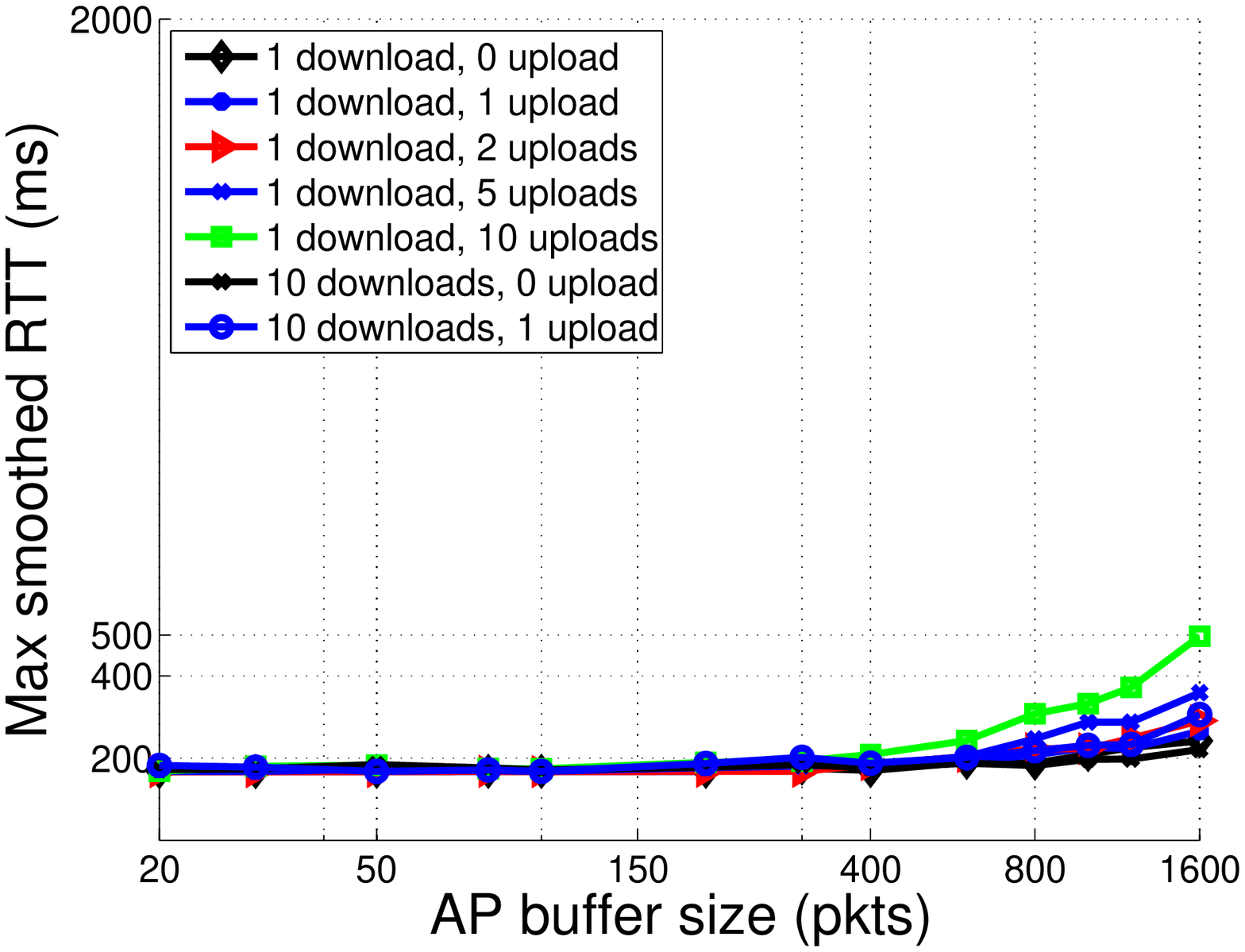}
              \label{fig_motivation_srtt_216}}
   \caption{Throughput efficiency and maximum smoothed round trip delays (max sRTT) for the topology in Fig. \ref{fig_topo_simu} when fixed size buffers are used. Here, the AP throughput efficiency is the ratio between the\DLaddition{download} throughput achieved using buffer sizes indicated on the x-axis and the maximum\DLaddition{download} throughput achieved using fixed size buffers.  Rates before and after the '/' are used physical layer data and basic rates. For the 216Mbps data, 8 packets are aggregated into each frame at the MAC layer to improve throughput efficiency in an 802.11n-like scheme.  The wired RTT is 200 ms. }
   \label{fig_motivation}
\end{figure*}

Wireless communication in 802.11 networks is \textit{time-varying} in nature, i.e., the mean service time and the distribution of service time at a wireless station vary in time.  The variations are primarily due to (i) changes in the number of active wireless stations and their load (i.e. offered load on the WLAN) and (ii) changes in the physical transmit rate used (i.e. in response to changing radio channel conditions). In the latter case, it is straightforward to see that the service time can be easily increased/decreased using low/high physical layer rates. To see the impact of offered load on the service time at a station, Fig. \ref{fig:802.11_0} plots the measured distribution of the MAC layer service time when there are 2 and 12 stations active. It can be seen that the mean service time changes by over an order of magnitude as the number of stations varies. Observe also from these measured distributions that there are significant fluctuations in the service time for a given fixed load. This is a direct consequence of the stochastic nature of the CSMA/CA contention mechanism used by the 802.11/802.11e MAC.

This time-varing nature directly affects buffering requirements. Figure \ref{fig_motivation} plots link utilization\footnote{Here the AP throughput percentage is the ratio between the actual throughput achieved using buffer sizes show on the x-axis and the maximum throughput using the buffer sizes shown on the x-axis.} and max sRTT (propagation plus smoothed queuing delay) vs buffer size for a range of WLAN offered loads and physical transmit rates.   We can make a number of observations.

First, it can be seen that as the physical layer transmit rate is varied from 1Mbps to 216Mbps, the minimum buffer size to ensure at least 90\% throughput efficiency varies from about 20 packets to about 800 packets.   No compromise buffer size exists that ensures both high efficiency and low delay across this range of transmit rates.  For example, a buffer size of 80 packets leads to RTTs exceeding 500ms (even when only a single station is active and so there are no competing wireless stations) at 1Mbps and throughput efficiency below 50\% at 216Mbps.   Note that the transmit rates in currently available draft 802.11n equipment already exceed 216Mbps (e.g. 300Mbps is supported by current Atheros chipsets) and the trend is towards still higher transmit rates.   Even across the restricted range of transmit rates 1Mbps to 54Mbps supported by 802.11a/b/g, it can be seen that a buffer size of 50 packets is required to ensure throughput efficiency above 80\% yet this buffer size induces delays exceeding 1000 and 3000 ms at transmit rates of 11 and 1 Mbps, respectively.

Second, delay is strongly dependent on the traffic load and the physical rates.   For example, as the number of competing stations (marked as ``uploads'' in the figure) is varied from 0 to 10, for a buffer size of 20 packets and physical transmit rate of 1Mbps the delay varies from 300ms to over 2000ms.   This reflects that the 802.11 MAC allocates available transmission opportunities equally on average amongst the wireless stations, and so the mean service time (and thus delay) increases with the number of stations.    In contrast, at 216Mbps the delay remains below 500ms for buffer sizes up to 1600 packets.

Our key conclusion from these observations is that there exists no fixed buffer size capable of ensuring both high throughput efficiency and reasonable delay across the range of physical rates and offered loads experienced by modern WLANs.   Any fixed choice of buffer size necessarily carries the cost of significantly reduced throughput efficiency and/or excessive queuing delays.

This leads naturally therefore to the consideration of adaptive approaches to buffer sizing, which dynamically adjust the buffer size in response to changing network conditions to ensure both high utilization of the wireless link while avoiding unnecessarily long queuing delays.

\section{Emulating BDP}\label{sec_eBDP}

We begin by considering a simple adaptive algorithm based on the classical BDP rule.   Although this algorithm cannot take advantage of statistical multiplexing opportunities, it is of interest both for its simplicity and because it will play a role in the more sophisticated $A^*$ algorithm developed in the next section.

As noted previously, and in contrast to wired networks,  in 802.11 WLANs the mean service time is generally time-varying (dependent on WLAN load and the physical transmit rate selected by a station). Consequently, there does not exist a fixed BDP value.   However, we note that a wireless station can measure its own packet service times by direct observation, i.e., by recording the time between a packet arriving at the head of the network interface queue $t_s$ and being successfully transmitted $t_e$ (which is indicated by receiving correctly the corresponding MAC ACK).  Note that this measurement can be readily implemented in real devices, e.g. by asking the hardware to raise an interrupt on receipt of a MAC ACK, and incurs only a minor computational burden.    Averaging these per packet service times yields the mean service time $T_{serv}$.   To accommodate the time-varying nature of the mean service time, this average can be taken over a sliding window.  In this paper, we consider the use of exponential smoothing $T_{serv}(k+1) = (1-W) T_{serv}(k) + W(t_e-t_s)$ to calculate a running average since this has the merit of simplicity and statistical robustness (by central limit arguments).  The choice of smoothing parameter $W$ involves a trade-off between accommodating time variations and ensuring the accuracy of the estimate -- this choice is considered in detail later.

Given an online measurement of the mean service time $T_{serv}$, the classical BDP rule yields the following eBDP buffer sizing strategy. Let $T_{max}$ be the target maximum queuing delay. Noting that $1/T_{serv}$ is the mean service rate, we select buffer size $Q_{eBDP}$ according to
$Q_{eBDP}=min(T_{max}/T_{serv},Q_{max}^{eBDP})$ where $Q_{max}^{eBDP}$ is the upper limit
on buffer size. This effectively regulates the buffer size to equal the current mean BDP.
The buffer size decreases when the service rate falls and increases when the service rate
rises, so as to maintain an approximately constant queuing delay of $T_{max}$ seconds.
We may measure the flows' RTTs to derive the value for $T_{max}$ in a similar way to measuring the mean service rate, but in the examples presented here we simply use a fixed value of 200ms since this is an approximate upper bound on the RTT of the majority of the current Internet flows.

We note that the classical BDP rule is derived from the behavior of TCP congestion control (in particular, the reduction of cwnd by half on packet loss) and assumes a constant service rate and fluid-like packet arrivals.  Hence, for example, at low service rates the BDP rule suggests use of extremely small buffer sizes.   However, in addition to accommodating TCP behavior, buffers have the additional role of absorbing short-term packet bursts and, in the case of wireless links, short-term fluctuations in packet service times.  It is these latter effects that lead to the steep drop-off in throughput efficiency that can be observed in Fig. \ref{fig_motivation}  when there are competing uploads (and so stochastic variations in packet service times due to channel contention, see Fig. \ref{fig:802.11_0}.) plus small buffer sizes.   We therefore modify the eBDP update rule to $Q_{eBDP}=min(T_{max}/T_{serv}+c,Q_{max}^{eBDP})$
where $c$ is an over-provisioning amount to accommodate short-term fluctuations in
service rate.  Due to the complex nature of the service time process at a wireless station (which is coupled to the traffic arrivals etc at other stations in the WLAN) and of the TCP traffic arrival process (where feedback creates coupling to the service time process), obtaining an analytic value for $c$ is intractable.  Instead, based on the measurements in Fig. \ref{fig_motivation} and others, we have found empirically that a value of $c=5$ packets works well across a wide range of network conditions. Pseudo-code for eBDP is shown in Algorithms \ref{algo_adt} and \ref{algo_mac}.


\begin{algorithm}[tb]
    \begin{algorithmic}[1]
        \STATE Set the target queuing delay $T_{max}$.
        \STATE Set the over-provision parameter $c$.
        \FOR {each incoming packet $p$}
            \STATE Calculate  $Q_{eBDP}=min(T_{max}/T_{serv}+c,Q_{max}^{eBDP})$
            where $T_{serv}$ is from MAC Algorithm \ref{algo_mac}.
            \IF { current queue occupancy $ < Q_{eBDP}$ }
                \STATE Put $p$ into queue
            \ELSE
                \STATE Drop $p$.
            \ENDIF
        \ENDFOR
    \end{algorithmic}
    \caption{Drop tail operation of the eBDP algorithm.}
    \label{algo_adt}
\end{algorithm}

\begin{algorithm}[tb]
    \begin{algorithmic}[1]
        \STATE Set the averaging parameter $W$.
        \FOR {each outgoing packet $p$}
            \STATE Record service start time $t_s$ for $p$.
            \STATE Wait until receive MAC ACK for $p$, record service end time $t_e$.
            \STATE Calculate service time of $p$: $T_{serv} = (1-W) T_{serv} + W(t_e-t_s) $.
        \ENDFOR
    \end{algorithmic}
    \caption{MAC operation of the eBDP algorithm.}
    \label{algo_mac}
\end{algorithm}

\begin{figure}[tb]
   \centering
   \subfigure[1 download, 0 upload]{\includegraphics[width=0.48\columnwidth]{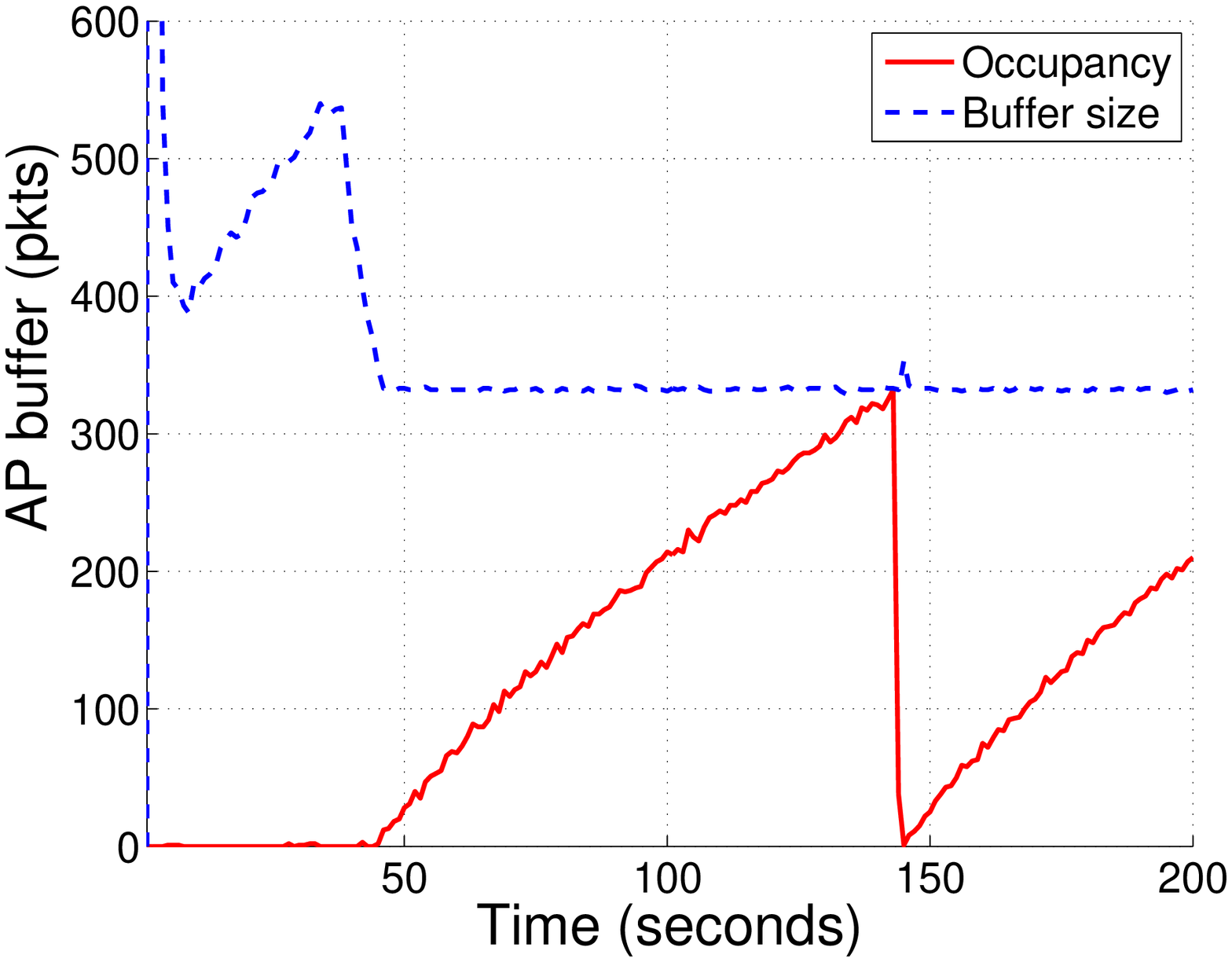}
  \label{fig_eBDP_histories_dl1}}
   \subfigure[1 download, 10 uploads]{\includegraphics[width=0.48\columnwidth]{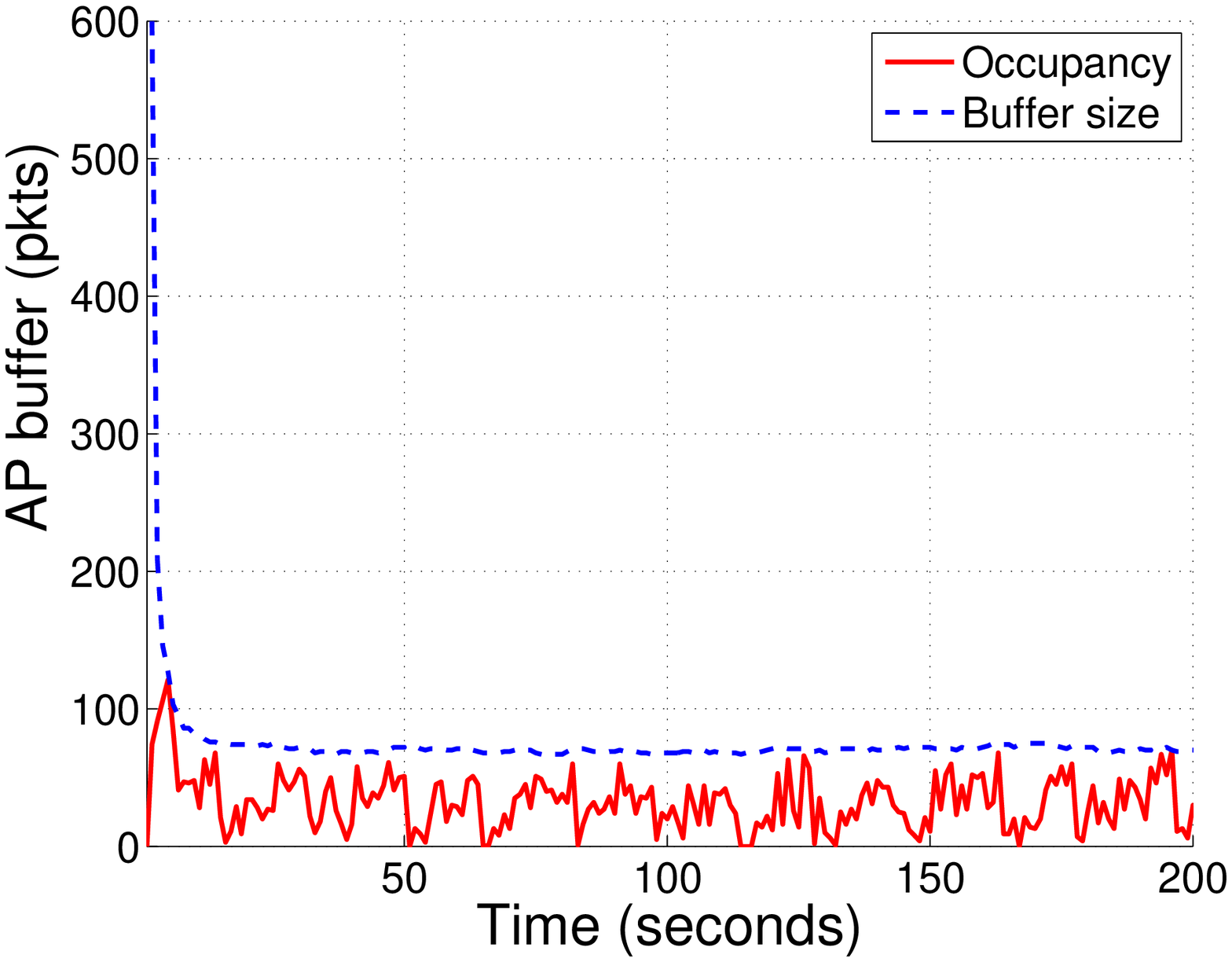}}
   \caption{Histories of buffer size and buffer occupancy with the eBDP algorithm.
   In (a) there is one download and no upload flows. In (b) there are 1 download and 10 upload flows. 54/6Mbps physical data/basic rates. }
   \label{fig_eBDP_histories}
\end{figure}

The effectiveness of this simple adaptive algorithm is illustrated in Fig.
\ref{fig_eBDP_histories}.  Fig. \ref{fig_eBDP_histories}(a) shows the buffer size and queue occupancy time histories when only a single station is active in a WLAN while Fig. \ref{fig_eBDP_histories}(b) shows the corresponding results when ten additional stations now also contend for channel access.    Comparing with Fig. \ref{fig_motivation_thru_54}, it can be seen that buffer sizes of 330 packets and 70 packets, respectively, are needed to yield 100\% throughput efficiency and eBDP selects buffer sizes which are in good agreement with these thresholds.

\begin{figure}[tb]
   \centering
   \subfigure[Throughput]{\includegraphics[width=0.48\columnwidth]{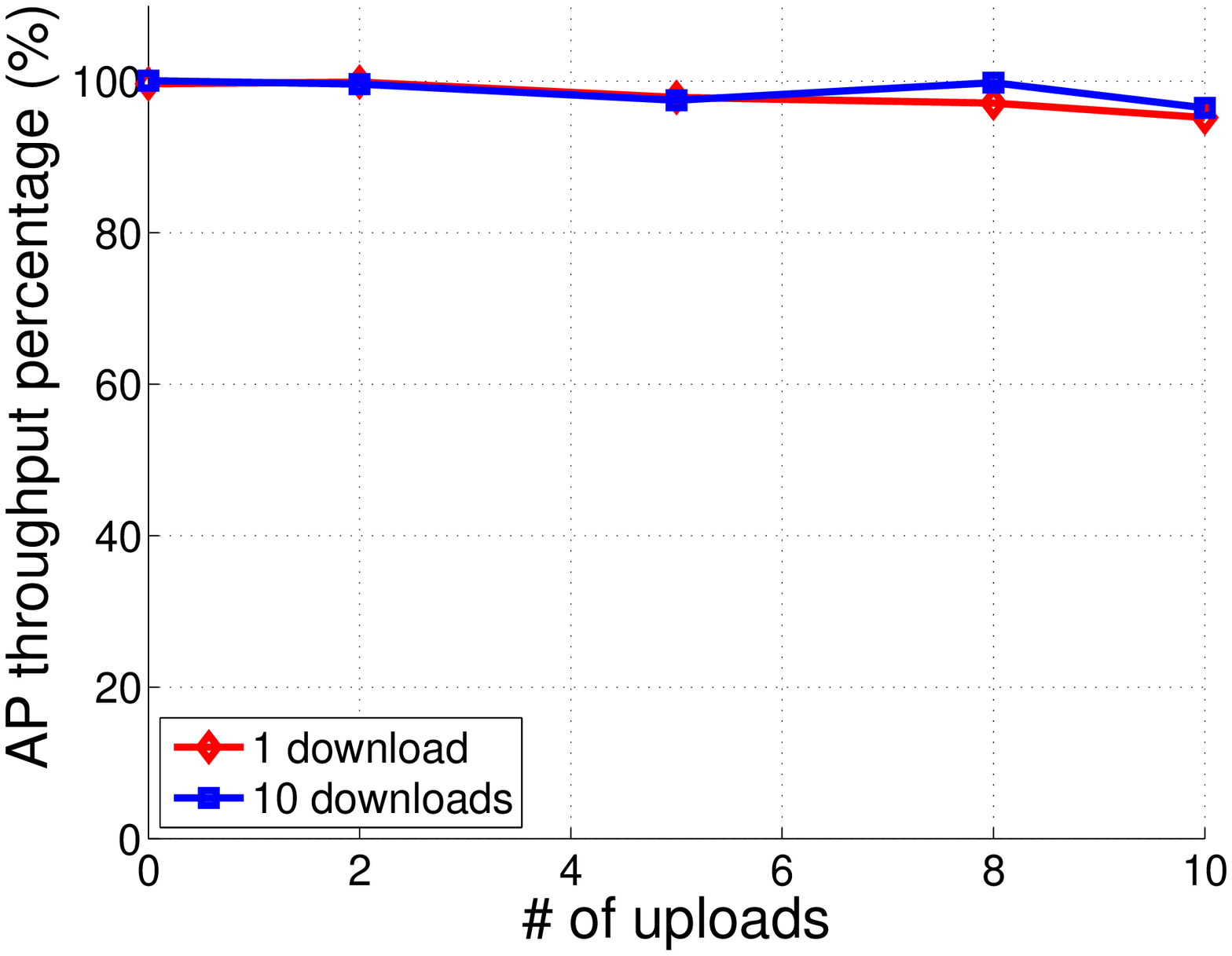}}
   \subfigure[Delay]{\includegraphics[width=0.48\columnwidth]{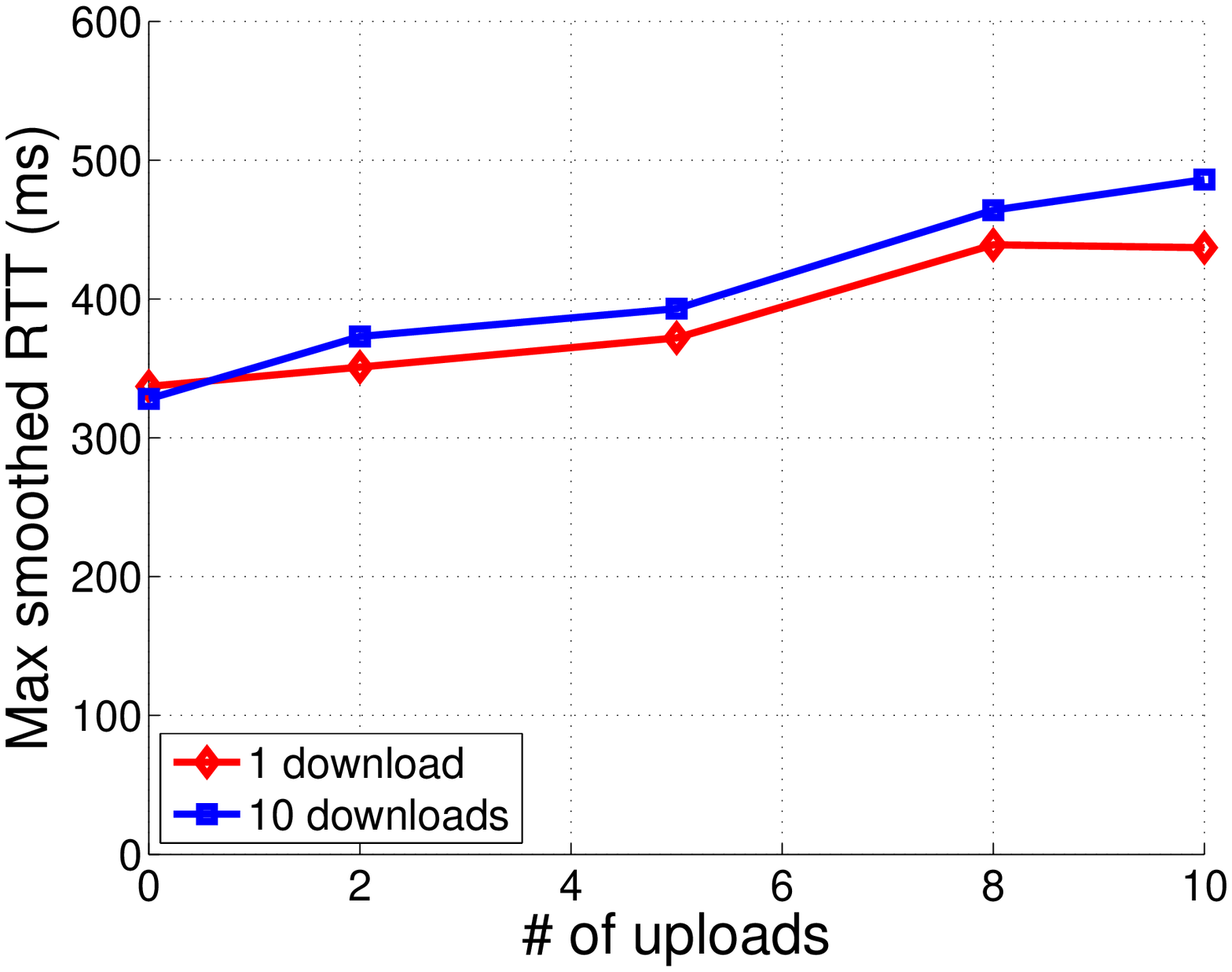}}
   \caption{Performance of the eBDP algorithm as the number of upload
    flows is varied. Data is shown for 1, 10 download flows
   and 0, 2, 5, 10 uploads.  Wired RTT 200ms.  Here the AP throughput percentage is the ratio between the throughput achieved using the eBDP algorithm and that by a fixed buffer size of 400 packets (i.e. the maximum achievable throughput in this case). }
   \label{fig_eBDP_varyuls}
\end{figure}

In Fig. \ref{fig_eBDP_varyuls} we plot the throughput efficiency (measured as the ratio of the achieved throughput to that with a fixed 400-packet buffer) and max smoothed RTT over a range of network conditions obtained using the eBDP algorithm.  It can be seen that the adaptive algorithm maintains high throughput efficiency across the entire range of operating conditions. This is achieved while maintaining the latency approximately constant at around 400ms (200ms propagation delay plus $T_{max}=200$ms queuing delay) -- the latency rises slightly with the number of uploads due to the over-provisioning parameter $c$ used to accommodate
stochastic fluctuations in service rate.

\begin{figure}[tb]
   \centering
   \subfigure[Throughput]{\includegraphics[width=0.48\columnwidth]{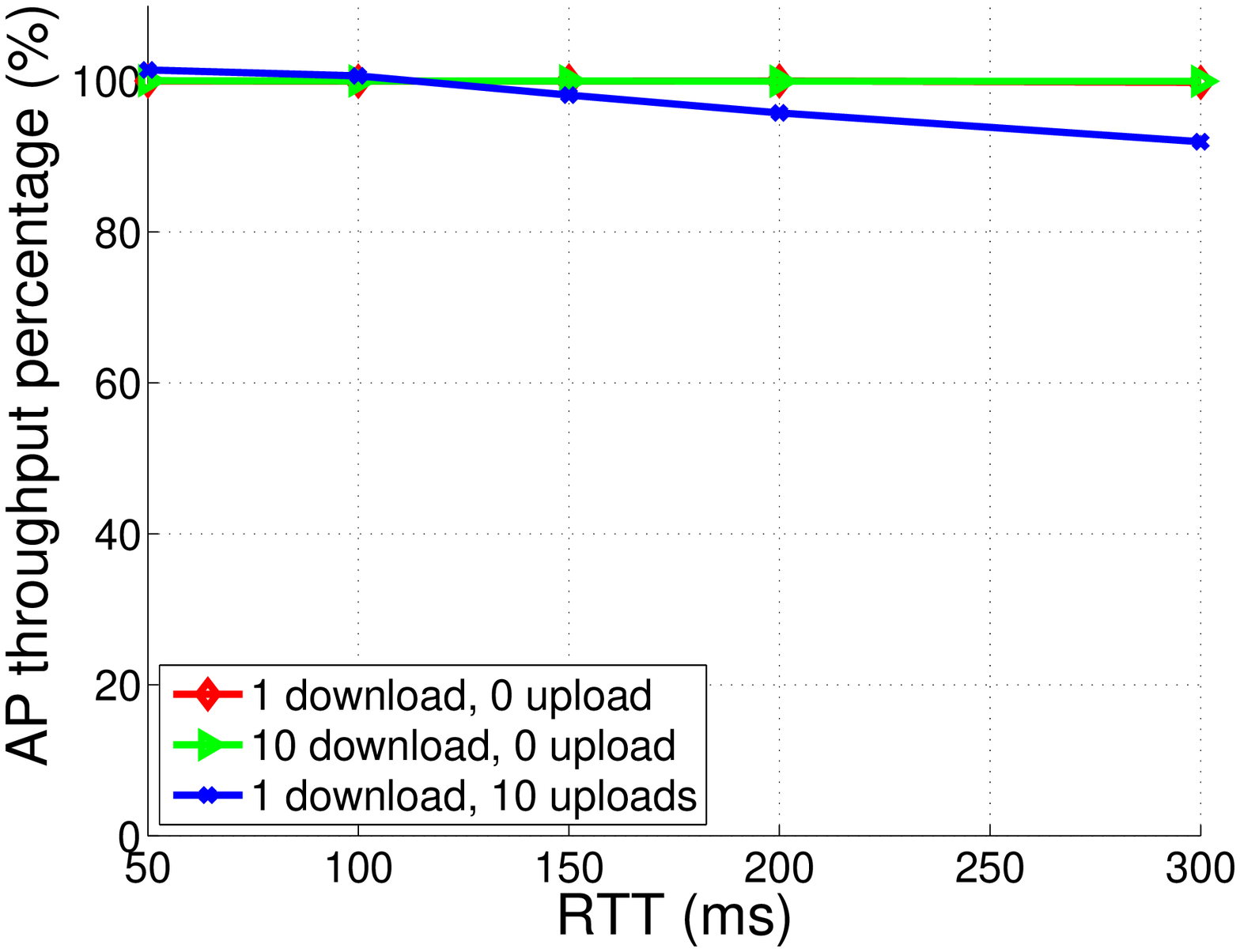}}
   \subfigure[Delay]{\includegraphics[width=0.48\columnwidth]{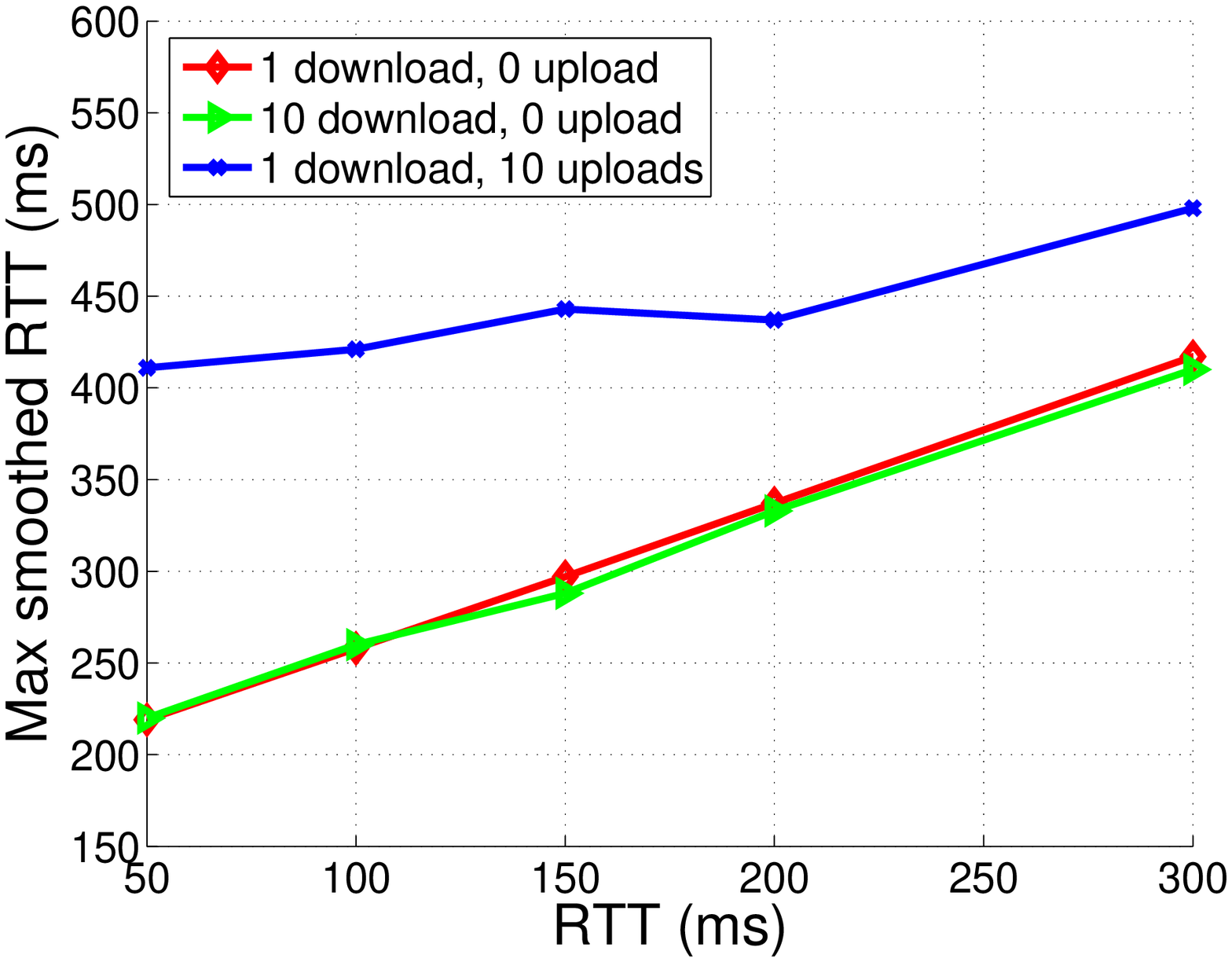}}
   \caption{Performance of the eBDP algorithm as the RTT of wired backhaul is varied.
   Data is shown for 1, 10 downloads and 0, 10 uploads.    Here the AP throughput percentage is the ratio between the throughput achieved using the eBDP algorithm and that by a fixed buffer size of 400 packets (i.e. the maximum achievable throughput in this case).}
   \label{fig_eBDP_varyrtt}
\end{figure}

While $T_{max}=200$ms is used as the target drain time in the eBDP algorithm, realistic
traffic tends to consist of flows with a mix of RTTs. Fig. \ref{fig_eBDP_varyrtt} plots
the results as we vary the RTT of the wired backhaul link while keeping $T_{max}=200$ms.
We observe that the throughput efficiency is close to 100\% for RTTs up to 200ms. For an RTT of 300ms, we observe a slight decrease in throughput when there is 1 download and 10
contending upload flows, which is to be expected since $T_{max}$ is less than the link
delay and so the buffer is less than the BDP. This could improved by measuring the
average RTT instead of using a fixed value, but it is not clear that the benefit is worth
the extra effort. We also observe that there is a difference between the max smoothed RTT
with and without upload flows. The RTT in our setup consists of the wired link RTT, the
queuing delays for TCP data and ACK packets and the MAC layer transmission delays for
TCP data and ACK packets. When there are no upload flows, TCP ACK packets can be
transmitted with negligible queuing delays since they only have to contend with the AP.
When there are upload flows however, stations with TCP ACK packets have to contend with
other stations sending TCP data packets as well. TCP ACK packets therefore can be delayed
accordingly, which causes the increase in RTT observed in Fig. \ref{fig_eBDP_varyrtt}.

\begin{figure}[tb]
   \centering
   {\includegraphics[width=0.5\columnwidth]{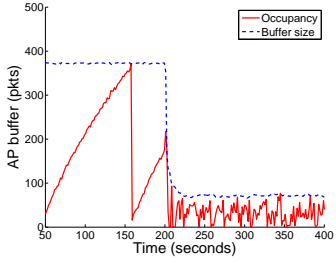}
   \label{fig_eBDP_converge_w_0.001}}
   \caption{Convergence of the eBDP algorithm following a change in network conditions.
   One download flow. At time 200s the number of upload flows is increased from 0 to 10. }
   \label{fig_eBDP_converge}
\end{figure}

Fig. \ref{fig_eBDP_converge} demonstrates the ability of the eBDP algorithm to respond to
changing network conditions.   At time 300s the number of uploads is increased from 0 to
10 flows.  It can be seen that the buffer size quickly adapts to the changed conditions
when the weight $W=0.001$. This roughly corresponds to averaging over the last 1000
packets\footnote{As per \cite{Chatfield_2004}, the current value is averaged over the
last $t$ observations for x\% percentage of accuracy where $x=1-(1-W)^t$, $t$ is the
number of updates (which are packets in our case). When $W=0.001$ and $t=1000$ we have
that $x=0.64.$}. When the number of uploads is increased at time 300s, it takes 0.6
seconds (current throughput is 13.5Mbps so $t=1000*8000/13.5*10^6=0.6$) to send 1000
packets, i.e., the eBDP algorithm is able to react to network changes roughly on a
timescale of 0.6 second.

\section{Exploiting Statistical Multiplexing: The A* Algorithm}\label{sec_astar}

\begin{figure}[tb]
   \centering
   \subfigure[Throughput]{\includegraphics[width=0.48\columnwidth]{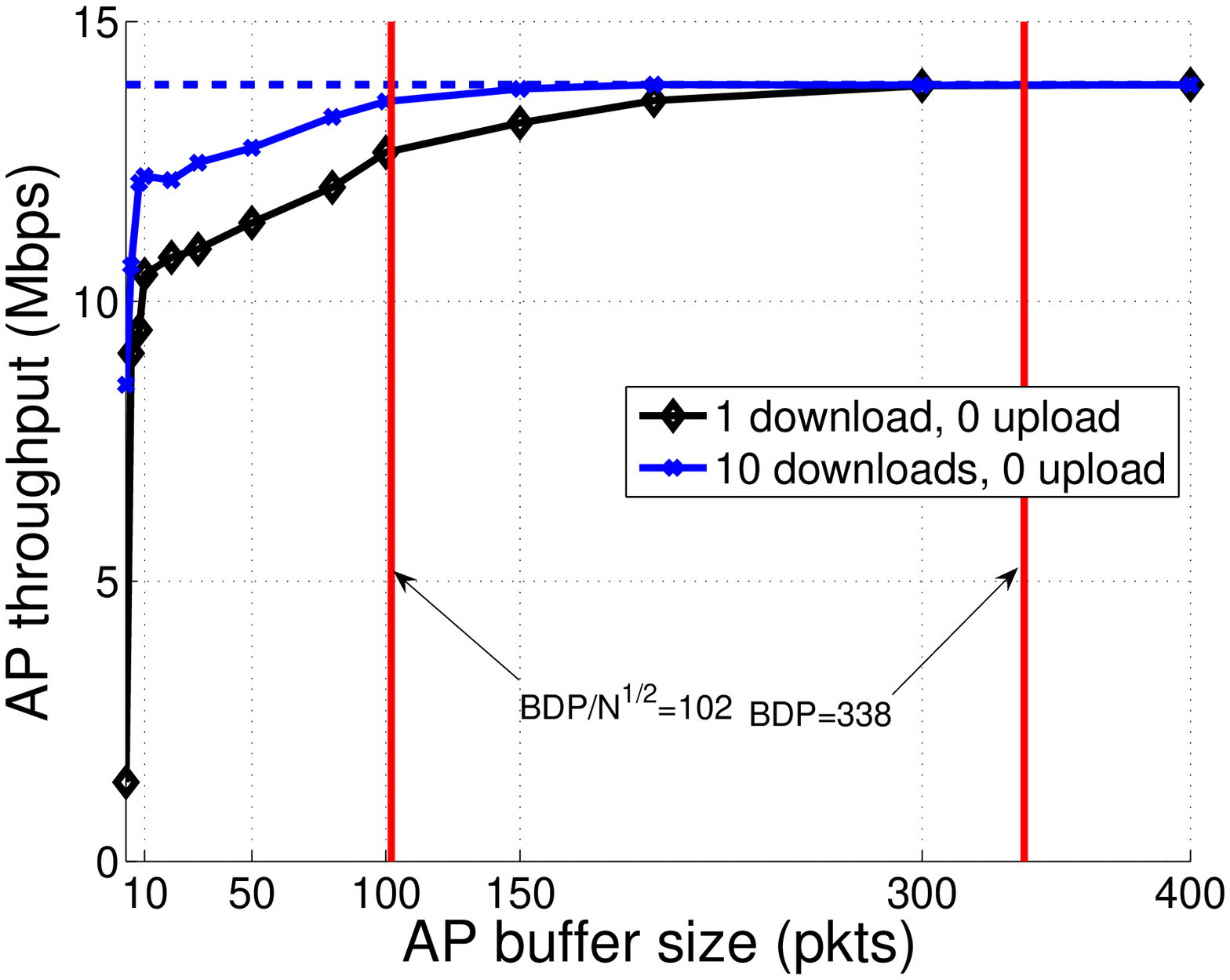}}
   \hfil
   \subfigure[Delay]{\includegraphics[width=0.48\columnwidth]{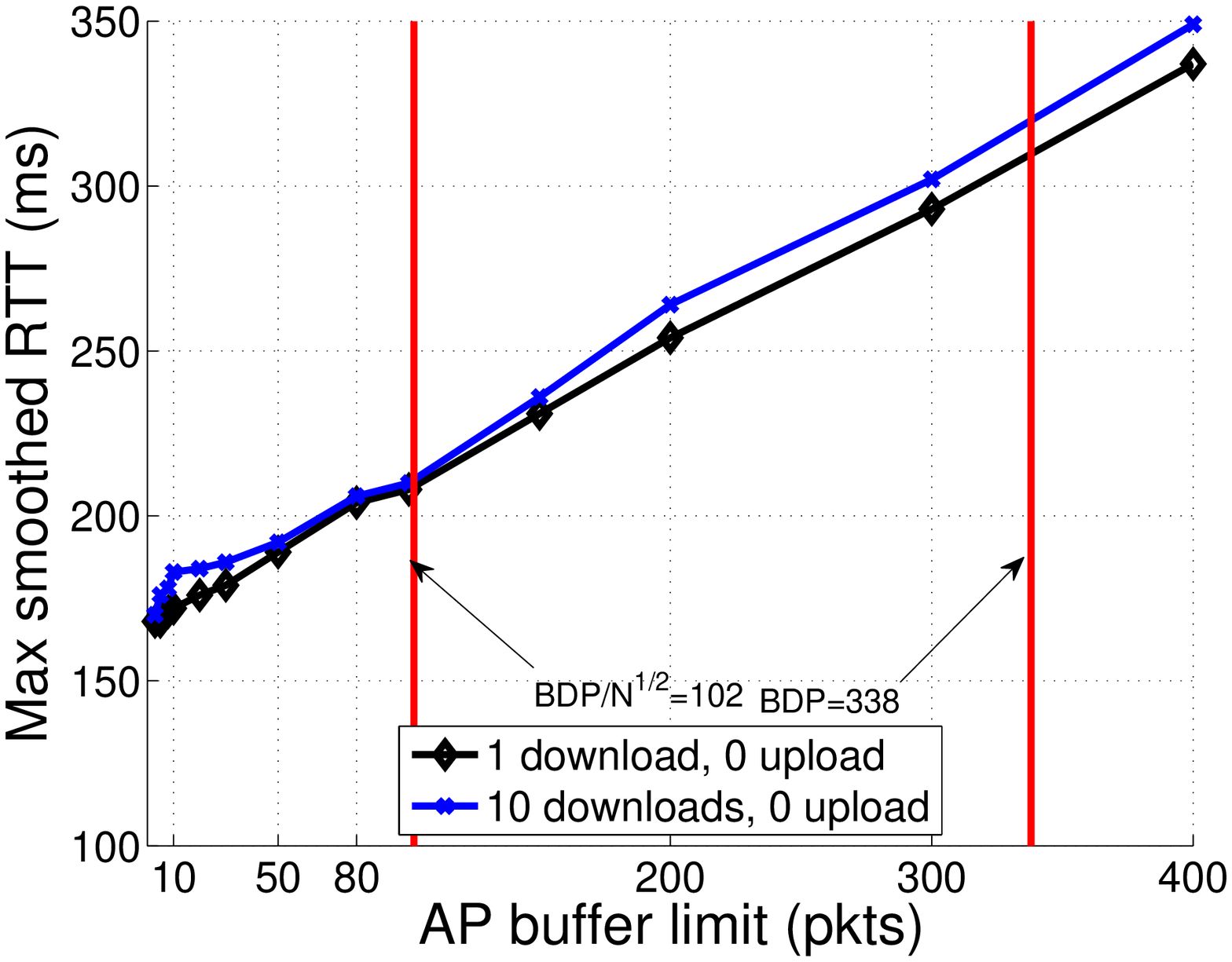}}
   \caption{Impact of statistical multiplexing.
      There are 1/10 downloads and no uploads.
      Wired RTT 200ms.}
   \label{fig_multiplexing}
\end{figure}

\begin{figure}[tb]
   \centering
   \subfigure[Time history]{\includegraphics[width=0.5\columnwidth]{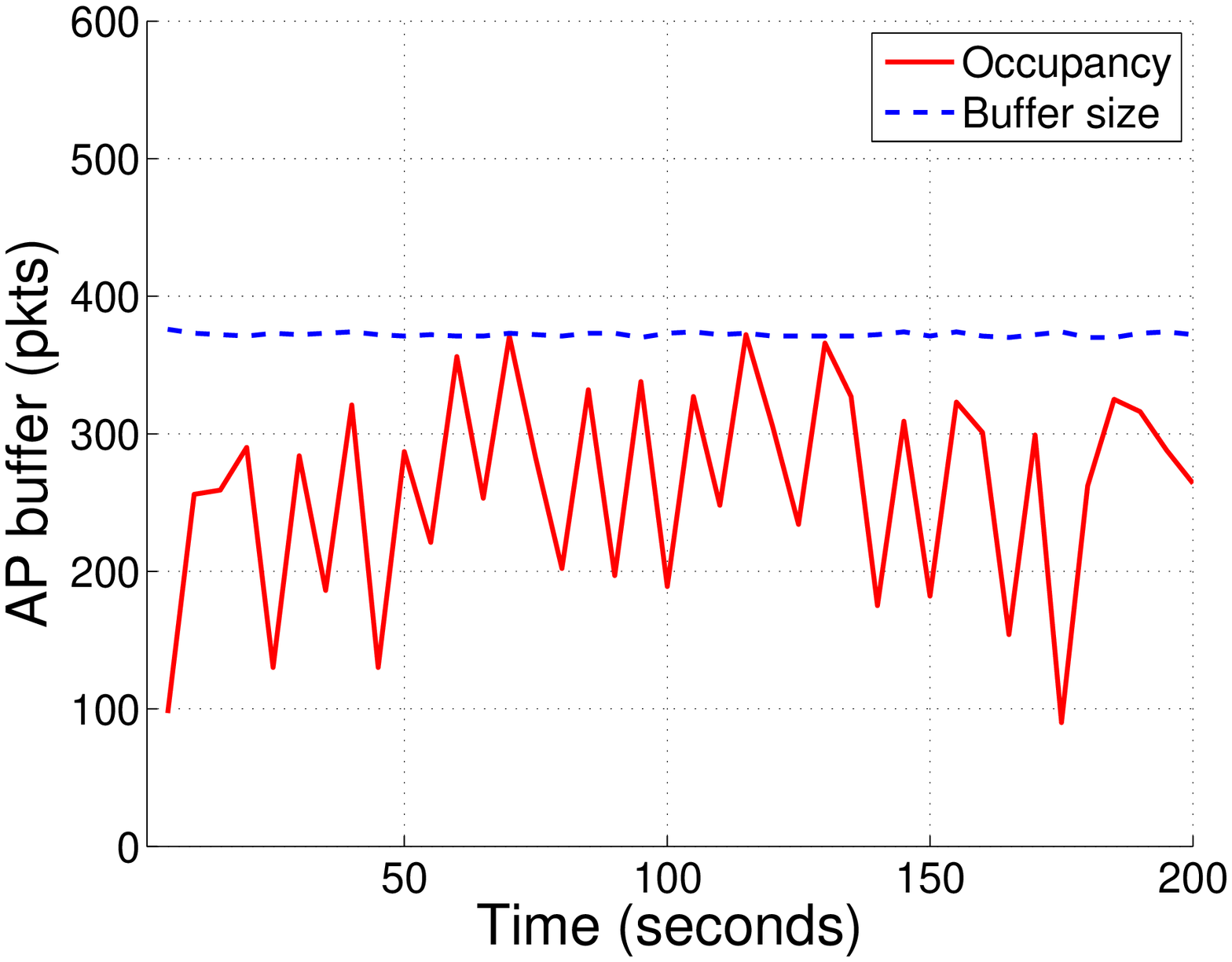}}
   \hfil
   \caption{Histories of buffer size and buffer occupancy with the eBDP algorithm
   when there are 10 downloads and no uploads.}
   \label{fig_eBDP_histories10}
\end{figure}

While the eBDP algorithm is simple and effective, it is unable to take advantage of the
statistical multiplexing of TCP cwnd backoffs when multiple flows share the same link. For example, it can be seen from Fig. \ref{fig_multiplexing} that while a buffer size of 338 packets is needed to maximize throughput with a single download flow, this falls to around 100 packets when 10 download flows share the link. However, in both cases the eBDP algorithm selects a
buffer size of approximately 350 packets (see Figs. \ref{fig_eBDP_histories_dl1} and
\ref{fig_eBDP_histories10}).   It can be seen from Fig. \ref{fig_eBDP_histories10} that as a result with the eBDP algorithm the buffer rarely empties when 10 flows share the link. That is, the potential exists to lower the buffer size without loss of throughput.

In this section we consider the design of a measurement-based algorithm (the ALT algorithm) that is capable of taking advantage of such statistical multiplexing opportunities.

\subsection{Adaptive Limit Tuning (ALT) Feedback Algorithm}

Our objective is to simultaneously achieve both efficient link utilization and low delays in the face of stochastic time-variations in the service time.  Intuitively, for efficient link utilization we need to ensure that there is a packet available to transmit whenever the station wins a transmission opportunity.  That is, we want to minimize the time that the station buffer lies empty, which in turn can be achieved by making the buffer size sufficiently large (under fairly general traffic conditions, buffer occupancy is a monotonically increasing function of buffer size \cite{Kumaran_stolyer_2003}.).   However, using large buffers can lead to high queuing delays, and to ensure low delays the buffer should be as small as possible.   We would therefore like to operate with the smallest buffer size that ensures sufficiently high link utilization.  This intuition suggests the following approach. We observe the buffer occupancy over an interval of time. If the buffer rarely empties, we decrease the buffer size. Conversely, if the buffer is observed to be empty for too long, we increase the buffer size.    Of course, further work is required to convert this basic intuition into a well-behaved algorithm suited to practical implementation.   Not only do the terms ``rarely'' , ``too long'' etc need to be made precise, but we note that an inner feedback loop is created whereby buffer size is adjusted depending on the measured link utilization, which in turn depends on the buffer size.   This new feedback loop is in addition to the existing outer feedback loop created by TCP congestion control, whereby the offered load is adjusted based on the packet loss rate, which in turn is dependent on buffer size.   Stability analysis of these cascaded loops is therefore essential.

We now introduce the following Adaptive Limit Tuning (ALT) algorithm.  The dynamics and stability of this algorithm will then be analyzed in later sections. \DLaddition{Define a queue occupancy threshold $q_{thr}$ and let $t_i(k)$ (referred to as the \emph{idle time}) be the duration of time that the queue spends at or below this threshold in a fixed observation interval $t$, and $t_b(k)$ (referred to as the \emph{busy time}) be the corresponding duration spent above the threshold.  Note that $t=t_i(k)+t_b(k)$ and the aggregate amount of idle/busy time $t_i$ and $t_b$ over an interval can be readily observed by a station.  Also, the link utilitisation is lower bounded by $t_b/(t_b+t_i)$.  Let $q(k)$ denote the buffer size during the $k$-th observation interval.}  The buffer size is then updated according to
\begin{equation}
    q(k+1)  =  q(k) + a_1 t_i(k) - b_1 t_b(k),
    \label{eqn_buffer_dynamic_per_interval}
\end{equation}
where $a_1$ and $b_1$ are design parameters. Pseudo-code for
this ALT algorithm is given in Algorithm \ref{algo_alt}.   This algorithm seeks to maintain a balance between the time $t_i$ that the queue is idle and the time $t_b$ that the queue is busy.  That is, it can be seen that when $a_1 t_i(k) = b_1 t_b(k)$, the buffer size is kept unchanged.  When the idle time is larger so that $a_1 t_i(k) > b_1 t_b(k)$, then the buffer size is increased.  Conversely, when the busy time is large enough that $a_1 t_i(k) < b_1 t_b(k)$, then the buffer size is decreased.

More generally, assuming $q$ converges to a stationary distribution (we discuss this in more detail
later), then in steady-state we have that $a_1E[t_i]=b_1E[t_b]$, i.e., $E[t_i]=\frac{b_1}{a_1}E[t_b]$ and the mean link utilization is therefore lower bounded by
\begin{equation}
    E[\frac{t_b}{t_i+t_b}] = \frac{E[t_b]}{t} =\frac{1}{1+b_1/a_1}.
    \label{eqn_lowerbound}
\end{equation}
where we have made use of the fact that  $t=t_i(k)+t_b(k)$ is constant.  It can therefore be seen that choosing $\frac{b_1}{a_1}$ to be small then ensures high utilization.   Choosing values for the parameters $a_1$ and $b_1$ is discussed in detail in Section \ref{sec:analysis},  but we note here that values of $a_1=10$ and $b_1=1$ are found to work well and unless otherwise stated are used in this paper.   With regard to the choice of observation interval $t$, this is largely determined by the time required to obtain accurate estimates of the queue idle and busy times.  In the reminder of this paper we find a value of $t=1$ second to be a good choice.

It is prudent to constrain the buffer size $q$ to lie between the minimum and the maximum values $q_{min}$ and $q_{max}$.  In the following, the maximum size $q_{max}$ and the minimum buffer size $q_{min}$ are set to be 1600 and 5 packets respectively.   

\begin{algorithm}[tb]
    \begin{algorithmic}[1]
        \STATE Set the initial queue size, the maximum buffer size $q_{max}$ and
            the minimum buffer size $q_{min}$.
        \STATE Set the increase step size $a_1$ and the decrease step size $b_1$.
        \FOR {Every $t$ seconds}
            \STATE Measure the idle time $t_i$.
            \STATE $q_{ALT}=q_{ALT}+a_1t_i-b_1(t-t_i)$.
            \STATE $q_{ALT} = min(max(q_{ALT}, q_{min}), q_{max})$ 
       \ENDFOR
    \end{algorithmic}
    \caption{: The ALT algorithm.}
    \label{algo_alt}
\end{algorithm}

\subsection{Selecting the Step Sizes for ALT}\label{sec:analysis}

\begin{figure}[tb]
   \centering
   \includegraphics[width=0.7\columnwidth]{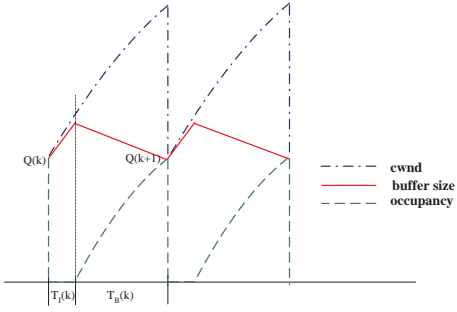}
   \caption{Illustrating evolution of the buffer size.}
   \label{fig_buffer}
\end{figure}

Define a \emph{congestion event} as an event where the sum of all senders' TCP cwnd decreases. This cwnd decrease can be caused by the response of TCP congestion control to a single packet loss, or multiple packet losses that are lumped together in one RTT. Define a \emph{congestion epoch} as the duration between two adjacent congestion events.

Let $Q(k)$ denote the buffer size at the $k$-th congestion event. Then,
\begin{equation}\label{eq:Q}
Q(k+1)=Q(k) + aT_I(k) - bT_B(k)
\end{equation}
where $T_I$ is the ``idle'' time, i.e., the duration in seconds when\DLaddition{the queue occupancy is below $q_{thr}$ 
during the $k$-th congestion epoch, and $T_B$ the ``busy'' time, i.e., the duration when
the queue occupancy is above $q_{thr}$.} This is illustrated in Fig. \ref{fig_buffer} for the case of a
single TCP flow.

Notice that $a=a_1$ and $b=b_1$ where $a_1$ and $b_1$ are parameters used in the ALT
algorithm. In the remainder of this section we
investigate conditions to guarantee convergence and stability of the buffer dynamics with
TCP traffic, which naturally lead to guidelines for the selection of $a_1$ and $b_1$. We
first define some TCP related quantities before proceeding.

Consider the case where TCP flows may have different round-trip times and drops need not
be synchronized. Let $n$ be the number of TCP flows sharing a link, $w_i(k)$ be the cwnd
of flow $i$ at the $k$-th congestion event, $T_i$ the round-trip propagation delay of
flow $i$. To describe the cwnd additive increase we define the following quantities: (i)
$\alpha_i$ is the rate in $packet/s$ at which flow $i$ increases its congestion
window\footnote{Standard TCP increases the flow congestion window by one packet per RTT,
in which case $\alpha_i \approx 1/T_i$.}, (ii) $\alpha_T=\sum_{i=1}^n\alpha_i$ is the
aggregate rate at which flows increase their congestion windows, in packets/s, and (iii)
$A_T=\sum_{i=1}^n\alpha_i/T_i$ approximates the aggregate rate, in $packets/s^2$, at
which flows increase their sending rate.   Following the $k$-th congestion event, flows
backoff their cwnd to $\beta_i(k)w_i(k)$. Flows may be unsynchronized, i.e., not all
flows need back off at a congestion event. We capture this with $\beta_i(k)=1$ if flow
$i$ does not backoff at event $k$. \DLaddition{We assume that the $\alpha_i$ are constant and that the $\beta_i(k)$ (i.e. the pattern of flow backoffs) are independent of the flow congestion windows $w_i(k)$ and the buffer size $Q(k)$ (this appears to be a good approximation in many practical situations, see \cite{shorten_ton_2006}).  }

\DLaddition{To relate the queue occupancy to the flow cwnds, we adopt a fluid-like approach and ignore sub-RTT burstiness.   We also assume that $q_{thr}$ is sufficiently small relative to the buffer size that we can approximate it as zero.    Considering now the idle time $T_I(k)$, on backoff after the $k$-th congestion event, if the queue occupancy does not fall below $q_{thr}$ then $T_I(k)=0$.  Otherwise, immediately after backoff the send rate of flow $i$ is $\beta_i(k)w_i(k)/T_i$ and we have that}
\begin{equation}
T_I(k)=\frac{E[B]-\sum_{i=1}^n \beta_i(k)w_i(k)/T_i}{A_T},
\end{equation} where $E[B]$ is the mean service rate of the considered
buffer.

At congestion event $k$ the aggregate flow throughput necessarily equals the link
capacity, i.e.,
$$
\sum_{i=1}^n \frac{w_i(k)}{T_i+Q(k)/E[B]}=E[B].
$$

We then have that
\DLaddition{
\begin{eqnarray*}
    \sum_{i=1}^n \frac{w_i(k)}{T_i}
    &=&        \sum_{i=1}^n \frac{w_i(k)}{T_i}\frac{T_i + Q(k)/E[B]}{T_i + Q(k)/E[B]} \\
    &=&        \sum_{i=1}^n \frac{w_i(k)}{T_i + Q(k)/E[B]}  +
               \\ && \frac{Q(k)}{E[B]} \sum_{i=1}^n\frac{w_i(k)}{T_i + Q(k)/E[B]}\frac{1}{T_i}
 \end{eqnarray*}
 Assume that the spread in flow round-trip propagation delays and congestion windows is small enough that $\sum_{i=1}^n (w_i(k)/(E[B]T_i + Q(k))(1/T_i)$ can be accurately approximated by $1/T_T$, where $T_T=\frac{n}{\sum_{i=1}^n\frac{1}{T_i}}$ is the harmonic mean of $T_i$.  Then  
\begin{eqnarray*}
\sum_{i=1}^n \frac{w_i(k)}{T_i}&\approx  E[B]+ \frac{Q(k)}{T_T},
\end{eqnarray*} 
and
%
}\begin{equation}\label{eq:1}
T_I(k) \approx \frac{(1-\beta_T(k))E[B]-\beta_T(k) Q(k)/T_T}{A_T}
\end{equation}
where $\beta_T(k)=\frac{\sum_{i=1}^n \beta_i(k) w_i(k)/T_i}{\sum_{i=1}^n w_i(k)/T_i}$ is the
effective aggregate backoff factor of the flows.    When flows are synchronized, i.e.,
$\beta_i=\beta \ \forall i$, then $\beta_T=\beta$. When flows are
unsynchronized
but have the same \emph{average}
backoff factor, i.e., $E[\beta_i]=\beta$, then $E[\beta_T]=\beta$ .

If the queue empties after backoff, the queue busy time $T_B(k)$ is directly given by
\begin{equation}\label{eq:2}
T_B(k)=Q(k+1)/ \alpha_T
\end{equation}
where $\alpha_T=\sum_{i=1}^n\alpha_i$ is the aggregate rate at which flows increase their
congestion windows, in packets/s.  Otherwise,
\begin{equation}\label{eq:2a}
T_B(k)=(Q(k+1)-q(k))/ \alpha_T
\end{equation}
where $q(k)$ is the buffer occupancy after backoff.   It turns out that for the analysis
of stability it is not necessary to calculate $q(k)$ explicitly. Instead, letting
$\delta(k)=q(k)/Q(k)$, it is enough to note that $0 \leq \delta(k)<1$.

Combining (\ref{eq:Q}), (\ref{eq:1}), (\ref{eq:2}) and (\ref{eq:2a}),
\[Q(k+1) = \left\{
    \begin{array}{l l}
    \lambda_e(k)Q(k) + \gamma_e(k)E[B]T_T, & q(k)\le q_{thr}\\
    \lambda_f(k)Q(k), &  \mbox{otherwise}
    \\ \end{array}
    \right.
\]
where
\begin{eqnarray*}
\lambda_e(k)&=&\frac{\alpha_T-a\beta_T(k)\alpha_T/(A_TT_T)}{\alpha_T+b} ,\\
\lambda_f(k)&=&\frac{\alpha_T+ b\delta(k)}{\alpha_T+b},\gamma_e(k)=a\frac{1-\beta_T(k)}{\alpha_T+b}\frac{\alpha_T}{A_TT_T}.
\end{eqnarray*}
\DLaddition{
Taking expectations,
\begin{align*}
E[Q(k+1)] & \\
=&E[\lambda_e(k)Q(k) + \gamma_e(k)E[B]T_T|q(k)\le q_{thr}]p_e(k)\\
& + E[\lambda_f(k)Q(k)|q(k)> q_{thr}](1-p_e(k))
\end{align*}
with $p_{e}(k)$ the probability that the queue $\le q_{thr}$ following the $k$-th congestion event.
Since the $\beta_i(k)$ are assumed independent of $Q(k)$ we may assume that $E[Q(k)|q(k)\le q_{thr}]=E[Q(k)|q(k)> q_{thr}]=E[Q(k)]$ and
\begin{align}\label{eq:average}
E[Q(k+1)]&=\lambda(k)E[Q(k)] + \gamma(k) E[B]T_T
\end{align}
where
\begin{align*}
\lambda(k)=&p_{e}(k)E[\lambda_e(k)|q(k)\le q_{thr}] \\
&+(1-p_{e}(k))E[\lambda_f(k)|q(k)> q_{thr}],\\
\gamma(k)=&p_e(k)E[\gamma_e(k)|q(k)\le q_{thr}]
\end{align*}
}

\subsection{A Sufficient Condition for Stability}\label{subsec_stability}

\begin{figure}[tb]
   \centering
   \subfigure[Instability]{{\includegraphics[width=.45\columnwidth]{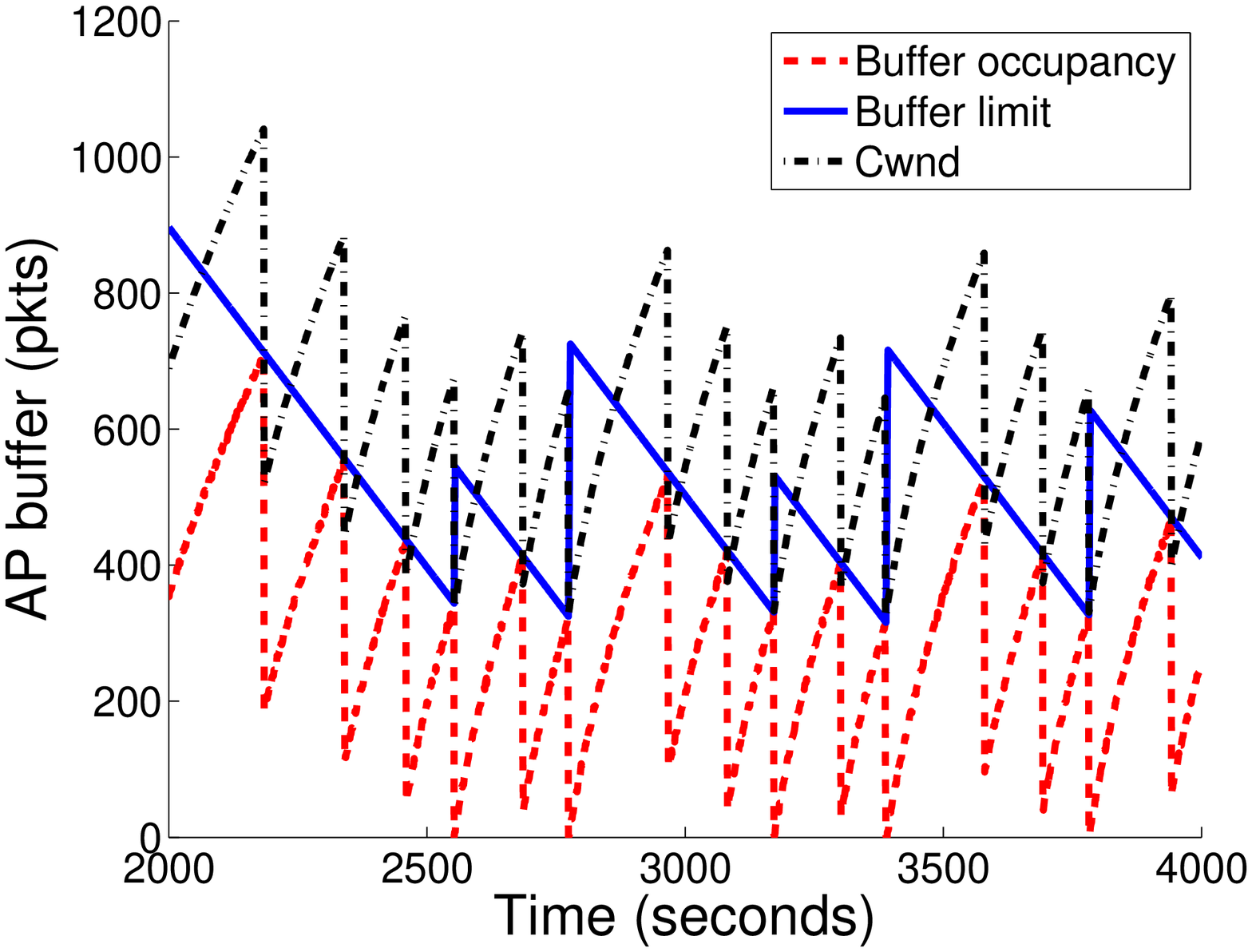}}
   \label{fig_validation_instability}}
   \subfigure[Stability]{{\includegraphics[width=.45\columnwidth]{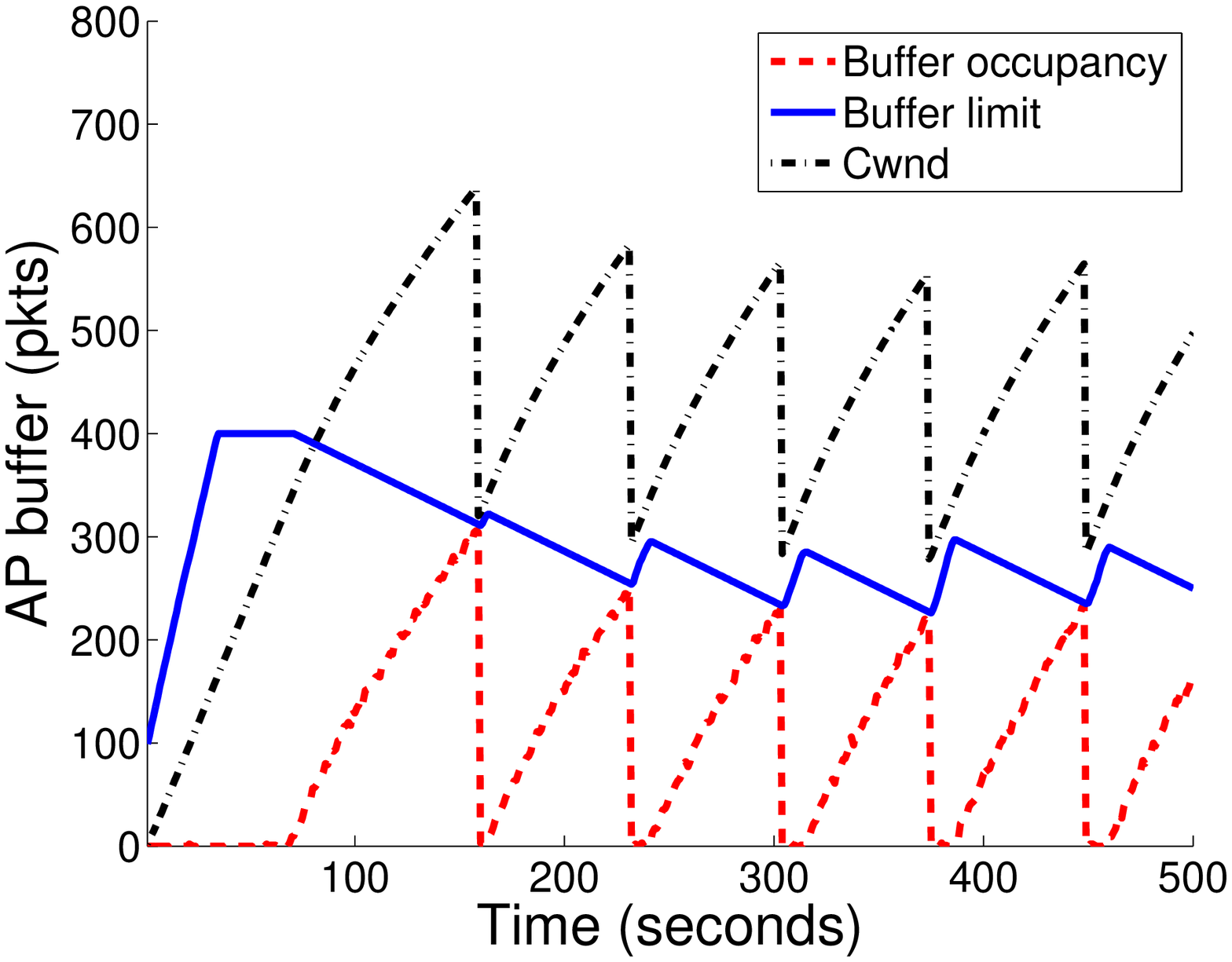}}
   \label{fig_validation_stability}}
   \caption{Instability and stability of the ALT algorithm.
   In (a), a=100, b=1, the maximum buffer size is 50000 packets.
   In (b), a=10, b=1, the maximum buffer size is 400 packets.
   In both figures, there is 1 download and no upload.}
   \label{fig_stability}
\end{figure}

Provided  $|\lambda(k)|<1$ the queue dynamics in (\ref{eq:average}) are
exponentially stable.  In more detail, $\lambda(k)$ is the convex combination of
$E[\lambda_e(k)]$ and $E[\lambda_f(k)]$\DLaddition{(where the conditional dependence of these expectations is understood, but omitted to streamline notation)}.  Stability is therefore guaranteed provided
$|E[\lambda_e(k)]|<1$ and $|E[\lambda_f(k)]|<1$.  We have that $0<E[\lambda_f(k)] <1$
when $b>0$ since $\alpha_T$ is non-negative and $0 \leq \delta(k)<1$.   The stability
condition is therefore that $|E[\lambda_e(k)]|<1$.

Under mild independence conditions,
$$
E[\lambda_e(k)]=\frac{\alpha_T-aE[\beta_T(k)]\alpha_T/(A_TT_T)}{\alpha_T+b}.
$$
Observe that,
$$
\frac{\alpha_T}{A_TT_T} = \frac{1}{n}\frac{(\sum_{i=1}^n 1/T_i)^2}{\sum_{i=1}^n1/T_i^2}
$$
when we use the standard TCP AIMD increase of one packet per RTT, in which case $\alpha_i
\approx 1/T_i$.  We therefore have that $1/n\le \alpha_T /(A_T T_T)\le1$. Also, when the
standard AIMD backoff factor of 0.5 is used, $0.5<E[\beta_T(k)]<1$. Thus, since $a>0$,
$b>0$, $\alpha_T>0$, it is sufficient that
$$
-1 < \frac{\alpha_T-a}{\alpha_T+b}\le E[\lambda_e(k)]\le\frac{\alpha_T}{\alpha_T+b}<1
$$
A sufficient condition (from the left inequality) for stability is then that $a <
2\alpha_T+b$. Using again (as in the eBDP algorithm) 200ms as the maximum RTT , a rough
lower bound on $\alpha_T$ is 5 (corresponding to 1 flow with RTT 200ms).  The stability
constraint is then that
\begin{equation}\label{eq:acond}
    a < 10+b.
\end{equation}

Fig. \ref{fig_validation_instability} demonstrates that the instability is indeed
observed in simulations. Here, $a=100$ and $b=1$ are used as example values, i.e., the
stability conditions are not satisfied. It can be seen that the buffer size at congestion
events oscillates around 400 packets rather than converging to a constant value. We note,
however, that in this example and others the instability consistently manifests itself in
a benign manner (small oscillations).  However, we leave detailed analysis of the onset
of instability as future work.

Fig. \ref{fig_validation_stability} shows the corresponding results with $a=10$ and
$b=1$, i.e., when the stability conditions are satisfied. It can be seen that the buffer
size at congestion events settles to a constant value, thus the buffer size time history
converges to a periodic cycle.

\subsection{Fixed point}\label{subsec_fixed_point}

When the system dynamics are stable\DLaddition{and $p_e=0$}, from (\ref{eq:average}) we have that
\begin{equation}\label{eq:limit}
\lim_{k\rightarrow \infty}E[Q(k)]=\frac{(1-E[\beta_T])}{b/a+E[\beta_T]}E[B]T_T.
\end{equation}
For synchronized flows with the standard TCP backoff factor of 0.5 (i.e.,
$E[\beta_T]=0.5$) and the same RTT, $\frac{(1-E[\beta_T])}{b/a+E[\beta_T]}E[B]T_T$
reduces to the BDP when $b/a=0$. This indicates that for high link utilization\DLaddition{we would like}
the ratio $b/a$ to be small.    Using (\ref{eq:1}), (\ref{eq:2}) and
(\ref{eq:limit}) we have that in steady-state the expected link utilization is lower
bounded by
\begin{equation}
   \frac{1}{1+\frac{b}{a}\frac{\alpha_T}{A_TT_T}}
    \geq \frac{1}{1+\frac{b}{a}}.
    \label{eqn_efficiency}
\end{equation}

\begin{figure}[tb]
   \centering
   \includegraphics[width=.5\columnwidth]{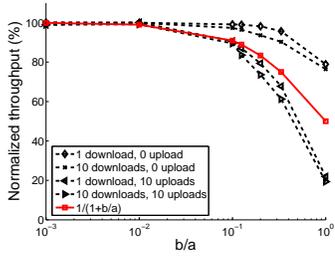}
   \caption{Impact of $b/a$ on throughput efficiency. The maximum buffer size is 400
   packets, and the minimum buffer size is 2 packets. }
   \label{fig_alt_efficiency}
\end{figure}

This lower bound is plotted in Fig. \ref{fig_alt_efficiency} together with the measured throughput
efficiency vs $b/a$ in a variety of traffic conditions. Note that in this figure the lower bound is violated by the measured data when $b/a>0.1$ and we have a large number of uploads.   At such large values of $b/a$ plus many contending stations, the target buffer sizes are extremely small and micro-scale burstiness means that TCP RTOs occur frequently.  It is this that
leads to violation of the lower bound (\ref{eqn_efficiency})\DLaddition{(since $p_e=0$ does not hold)}.  However, this corresponds to an
extreme operating regime and for smaller values of $b/a$ the lower bound is
respected.   It can be seen from Fig. \ref{fig_alt_efficiency} that the
efficiency decreases when the ratio of $b/a$ increases. In order to ensure throughput
efficiency $\geq$ 90\% it is required that
\begin{equation}\label{eqn_fixed_point}
    \frac{b}{a} \leq 0.1.
\end{equation}
Combined with the stability condition in inequality (\ref{eq:acond}),
we have that $a=10$, $b=1$ are feasible integer values, that is, we choose $a_1=10$ and $b_1=1$ for the A* algorithm.

\subsection{Convergence rate}\label{subsec_convergence_rate}

\begin{figure}[tb]
   \centering
   \subfigure[The ALT algorithm]{{\includegraphics[width=.45\columnwidth]{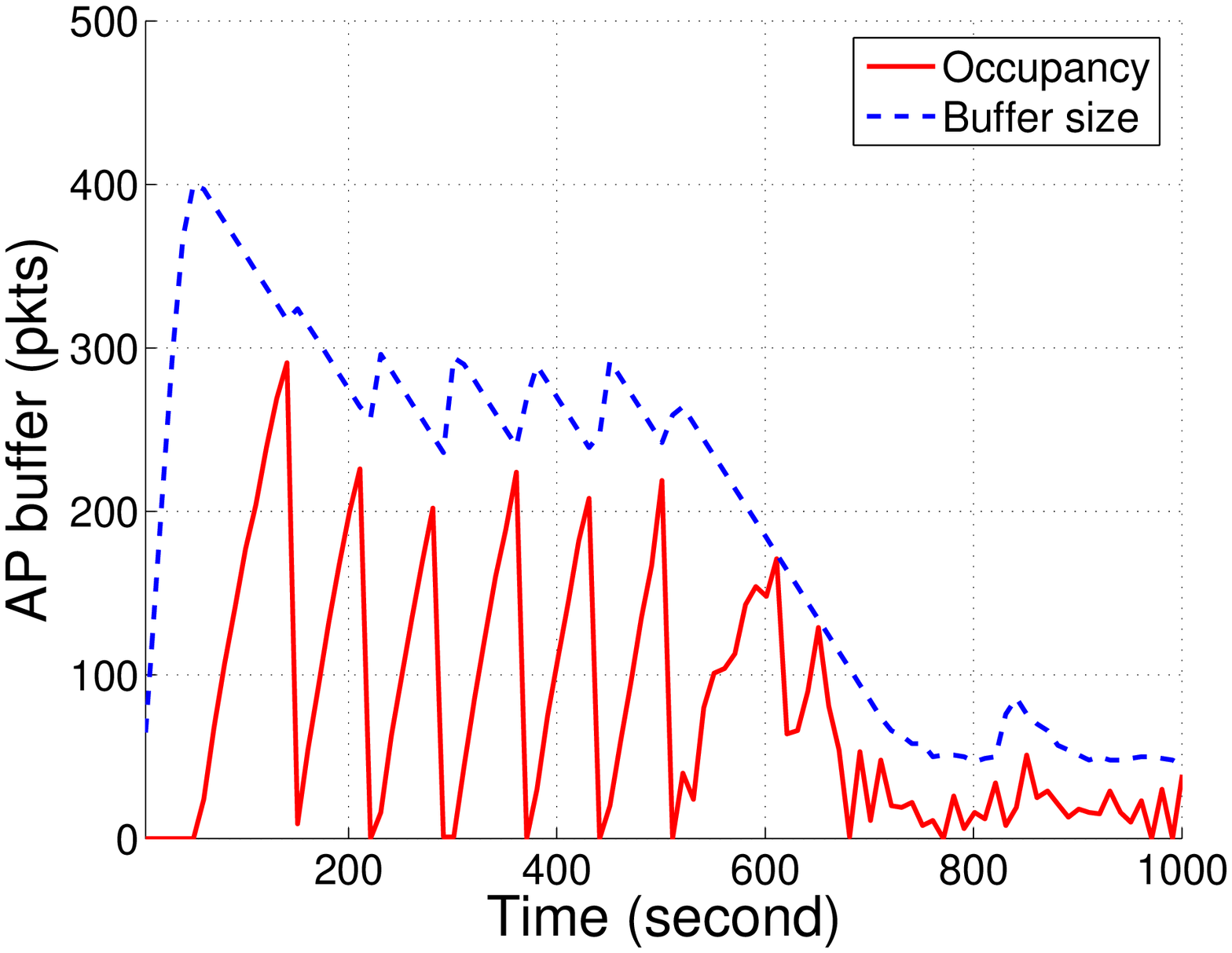}}
   \label{fig_alt_converge}}
   \subfigure[The A* algorithm]{{\includegraphics[width=.45\columnwidth]{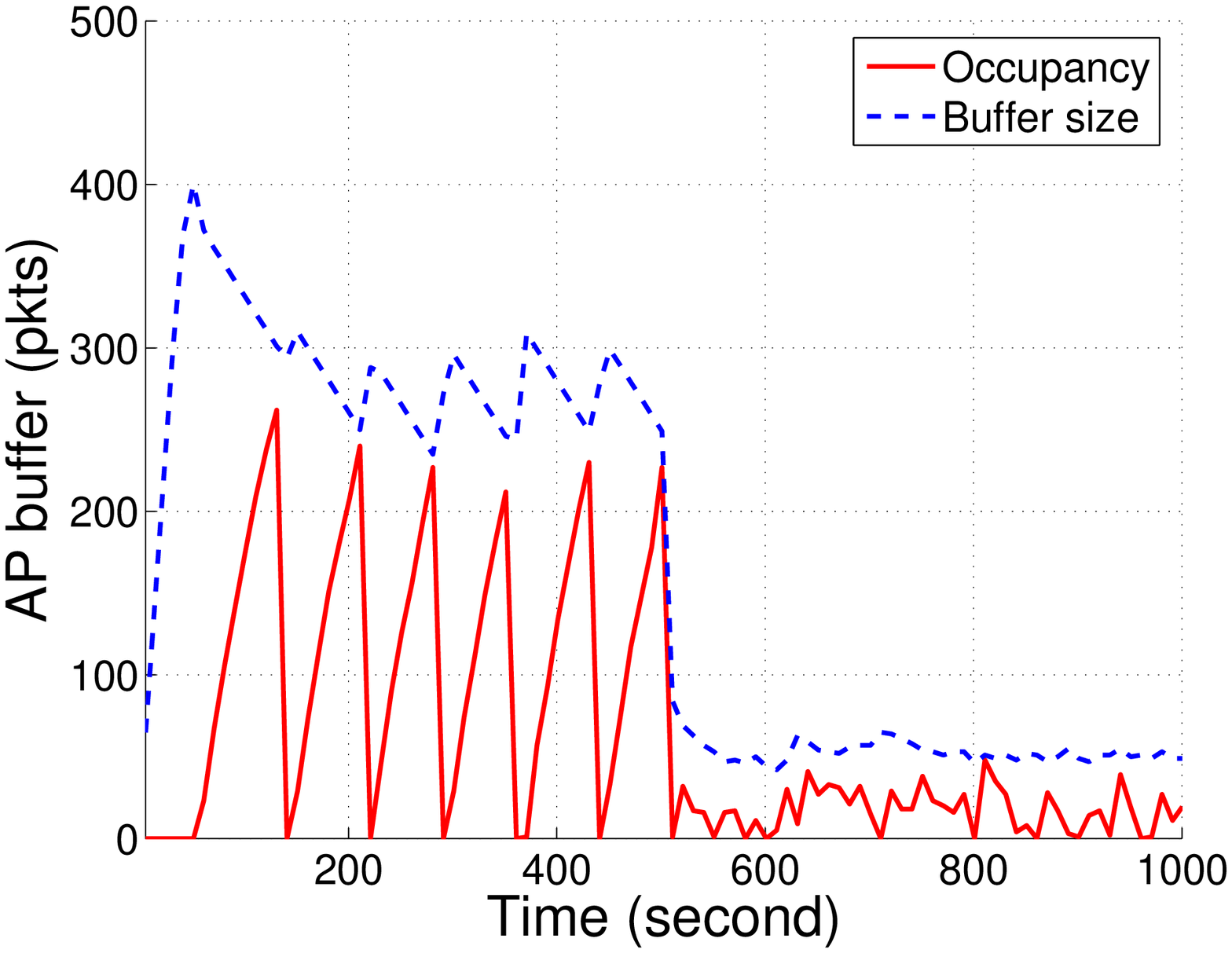}}
   \label{fig_alt_converge_combine}}
   \caption{Convergence rate of the ALT and A* algorithms.
   One download flow, $a=10$, $b=1$.
   At time 500s the number of upload flows is increased from 0 to 10.}
\end{figure}

In Fig. \ref{fig_alt_converge} we illustrate the convergence rate of the ALT algorithm.
There is one download, and at time 500s the number of upload flows is increased from 0 to
10. It can be seen that the buffer size limit converges to its new value in around 200
seconds or 3 minutes. In general, the convergence rate is determined by the product
$\lambda(0)\lambda(1)...\lambda(k)$. In this example, the buffer does not empty after
backoff and the convergence rate is thus determined by
$\lambda_f(k)=\frac{\alpha_T+b\delta(k)}{\alpha_T+b}$. To achieve fast convergence, we
require small $\lambda_f(k)$ so that $Q(k+1)=\lambda_f(k)Q(k)$ is decreased quickly to
the desired value. We thus need large $b$ to achieve fast convergence. However, $b=1$ is
used here in order to respect the stability condition in (\ref{eq:acond}) and the throughput efficiency condition in (\ref{eqn_fixed_point}).  Note that
when conditions change such that the buffer size needs to increase, the convergence rate
is determined by the $a$ parameter.  This has a value of $a=10$ and thus the algorithm
adapts much more quickly to increase the buffer than to decrease it and the example in
Fig. \ref{fig_alt_converge} is essentially a worst case.  In the next section, we address
the slow convergence by combining the ALT and the eBDP algorithms to create a hybrid algorithm.

\subsection{Combining eBDP and ALT: The A* Algorithm}\label{subsec_final}

We can combine the eBDP and ALT algorithms by using the mean packet service time to calculate $Q_{eBDP}$ as per the eBDP algorithm (see Section \ref{sec_eBDP}), and the idle/busy times to calculate $q_{ALT}$ as per the ALT algorithm.   We then select the buffer size as $\min\{Q_{eBDP},q_{ALT}\}$ to yield a hybrid  algorithm, referred to as the A* algorithm, that combines the eBDP and the ALT algorithms.

\begin{figure}[tb]
   \centering
   \subfigure[10 downloads only]{\includegraphics[width=0.48\columnwidth]{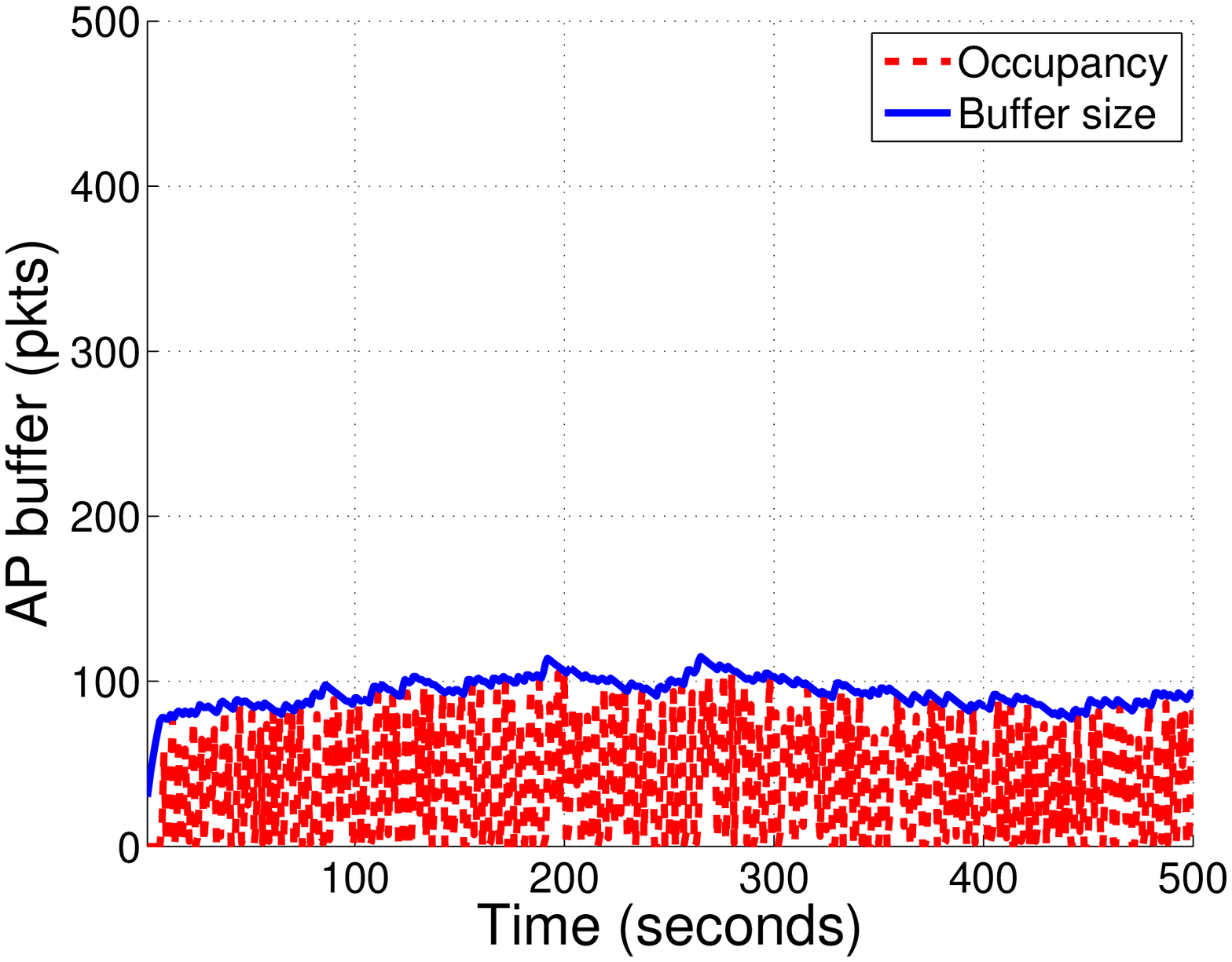}\label{fig_astar_histories_0ul_10dl}}
   \subfigure[10 downloads, with 10 uploads starting at time 500s]{\includegraphics[width=0.48\columnwidth]{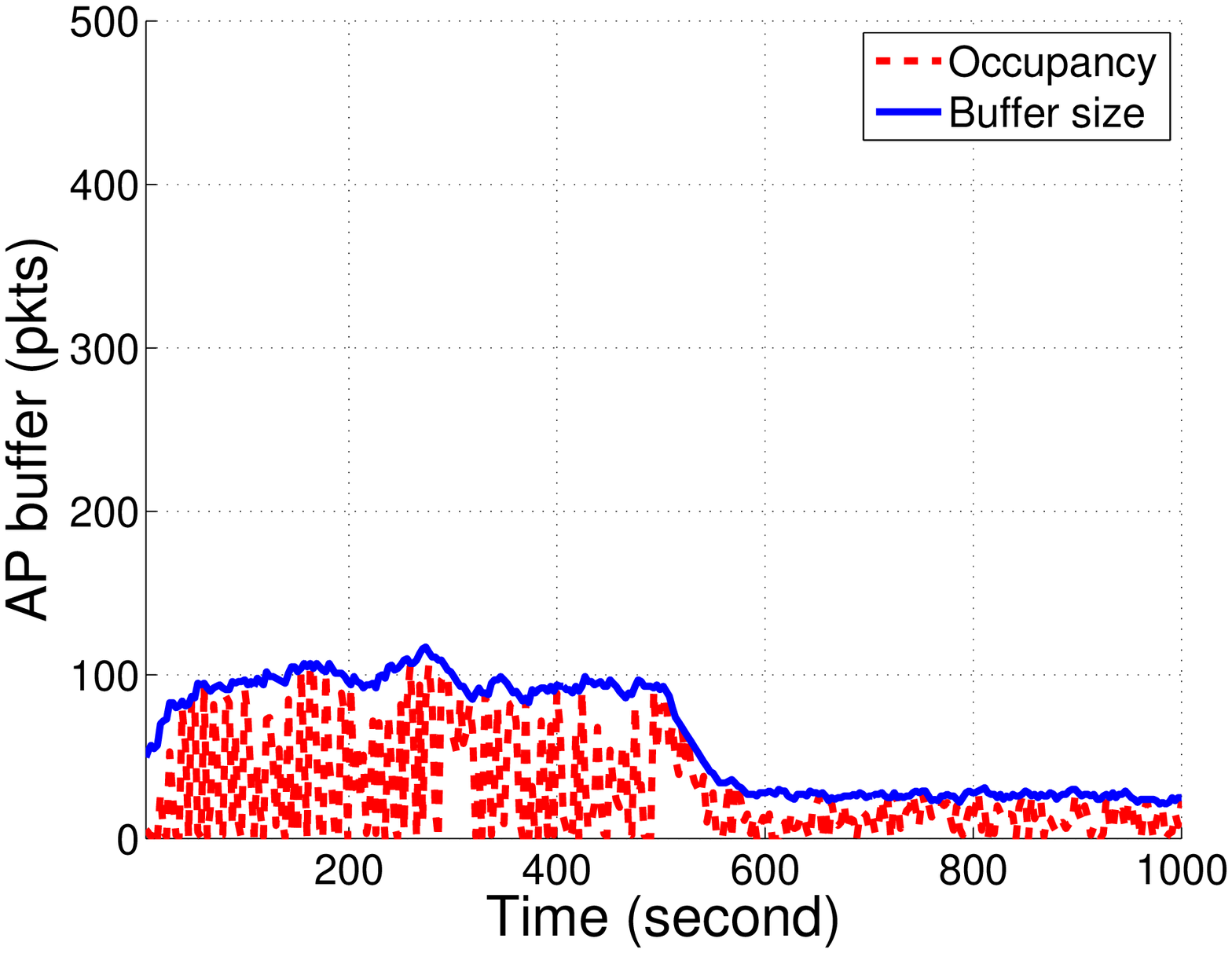}\label{fig_astar_histories_10ul_10dl}}
   \caption{Buffer time histories with the A* algorithm, a=10, b=1.}
   \label{fig_astar_histories}
\end{figure}

When channel conditions change, the A* algorithm uses the eBDP measured service
time to adjust the buffer size promptly. The convergence rate depends on the smoothing
weight $W$. As calculated in Section \ref{sec_eBDP}, it takes around 0.6 second for
$Q_{eBDP}$ to converge. The A* algorithm can further use the ALT algorithm to fine tune
the buffer size to exploit the potential reduction due to statistical multiplexing. The
effectiveness of this hybrid approach when the traffic load is increased suddenly is
illustrated in Fig. \ref{fig_alt_converge_combine} (which can be directly compared with Fig. \ref{fig_alt_converge}).   Fig. \ref{fig_astar_histories_10ul_10dl} shows the corresponding time histories for 10 download flows and a changing number of competing uploads.

\subsection{Performance}\label{subsec_performance}

\begin{figure*}[tb]
   \centering
   \subfigure[1/1Mbps, throughput]{\includegraphics[width=0.45\columnwidth]{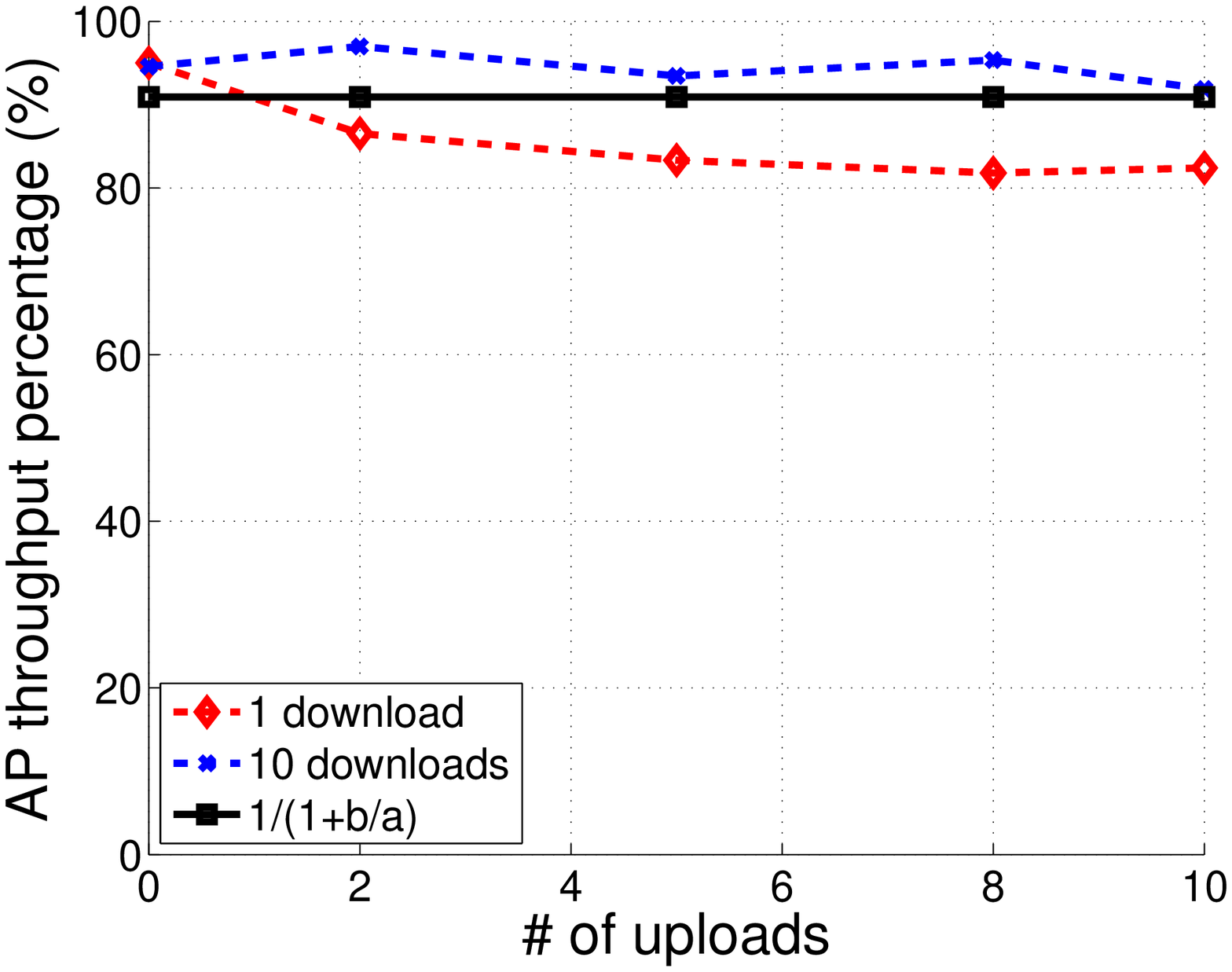}}
   \subfigure[1/1Mbps, delay]{\includegraphics[width=0.47\columnwidth]{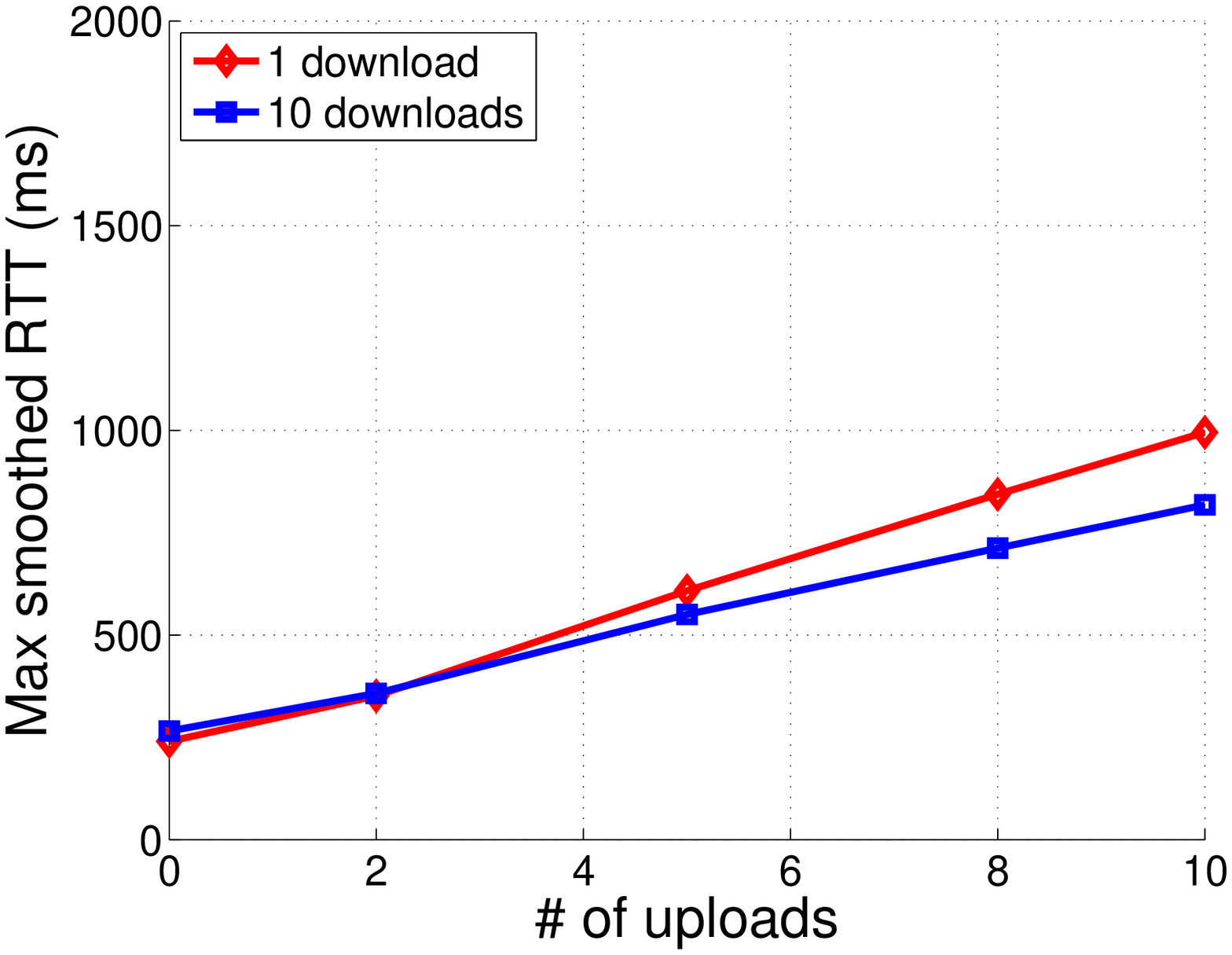}}
   \subfigure[11/1Mbps, throughput]{\includegraphics[width=0.45\columnwidth]{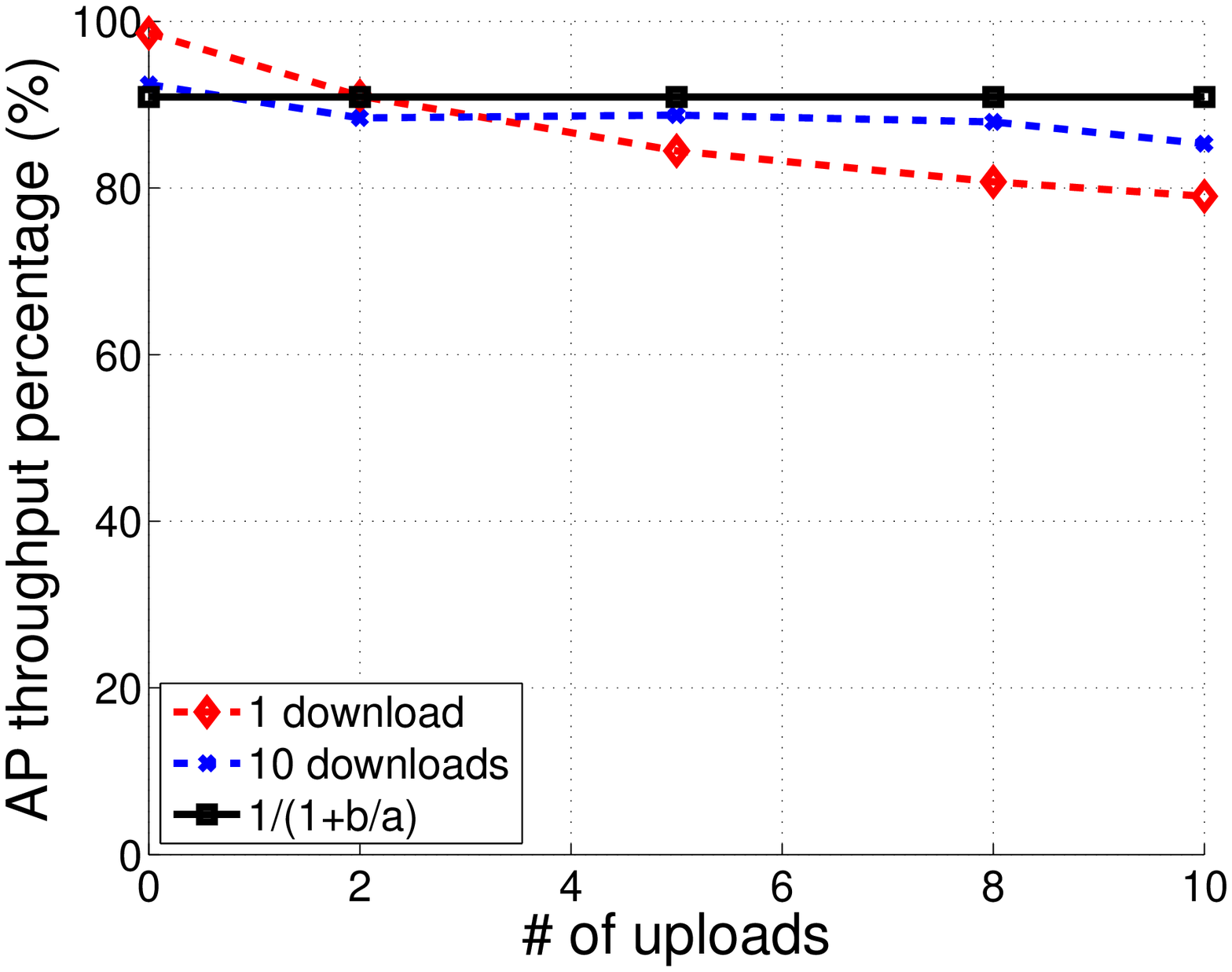}}
   \subfigure[11/1Mbps, delay]{\includegraphics[width=0.47\columnwidth]{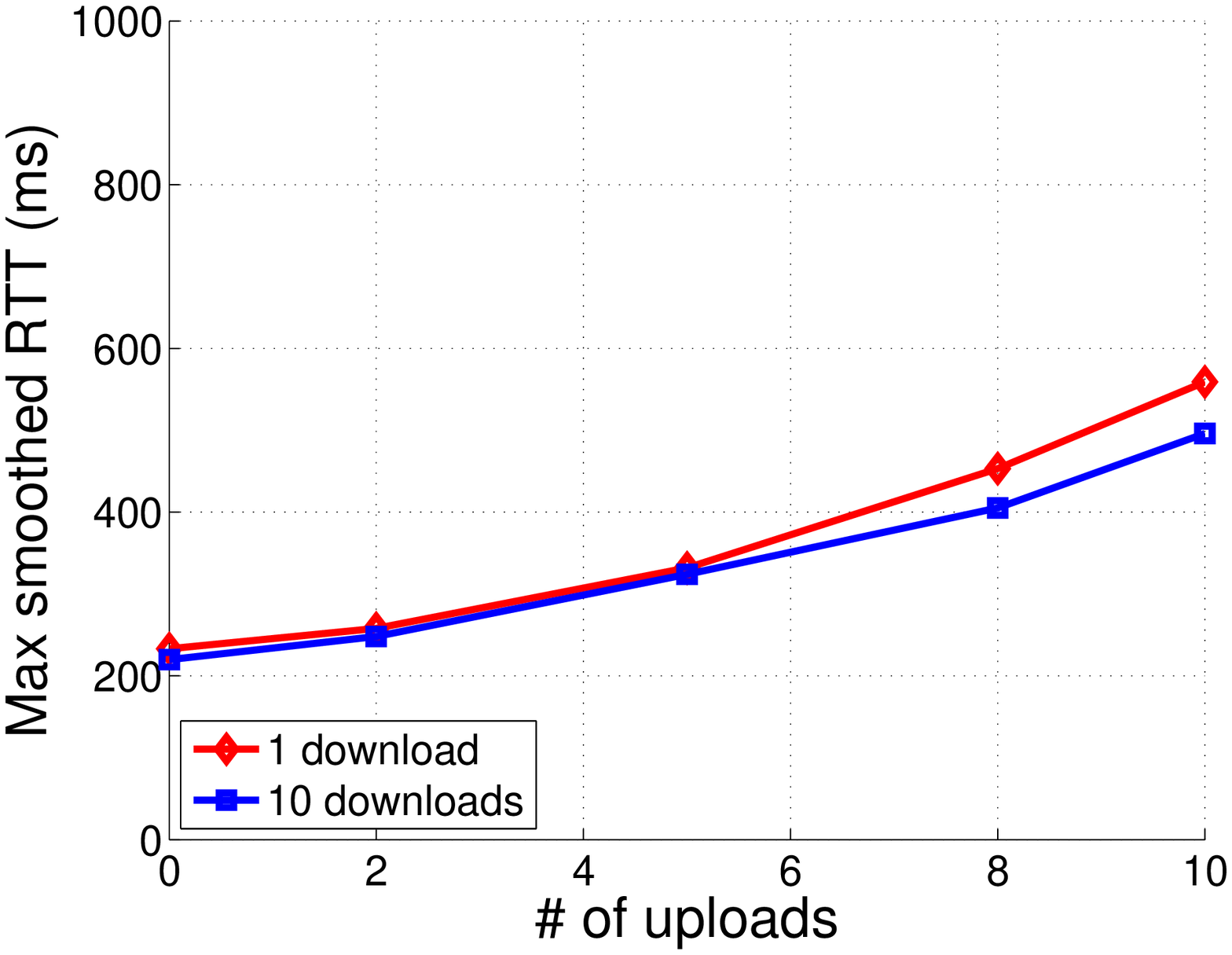}}
    \subfigure[54/6Mbps, throughput]{\includegraphics[width=0.46\columnwidth]{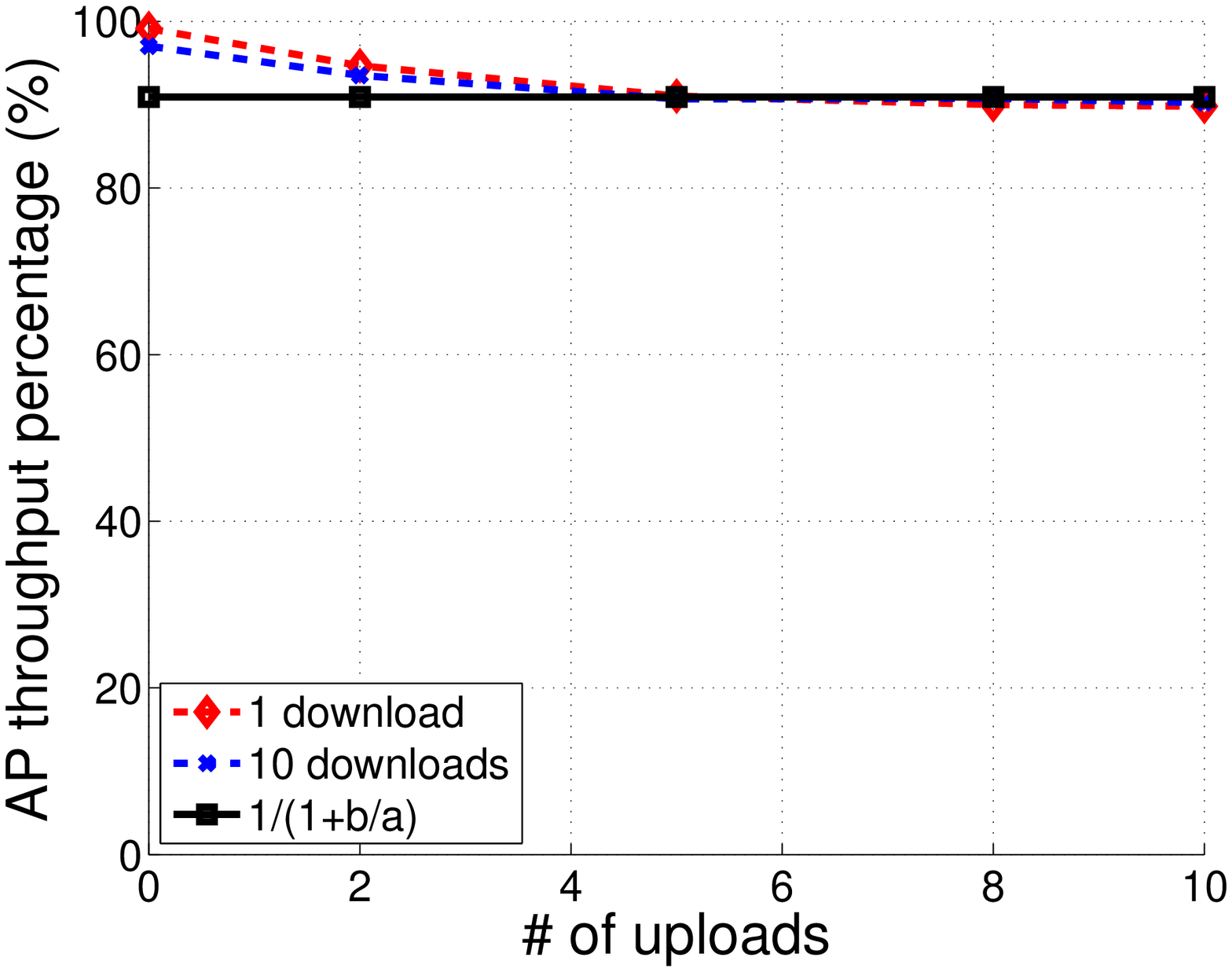}\label{fig_alt_varyul_thru}}
   \subfigure[54/6Mbps, delay]{\includegraphics[width=0.47\columnwidth]{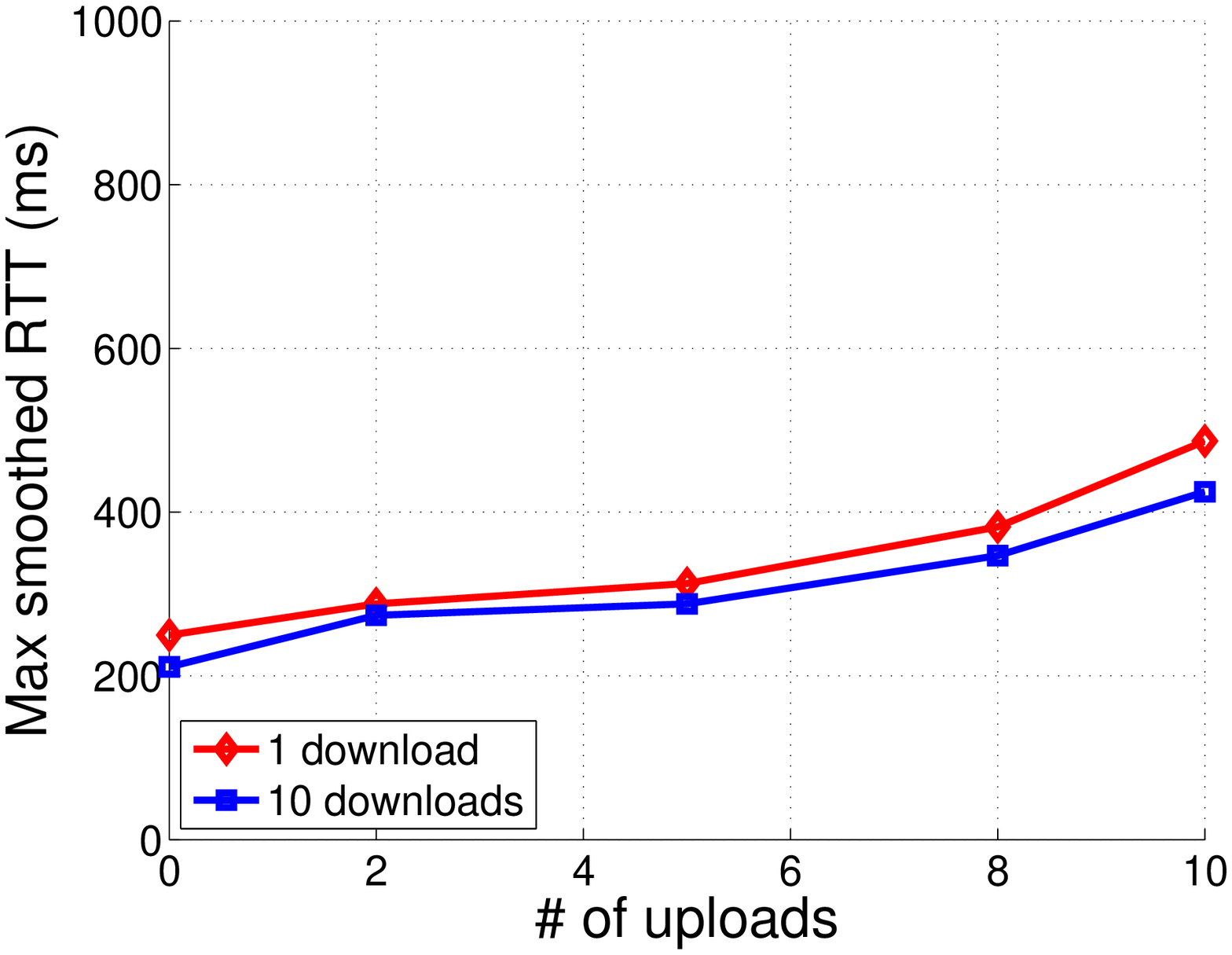}\label{fig_alt_varyul_srtt}}
   \subfigure[216/54Mbps, throughput]{\includegraphics[width=0.45\columnwidth]{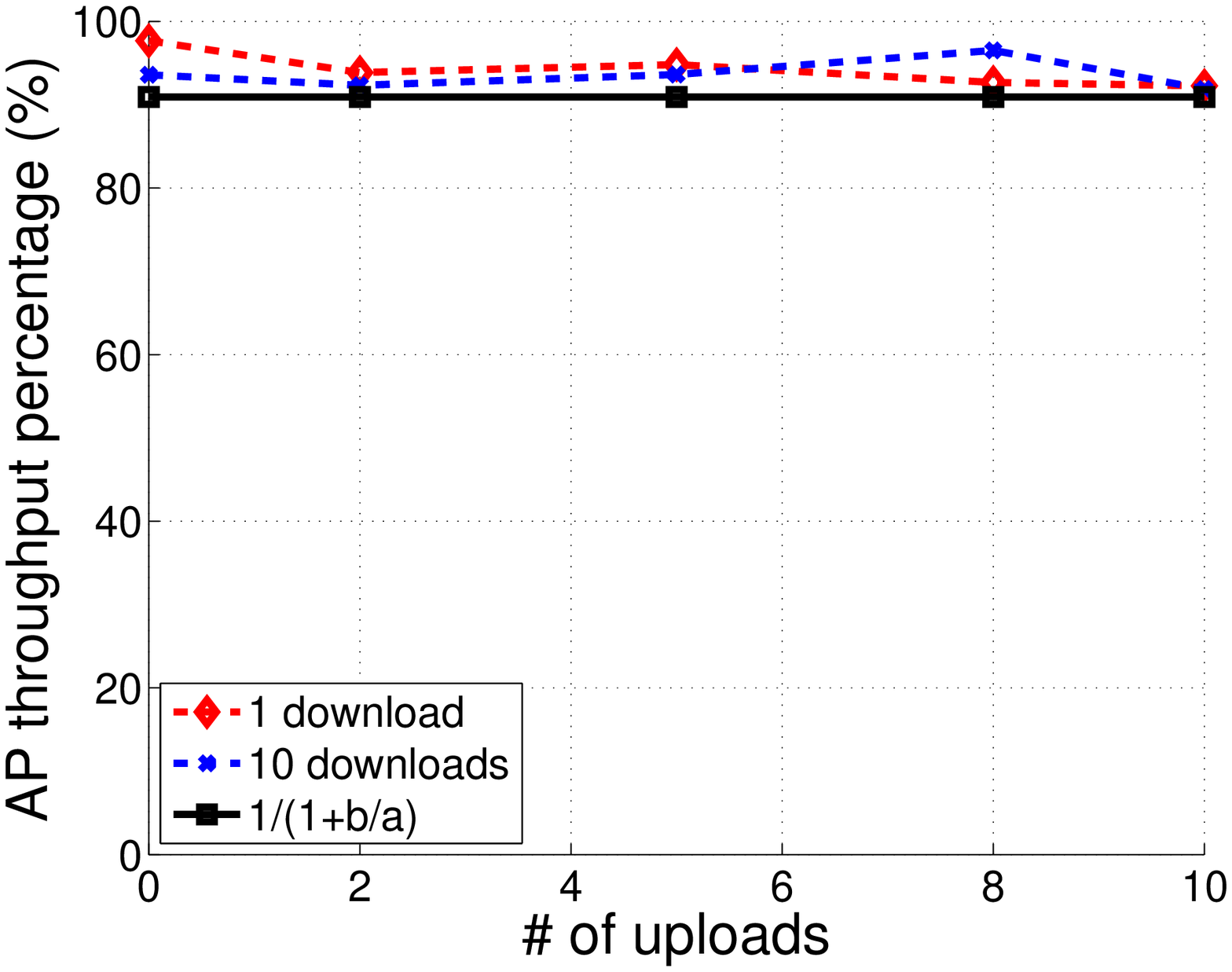}}
   \subfigure[216/54Mbps, delay]{\includegraphics[width=0.47\columnwidth]{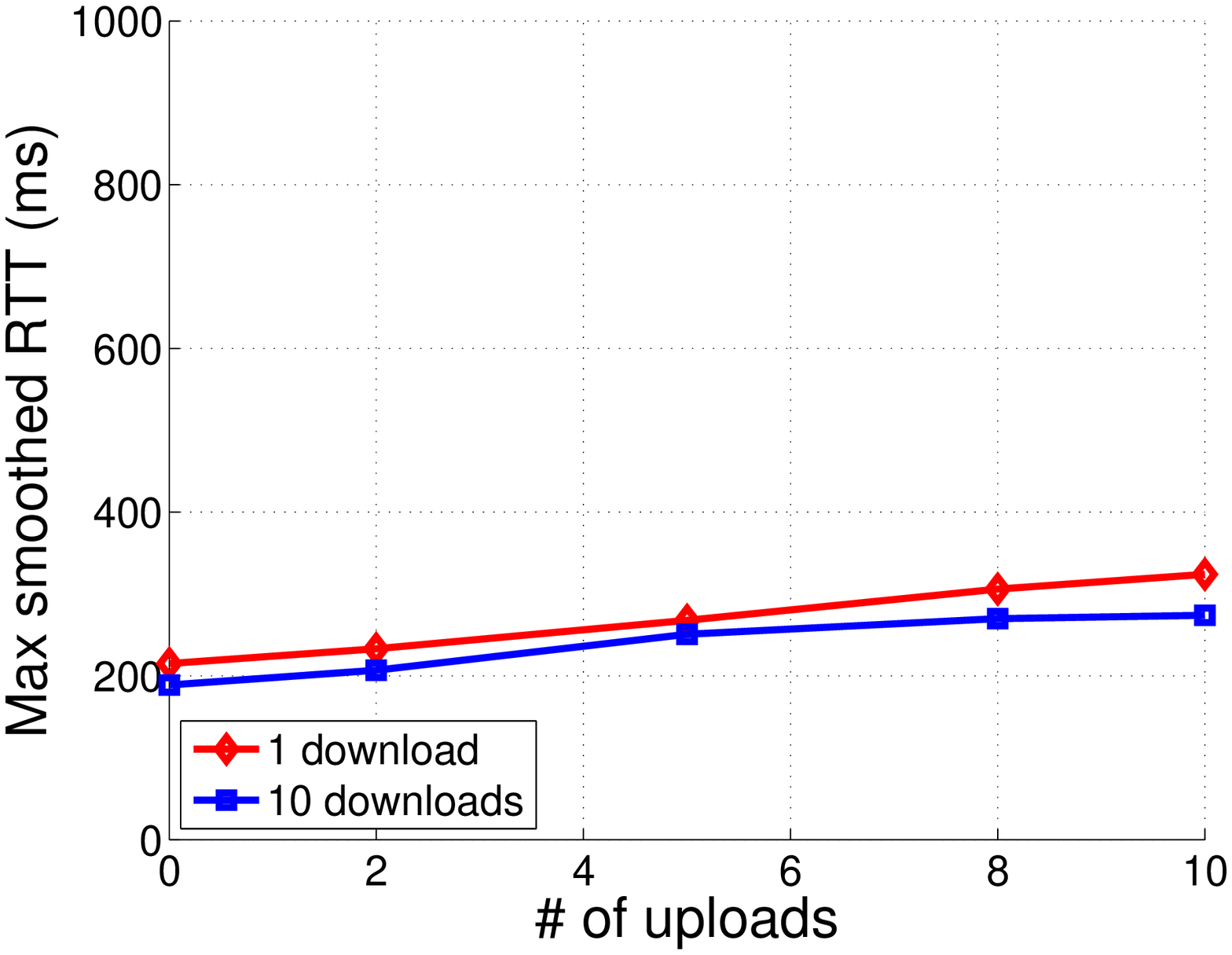}}
   \caption{Throughput efficiency and maximum smoothed round trip delays (max sRTT) for the topology in Fig. \ref{fig_topo_simu} when the A* algorithm is used. Here, the AP throughput efficiency is the ratio between the throughput achieved using the A* algorithm and the maximum throughput achieved using fixed size buffers. Rates before and after the '/' are used physical layer data and basic rates. For the 216Mbps data, 8 packets are aggregated into each frame at the MAC layer to improve throughput efficiency in an 802.11n-like scheme.  The wired RTT is 200 ms.}
   \label{fig_216m}
\end{figure*}

The basic impetus for the design of the A* algorithm is to exploit the
possibility of statistical multiplexing to reduce buffer sizes. Fig.
\ref{fig_astar_histories_0ul_10dl} illustrates the performance of the A* algorithm when there are
10 downloads and no upload flows.  Comparing with the results in Fig.
\ref{fig_eBDP_histories10} using fixed size buffers, we can see that the A* algorithm can
achieve significantly smaller buffer sizes (i.e., a reduction from more than 350 packets to 100
packets approximately) when multiplexing exists.  Fig.~\ref{fig_216m} summarizes the throughput and delay performance of the A* algorithm for a range of network conditions (numbers of uploads and downloads) and physical transmit rates ranging from  1Mbps to 216Mbps.  This can be compared with Fig. \ref{fig_motivation}.  It can be seen that in comparison with the use of a fixed buffer size the A* algorithm is able to achieve high throughput efficiency across a wide range of operating conditions while minimizing queuing delays.

\begin{figure}[tb]
   \centering
   \subfigure[Throughput efficiency]{\includegraphics[width=0.48\columnwidth]{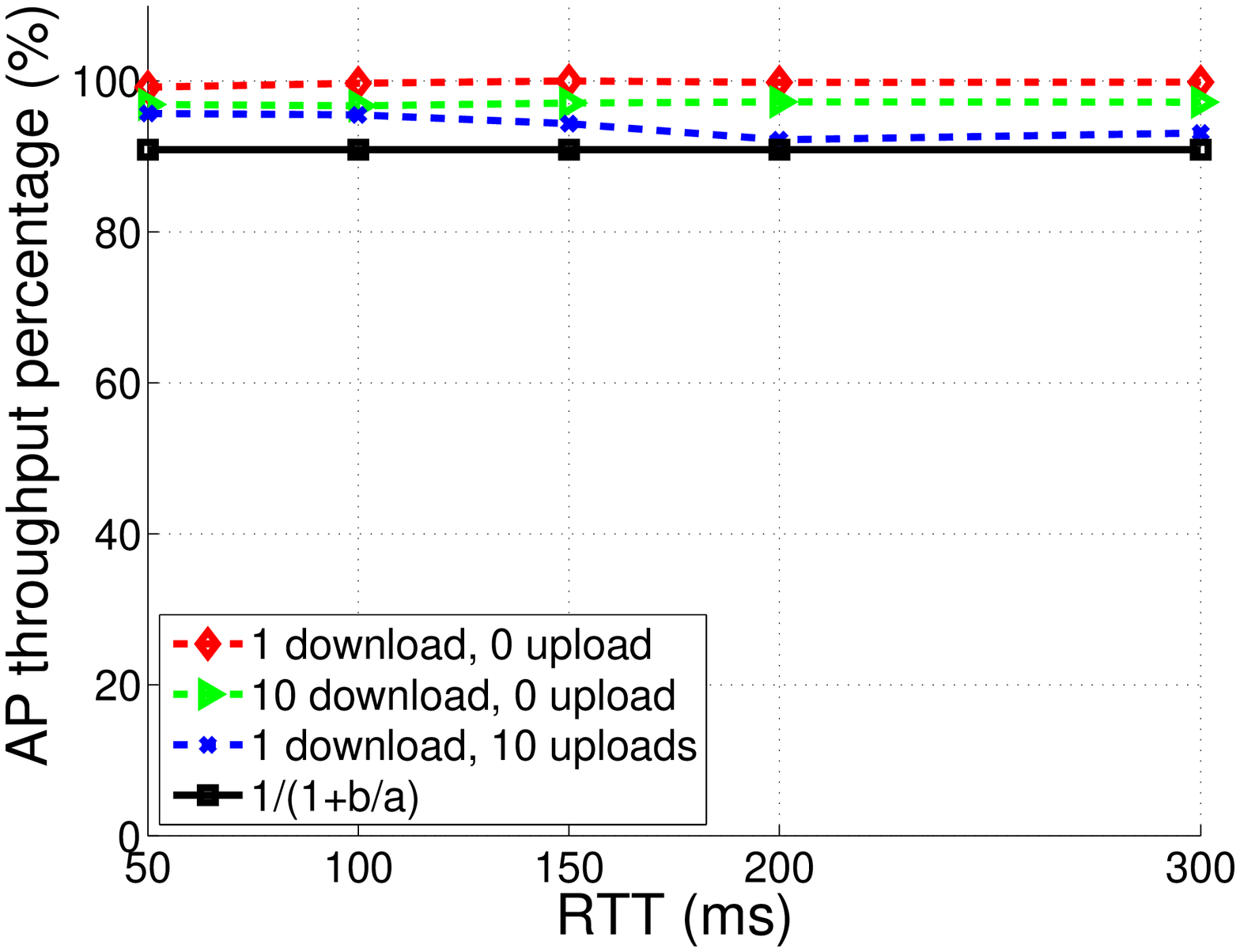}}
   \subfigure[Delay]{\includegraphics[width=0.48\columnwidth]{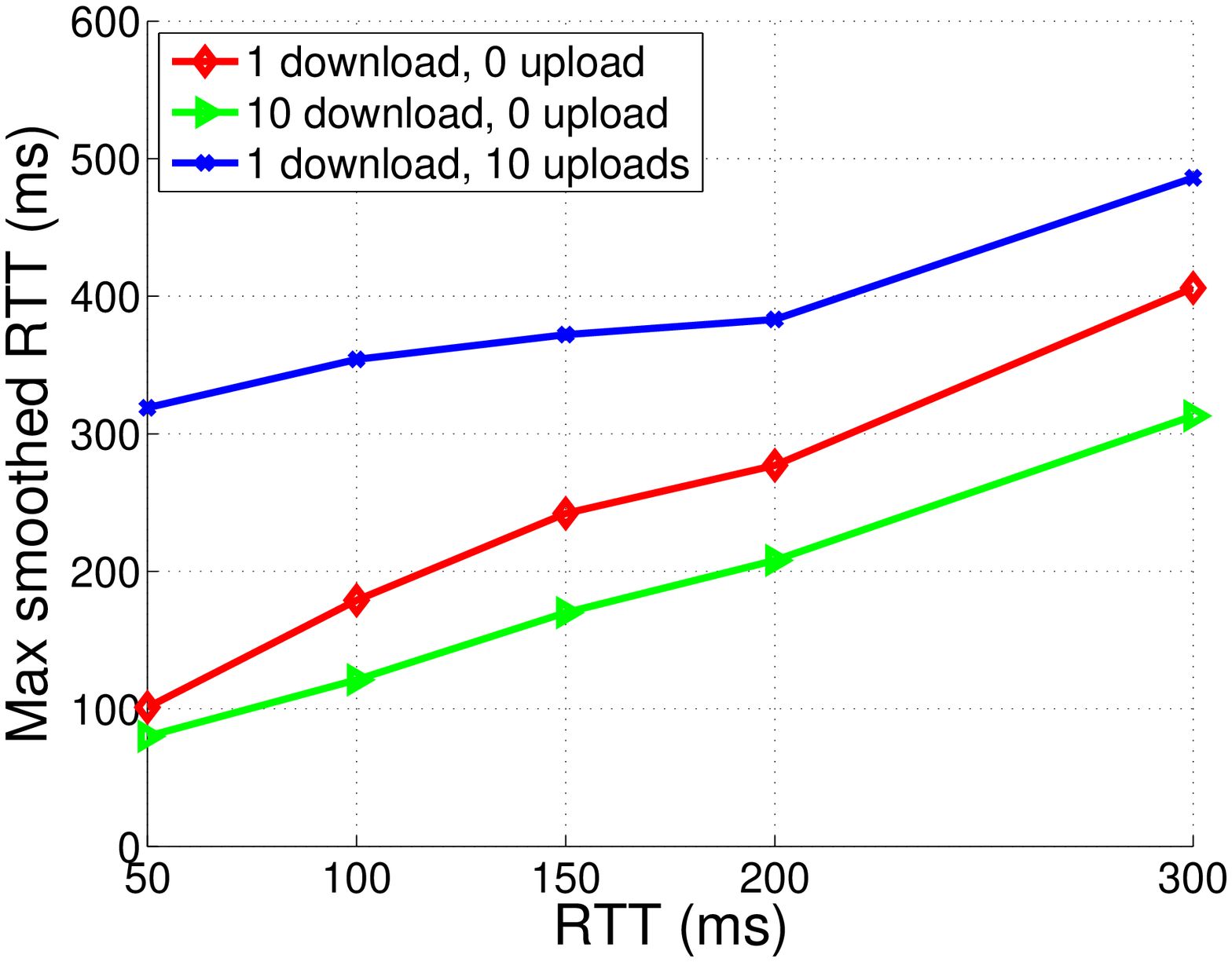}}
   \caption{Performance of the A* algorithm as the wired RTT is varied. Physical layer data and basic rates are 54 and 6 Mbps. Here the AP throughput percentage is the ratio between the throughput achieved using the A* algorithm and the maximum throughput using fixed size buffers.}
   \label{fig_alt_rtt}
\end{figure}

In Fig. \ref{fig_alt_rtt} we further evaluate the
A* algorithm when the wired RTTs are varied from 50-300ms and the number of uploads is varied
from 0-10. Comparing these with the results (Fig. \ref{fig_eBDP_varyuls} and
\ref{fig_eBDP_varyrtt}) of the eBDP algorithm we can see that the A* algorithm is capable
of exploiting the statistical multiplexing where feasible. In particular, significantly
lower delays are achieved with 10 download flows whilst maintaining comparable throughput
efficiency.

\subsection{Impact of Channel Errors}
\begin{figure}[tb]
   \centering
   \subfigure[Throughput]{\includegraphics[width=0.48\columnwidth]{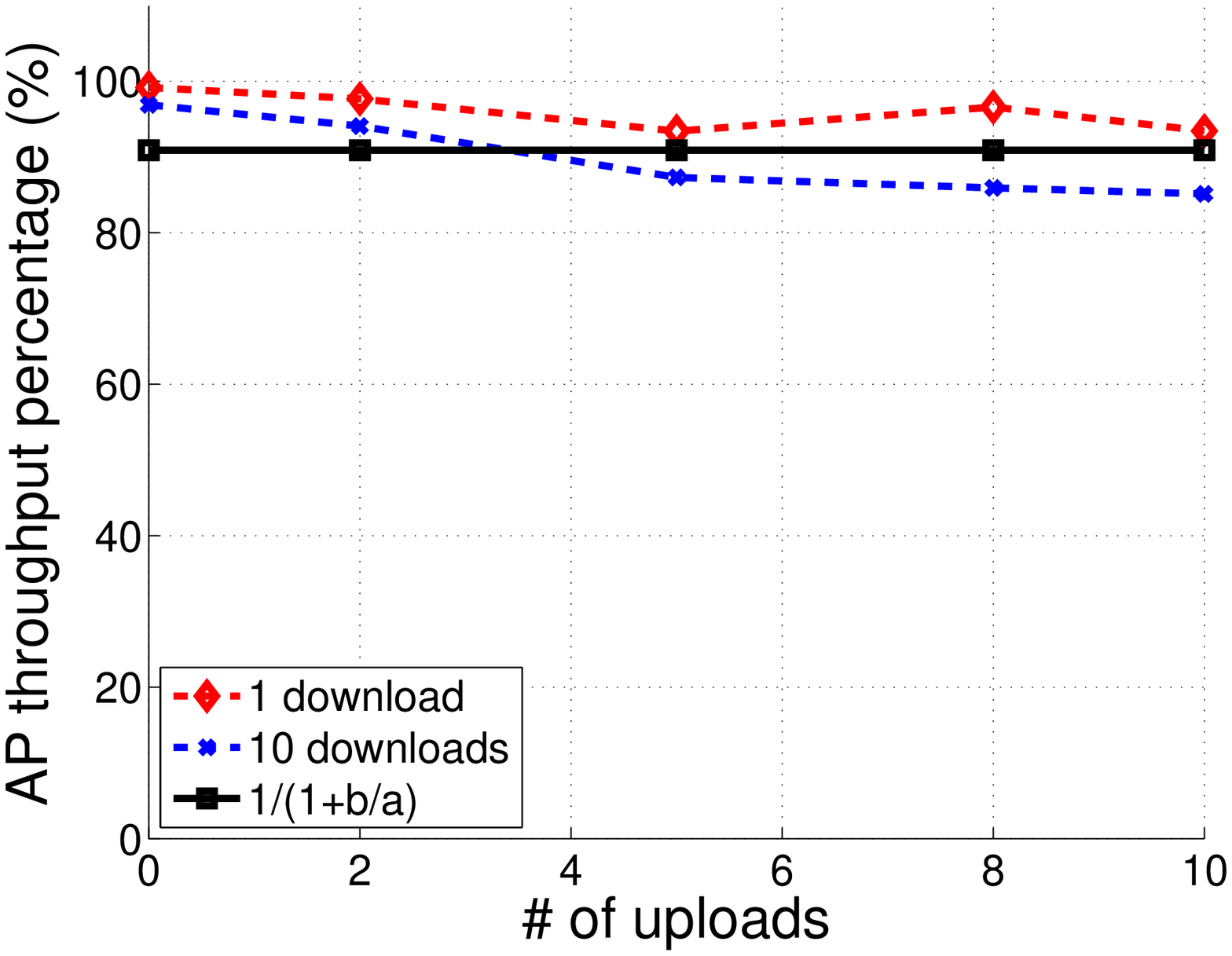}}
   \subfigure[Delay]{\includegraphics[width=0.48\columnwidth]{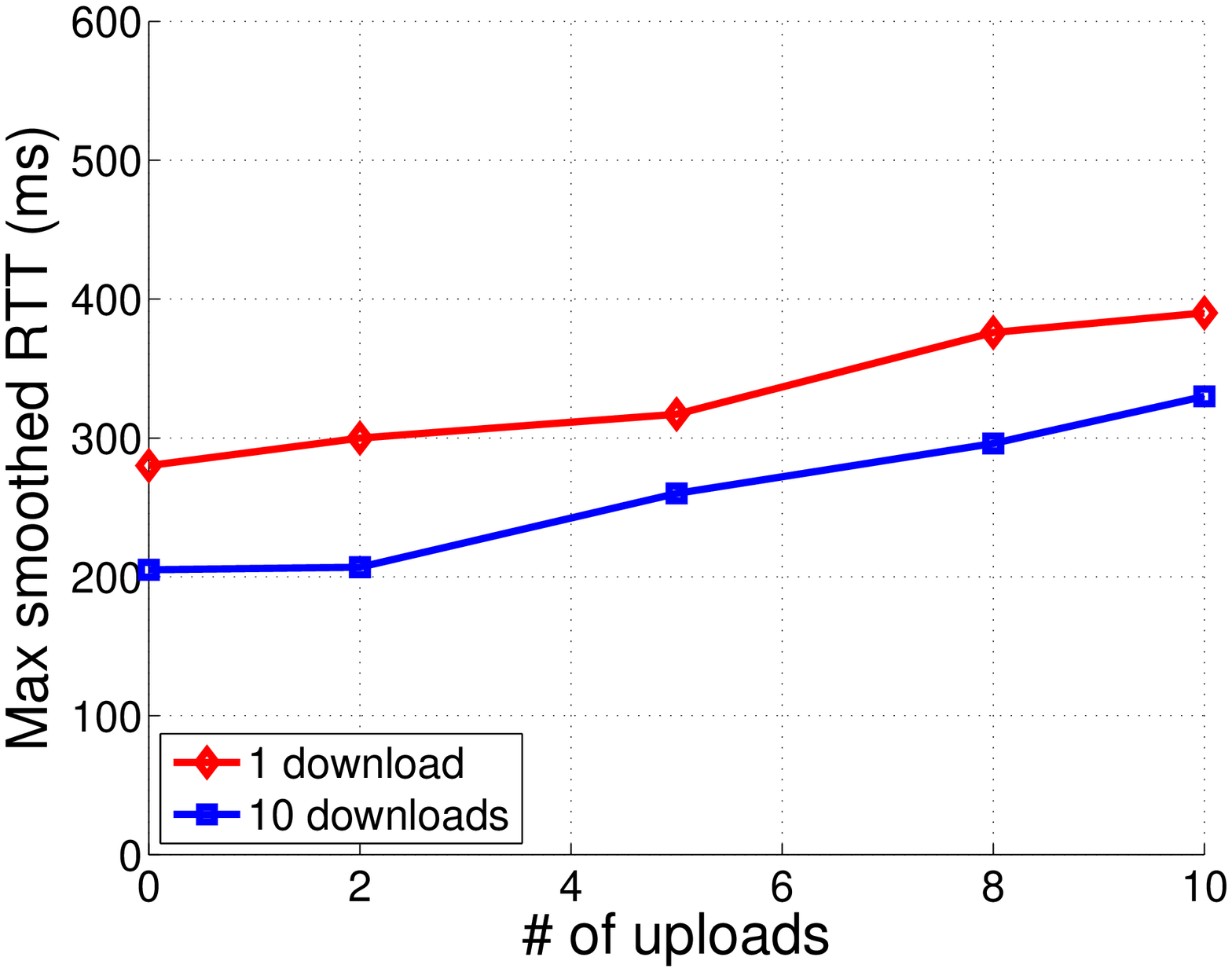}}
   \caption{Performance of the A* algorithm when the channel has a BER of $10^{-5}$. Physical layer data and basic rates are 54 and 6 Mbps. Here the AP throughput percentage is the ratio between the throughput achieved using the A* algorithm and the maximum throughput using fixed size buffers.}
   \label{fig_channelerror}
\end{figure}

In the foregoing simulations the channel is error free and packet losses are solely due to buffer overflow and MAC-layer collisions.  In fact, channel errors have only a minor impact on the effectiveness of buffer sizing algorithms as errors play a similar role to collisions with regard to their impact on link utilization. We support this claim first using a simulation example with a channel having an i.i.d noise inducing a bit error rate (BER) of $10^{-5}$.  Results are shown in Fig. \ref{fig_channelerror} where we can see a similar trend as in the cases when the medium is error free (Figs. \ref{fig_alt_varyul_thru} \ref{fig_alt_varyul_srtt}).

We further confirm this claim in our test-bed implementations where tests were conducted in 802.11b/g channels and noise related losses were observed. See Section \ref{sec_expt} for details.

\subsection{DCF Operation}

The proposed buffer sizing algorithms are still valid for DCF since link utilization and delay considerations remain applicable, as is the availability of service time (for the eBDP algorithm) and idle/busy time measurements (for the ALT algorithm). In particular, if the considered buffer is heavily backlogged, to ensure low delays, the buffer size should be reduced. If otherwise the buffer lies empty, it may be due to that the current buffer size is too small which causes the TCP source backs off after buffer overflow. To accommodate more future packets, the buffer size can be increased. Note that increasing buffer sizes in this case would not lead to high delays but has the potential to improve throughput. This tradeoff between the throughput and the delays thus holds for both EDCA and DCF.

\begin{figure}[tb]
   \centering
   \subfigure[Throughput]{\includegraphics[width=0.48\columnwidth]{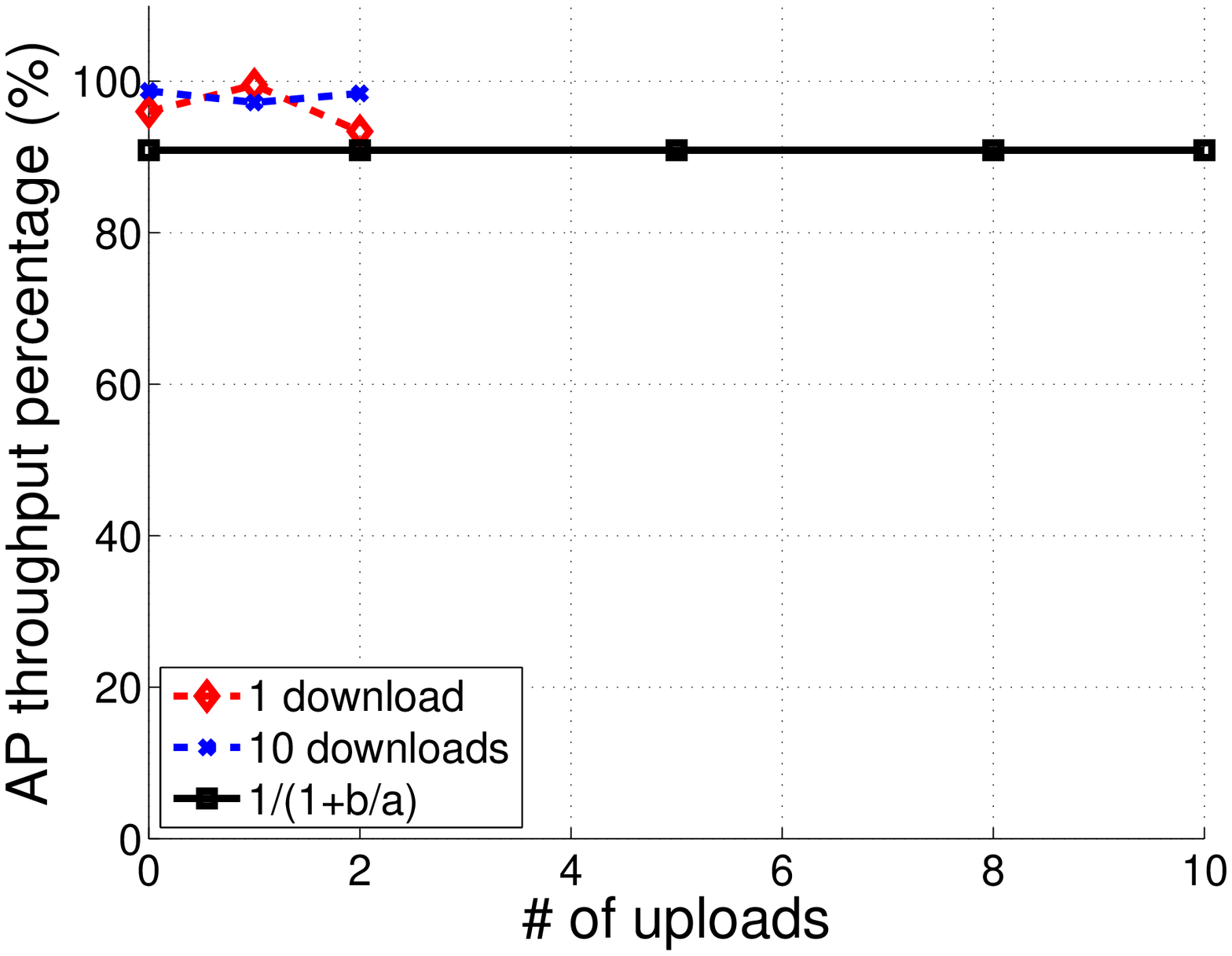}}
   \subfigure[Delay]{\includegraphics[width=0.48\columnwidth]{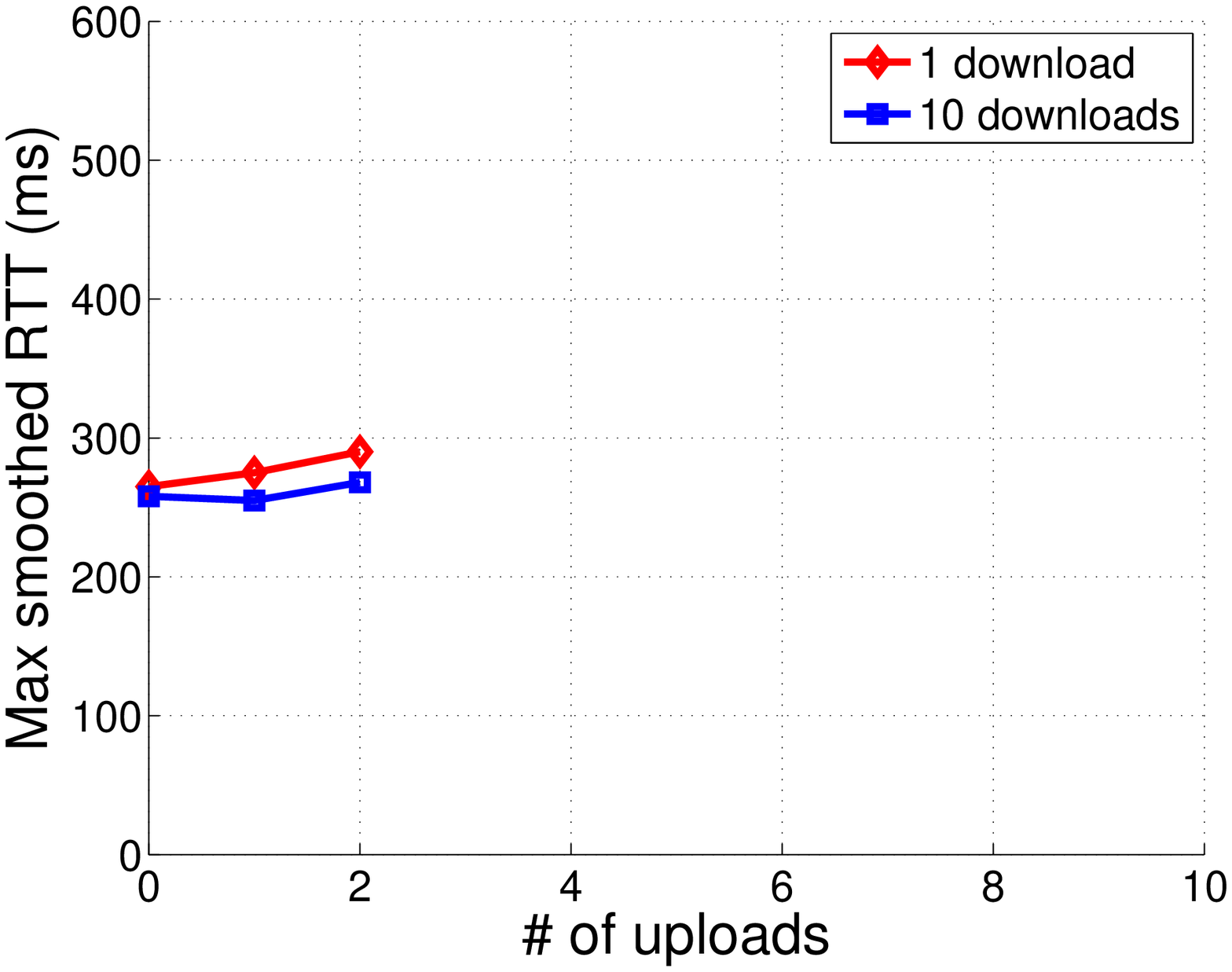}}
   \caption{Performance of the A* algorithm for 802.11 DCF operation when there are both upload and download flows in the network. Here the AP throughput percentage is the ratio between the throughput achieved using the A* algorithm and the maximum throughput using fixed size buffers. Physical layer data and basic rates used are 54 and 6 Mbps.}
   \label{fig_dcfvsastar}
\end{figure}

However, the DCF allocates roughly equal numbers of transmission opportunities to stations.  A consequence of using DCF is thus that when the number of upload flows increases, the uploads may produce enough TCP ACK packets to keep the AP's queue saturated. In fact, once there are two upload flows, TCP becomes unstable due to repeated timeouts (see \cite{David_winmee_2008} for a detailed demonstration), causing the unfairness issue discussed in Section \ref{subsec_tcp_unfairness}. Therefore, we present results for up to two uploads in Fig. \ref{fig_dcfvsastar}, as this is the greatest number of upload flows where TCP with DCF can exhibit stable behavior using both fixed size buffers and the A* algorithm. Note that in this case using the A* algorithm on upload stations can also decrease the delays and maintain high throughput efficiency if their buffers are frequently backlogged.

\begin{figure}[tb]
   \centering
   \subfigure[Throughput]{\includegraphics[width=0.48\columnwidth]{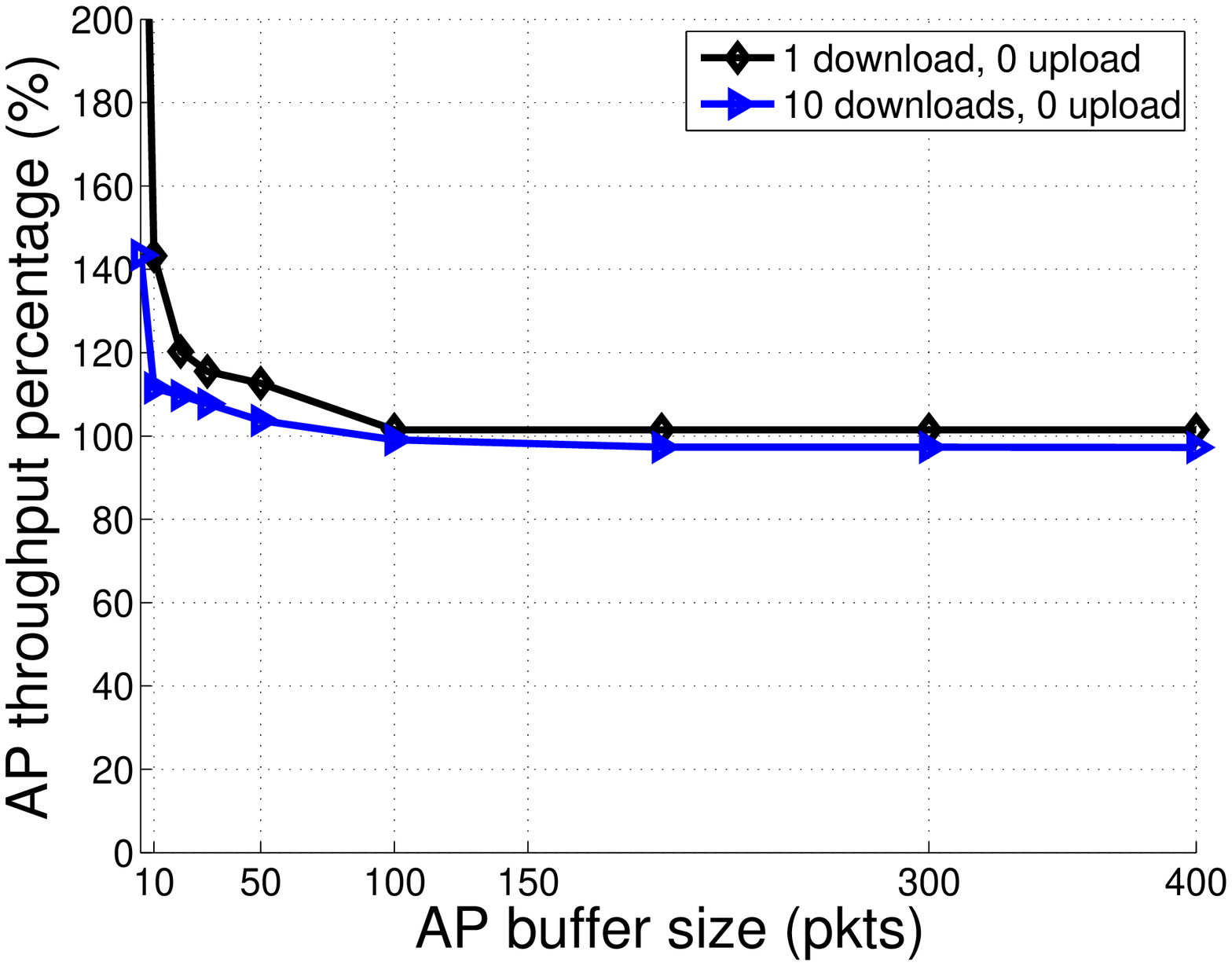}}
   \subfigure[Delay]{\includegraphics[width=0.48\columnwidth]{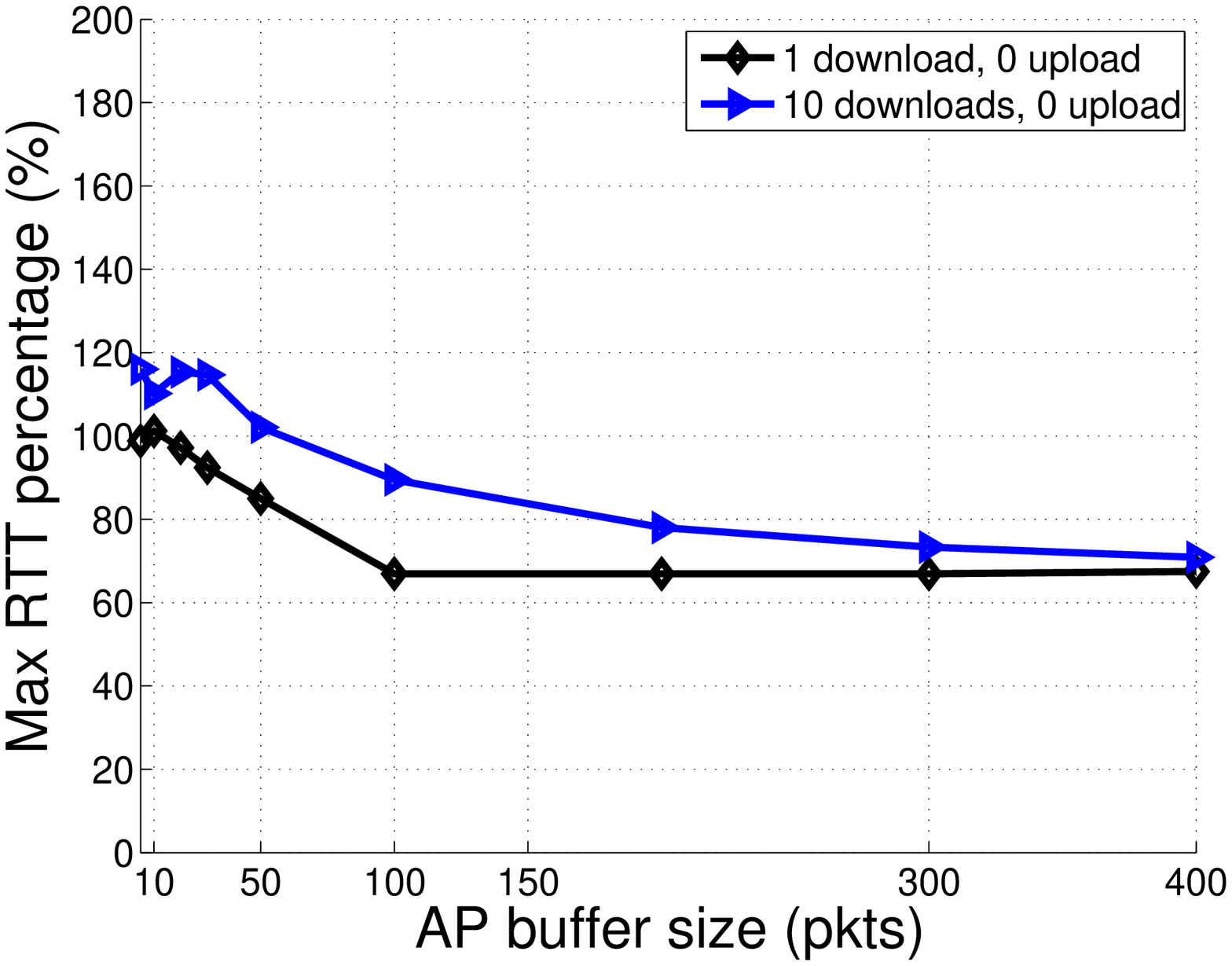}}
   \caption{Performance of the A* algorithm for 802.11 DCF operation when there are download flows only in the network. Here we illustrate the percentage (between the results achieved using the A* algorithm and those using varied AP buffer sizes as shown on the x-axis) of both throughput and delays. Physical layer data and basic rates used are 54 and 6 Mbps. }
   \label{fig_dcf}
\end{figure}

We also present results when there are download flows only (so the unfairness issue does not exist). Fig. \ref{fig_dcf} illustrates the throughput and delay performance achieved using the A* algorithm and fixed 400-packet buffers. As in the EDCA cases, we can see that the A* algorithm is able to maintain a high throughput efficiency with comparatively low delays.

Note that DCF is also used in the production WLAN test where the A* algorithm is observed to perform well (see Section \ref{sec_intr}).

\subsection{Rate Adaptation}
We did not implement rate adaptation in our simulations.  However, we did implement the
A* algorithm in the Linux MadWifi driver which includes rate adaptation algorithms.  We
tested the A* algorithm in the production WLAN of the Hamilton Institute with the default
SampleRate algorithm enabled. See Section \ref{sec_intr}.

\section{Experimental Results} \label{sec_expt}

We have implemented the proposed algorithms in the Linux MadWifi driver, and in this section we
present tests on an experimental testbed located in an office environment and introduce
results illustrating operation with complex traffic that includes both TCP and UDP, a mix
of uploads and downloads, and a mix of connection sizes.

\subsection{Testbed Experiment}
\begin{figure}[tb]
    \centering
    \includegraphics[width=\columnwidth]{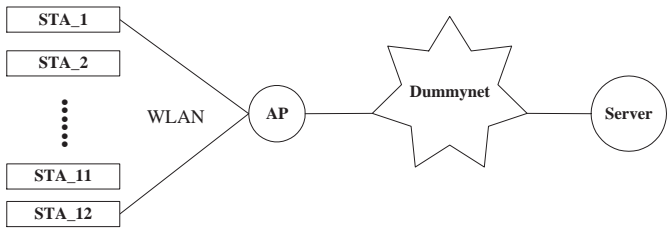}
    \caption{Topology used in experimental tests.   }
    \label{fig_topo}
\end{figure}

The testbed topology is shown in Fig. \ref{fig_topo}. A wired network is emulated using
a desktop PC running dummynet software on FreeBSD 6.2 which enables link rates and
propagation delays to be controlled. The wireless AP and the server are connected to the
dummynet PC by 100Mbps Ethernet links.  Routing in the network is statically configured.
Network management is carried out using ssh over a wired control plane to avoid affecting
wireless traffic.

In the WLAN, a desktop PC is used as the AP and 12 PC-based embedded Linux boxes based on
the Soekris net4801 are used as client stations. All are equipped with an Atheros
802.11b/g PCI card with an external antenna. All nodes run a Linux 2.6.21.1 kernel and a
MadWifi wireless driver (version r2366) modified to allow us to adjust the 802.11e
$CW_{min}$, $CW_{max}$ and $AIFS$ parameters as required. Specific vendor features on the
wireless card, such as turbo mode, rate adaptation and multi-rate retries, are disabled.
All of the tests are performed using a transmission rate of 11Mbps (i.e., we use an
802.11b PHY) with RTS/CTS disabled and the channel number explicitly set. Channel 1 has
been selected to carry out the experiments. The testbed is not in an isolated radio
environment, and is subject to the usual impairments seen in an office environment. Since
the wireless stations are based on low power embedded systems, we have tested these
wireless stations to confirm that the hardware performance (especially the CPU) is not a
bottleneck for wireless transmissions at the 11Mbps PHY rate used. The configuration of
the various network buffers and MAC parameters is detailed in
Table~\ref{tab:paramsummary}.

\begin{table}[tb]
    \begin{center}
            \begin{tabular}{|l|l|}
            \hline
            \textbf{Parameters}      & \textbf{Values}  \\
            \hline
            Interface tx queue & 2 packets  \\
            \hline
            Dummynet queue     & 100 packets  \\
            \hline
            MAC Preamble  & long \\
            \hline
            MAC data rate  & 11Mbps \\
            \hline
            MAC ACK rate & 11Mbps\\
            \hline
            MAC retries & 11 \\
            \hline
            \end{tabular}
    \end{center}
    \caption{Testbed parameters summary.} \label{tab:paramsummary}
\end{table}

Although both SACK enabled TCP NewReno and TCP CUBIC with receiver buffers of 4096KB have been
tested, here we only report the results for the latter as CUBIC is now the default
congestion control algorithm used in Linux.  Default values of Linux Kernel 2.6.21.1 are
used for all the other TCP parameters. We put TCP ACK packets into a high priority queue
(we use the WME\_AC\_VO queue of MadWifi as an example) which is assigned parameters of
$CW_{min}=3$, $CW_{max}=7$ and $AIFS=2$. TCP data packets are collected into a lower
priority queue (we use the WME\_AC\_VI queue) which is assigned $CW_{min}=31$,
$CW_{max}=1023$ and $AIFS=6$. We use iperf to generate TCP traffic and results are
collected using both iperf and tcpdump.

\subsection{Traffic Mix}

We configure the traffic mix on the network to capture the complexity of real networks in
order to help gain greater confidence in the practical utility of the proposed buffer
sizing approach.   With reference to Fig. \ref{fig_topo} showing the network topology, we
create the following traffic flows:

\begin{itemize}

\item \emph{TCP uploads}. One long-lived TCP upload from each of STAs 1, 2 and 3 to the
server in the wired network. STAs 2 and 3 always use a fixed 400-packet buffer, while STA
1 uses both a fixed 400-packet buffer and the A* algorithm.

\item \emph{TCP downloads}. One long-lived TCP download from the wired server to each of
STAs 4, 5 and 6.

\item \emph{Two way UDP}.  One two-way UDP flow from the wired server to STA 7. The packet size used is 64 bytes and the mean inter-packet interval is 1s.  Another UDP flow from the wired server to STA 8 with the used packet size of 1000 bytes and the mean inter-packet interval of 1s.

\item \emph{Mix of TCP connection sizes}.  These flows mimic web traffic\footnote{Note that in the production WLAN test, we used real web traffic.}. A short TCP download from the wired server to STA 9, the connection size of which is 5KB (approximately 3 packets).   A slightly longer TCP download from the wired server to STA 10 with a connection size of 20KB (approximately 13 packets) and another to STA 11 (connection size 30KB, namely, around 20 packets).  A fourth connection sized 100KB from the server to STA 12.  For each size of these connection, a new flow is started every 10s to allow collection of statistics on the mean completion time.

\end{itemize}

\subsection{Results}

\begin{figure}[tb]
    \centering
    \includegraphics[width=0.5\columnwidth]{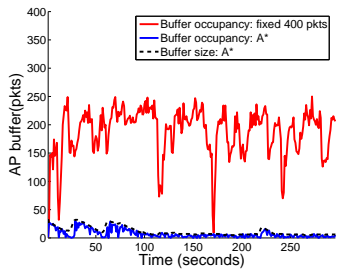}
     \caption{Buffer size and occupancy time histories measured at the AP with fixed 400-packet buffers and the A* algorithm.}
    \label{fig_cubic}
\end{figure}

Fig. \ref{fig_cubic} shows example time histories of the buffer size and occupancy at the AP with a fixed buffer size
of 400 packets and when the A* algorithm is used for dynamic buffer sizing.  Note that in this example the 400 packet
buffer never completely fills. Instead the buffer occupancy has a peak value of around 250 packets.  This is due to
non-congestive packet losses caused by channel noise (the testbed operates in a real office environment with
significant interference, because there are bluetooth devices and WLANs working in channel 1.) which prevent the TCP congestion window from growing to completely fill the buffer. Nevertheless, it can be seen that the buffer rarely empties and thus it is sufficient to provide an indication of the throughput when the wireless link is fully utilized.

We observe that while buffer histories are very different with a fixed size buffer and the A* algorithm, the throughput is very similar in these two cases (see Table \ref{tab_cubic}).

One immediate benefit of using smaller buffers is thus a reduction in network delays.
Table \ref{tab_delay} shows the measured delays experienced by the UDP flows sharing the WLAN
with the TCP traffic.  It can be seen that for STA 8 both the mean and the maximum delays are
significantly reduced when the A* algorithm is used. This potentially has major
implications for time sensitive traffic when sharing a wireless link with data traffic. Note that the queuing delays from STA 7 are for traffic passing through the high-priority traffic class used for TCP ACKs, while the measurements from STA 8 are for traffic in the same class as TCP data packets.  For the offered loads used, the service rate of the high-priority class is sufficient to avoid queue buildup and this is reflected in the measurements.

\begin{figure}[tb]
    \centering
    \includegraphics[width=0.5\columnwidth]{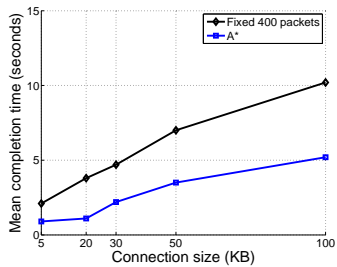}
    \caption{Measured completion time vs connection size. Results are averages of multiple runs.}
    \label{fig_motivation_short}
\end{figure}

The reduction in network delay not only benefits UDP traffic, but also short-lived TCP
connections.   Fig. \ref{fig_motivation_short} shows the measured completion time vs
connection size for TCP flows.  It can be seen that the completion time is consistently lower by a factor of at least two when A* dynamic buffer sizing is used. Since the majority
of internet flows are short-lived TCP connections (e.g., most web traffic), this
potentially translates into a significant improvement in user experience.

Note that STA's 2 and 3 in the A* column of Table \ref{tab_cubic} use fixed size buffers rather than the A* algorithm.   The results shown are the throughput they achieve when other stations run the A* algorithm. It can be seen that the A* algorithm does not significantly impact STAs 2 and 3, confirming that A* can support incremental roll-out without negatively impacting legacy stations that are using fixed size buffers.

\begin{table}[tb]
    \begin{center}
        {\footnotesize
            \begin{tabular}{|l|l|l|}
            \hline
                                    & \textbf{Fixed 400 packets} & \textbf{A*}  \\
            \hline
            Throughput of STA 1     & 1.36Mbps                  & 1.33Mbps  \\
            \hline
            Throughput of STA 2     & 1.29Mbps                  & 1.30Mbps  \\
            \hline
            Throughput of STA 3     & 1.37Mbps                  & 1.33Mbps  \\
            \hline
            Throughput of STA 4     & 0.35Mbps                  & 0.41Mbps  \\
            \hline
            Throughput of STA 5     & 0.39Mbps                  & 0.39Mbps  \\
            \hline
            Throughput of STA 6     & 0.52Mbps                  & 0.42Mbps  \\
            \hline
            \end{tabular}
        }
    \end{center}
    \caption{Measured throughput.} \label{tab_cubic}
\end{table}

\begin{table}[tb]
\centering
    \begin{footnotesize}
            \begin{tabular}{|l|l|l|}
            \hline
                           & \textbf{Fixed 400 packets}    & \textbf{A*}  \\
            \hline
                           & mean (max)                 & mean (max)   \\
            \hline
            RTT to STA 7   & 201ms (239ms)              & 200ms (236ms)  \\
            \hline
            RTT to STA 8   & 1465ms  (2430ms)           & 258ms  (482ms) \\
            \hline
            \end{tabular}
            \caption{Measured delays of the UDP flows. STA 7's traffic is prioritized to avoid queue buildup and this is reflected in the measurements.}
            \label{tab_delay}
    \end{footnotesize}

\end{table}

\section{Related Work}\label{sec_related_work}

The classical approach to sizing Internet router buffers is the BDP rule proposed in
\cite{Vallamizar_CCR_1994}. Recently, in \cite{Appenzeller_SIGCOMM_2004} it is argued
that the BDP rule may be overly conservative on links shared by a large number of flows.
In this case it is unlikely that TCP congestion window sizes (cwnd) evolve synchronously
and due to statistical multiplexing of cwnd backoff, the combined buffer requirement can
be considerably less than the BDP. The analysis in \cite{Appenzeller_SIGCOMM_2004}
suggests that it may be sufficient to size buffers as $BDP/\sqrt{n}$.  This work is
extended in \cite{Raina_NGI_2005}, \cite{Enachescu_INFOCOM_2006} and
\cite{Wischik_CCR_2005} to consider the performance of TCP congestion control with many
connections under the assumption of small, medium and large buffer sizes. Several authors
have pointed out that the value $n$ can be difficult to determine for realistic traffic
patterns, which not only include a mix of connections sizes and RTTs, but can also be
strongly time-varying \cite{Dhamdhere_ccr_2006}, \cite{Vu-Brugier_CCR_2007}. In
\cite{Vu-Brugier_CCR_2007}, it is observed from measurements on a production link that
traffic patterns vary significantly over time, and may contain a complex mix of flow
connection lengths and RTTs. It is demonstrated in
\cite{Dhamdhere_ccr_2006}\cite{Vu-Brugier_CCR_2007} that the use of very small buffers
can lead to an excessive loss rate.  Motivated by these observations, in
\cite{Rade_Letters_2006} \cite{Kellett_CDC_2006} a measurement-based adaptive buffer size
tuning method is proposed.  However, this approach is not applicable to WLANs since it
requires a priori knowledge of the link capacity or line rate, which in WLANs is
time-varying and load dependent.  \cite{Zhang_Iwqos_2008} introduces another adaptive buffer sizing algorithm based on control theory for Internet core routers. \cite{Prasad_ton_toappear,Lakshmikantha_infocom_2008} consider the role of the output/input capacity ratio at a network link in determining the required buffer size. \cite{Beheshti_imc_2008} experimentally investigates the analytic results reported in \cite{Appenzeller_SIGCOMM_2004}, \cite{Raina_NGI_2005}, \cite{Enachescu_INFOCOM_2006} and \cite{Wischik_CCR_2005}.   \cite{Eun_Ton_2008} considers sizing buffers managed with active queues management techniques.

The foregoing work is in the context of wired links, and to our knowledge the question of
buffer sizing for 802.11 wireless links has received almost no attention in the
literature. Exceptions include \cite{Malone_BufferSizing_voip}
\cite{Pilosof_INFOCOM_2003} \cite{Thottan_wicon_2006}. Sizing of buffers for voice
traffic in WLANs is investigated in \cite{Malone_BufferSizing_voip}.  The impact of fixed
buffer sizes on TCP flows is studied in \cite{Pilosof_INFOCOM_2003}. In
\cite{Thottan_wicon_2006}, TCP performance with a variety of AP buffer sizes and 802.11e
parameter settings is investigated.  In \cite{Li_commletter_2007}
\cite{Li_chinacom_2008}, initial investigations are reported related to the eBDP
algorithm and the ALT algorithm of the A* algorithm. We substantially extend the previous
work in this paper with theoretical analysis, experiment implementations in both testbed
and a production WLAN, and additional NS simulations.

\section{Conclusions } \label{sec_concl}

We consider the sizing of network buffers in 802.11 based wireless networks.   Wireless
networks face a number of fundamental issues that do not arise in wired networks.  We
demonstrate that the use of fixed size buffers in 802.11 networks inevitably leads to
either undesirable channel under-utilization or unnecessary high delays.  We present two
novel buffer sizing algorithms that achieve high throughput while maintaining low delay across a wide range of network conditions.   Experimental measurements demonstrate the utility of the proposed algorithms  in a real environment with real traffic.

The source code used in the NS-2 simulations and the experimental implementation in
MadWifi can be downloaded from www.hamilton.ie/tianji\_li/buffersizing.html.


\section*{Appendix}

In the production WLAN of the Hamilton Institute, the AP is equipped with an Atheros 802.11/a/b/g PCI card and an external antenna. The operating system is a recent Fedora 8 (kernel version 2.6.24.5). The latest MadWifi driver version (0.9.4) is used, in which the buffer size is fixed at 250 packets. The AP is running in 802.11g mode with the default rate adaptation algorithm enabled (i.e., SampleRate \cite{Bicket_msthesis_2005}). All data traffic is processed via the Best
Effort queue, i.e., MadWifi is operating in 802.11 rather than 802.11e mode. A
mix of Windows/Apple MAC/Linux laptops and PCs use the WLAN from time to time.

\begin{figure}[b]
    \centering
    \includegraphics[width=0.7\columnwidth]{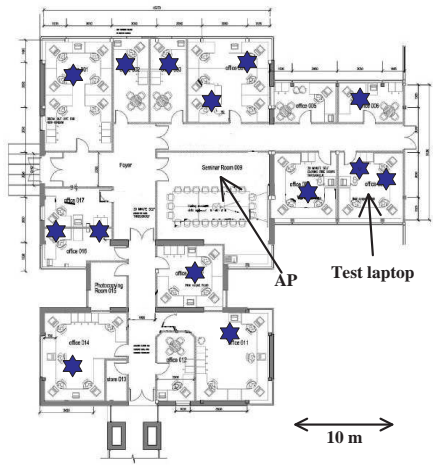}
    \caption{WLAN of the Hamilton Institute. Stars represent users' approximate locations.}
    \label{fig_hi_topo}
\end{figure}


\begin{IEEEbiography}{Tianji Li}
received the M.Sc. (2004) degree in networking and
distributed computation from \'{E}cole Doctorale STIC,
Universit\'{e} de Nice-Sophia Antipolis, France, and the  Ph.D. (2008) degree from the Hamilton
Institute, National University of Ireland Maynooth, Ireland, where he is currently a research fellow. He is interested in improving performance for computer and telecommunication networks.
\end{IEEEbiography}

\begin{IEEEbiography}{Douglas Leith}
graduated from the University of Glasgow in 1986 and was awarded his PhD, also from the
University of Glasgow, in 1989. In 2001, Prof. Leith moved to the National University of
Ireland, Maynooth to assume the position of SFI Principal Investigator and to establish
the Hamilton Institute (www.hamilton.ie) of which he is Director.  His current research
interests  include the analysis and design of network congestion control and distributed
resource allocation in wireless networks.
\end{IEEEbiography}

\begin{IEEEbiography}{David Malone}
received B.A.(mod), M.Sc. and Ph.D. degrees in mathematics from Trinity College Dublin.
During his time as a postgraduate, he became a member of the FreeBSD development team. He
is a research fellow at Hamilton Institute, NUI Maynooth, working on wireless networking.
His interests include wavelets, mathematics of networks, IPv6 and systems administration.
He is a co-author of O'Reilly's "IPv6 Network Administration".
\end{IEEEbiography}

\end{document}